\RequirePackage{lineno} 

\documentclass[twocolumn,showpacs,preprintnumbers,amsmath,amssymb,nofootinbib]{revtex4}

\usepackage{graphicx}
\usepackage{dcolumn}
\usepackage{bm}

\usepackage{epstopdf}
\def\bea{\begin{eqnarray}}
\def\eea{\end{eqnarray}}

\def\pp{\mbox{$p$-$p$}}
\def\pa{\mbox{$p$-$A$}}

\def\ha{\mbox{$h$-$A$}}
\def\auau{\mbox{Au-Au}}

\def\pbpb{\mbox{Pb-Pb}}
\def\aa{\mbox{$A$-$A$}}
\def\nn{\mbox{$N$-$N$}}

\def\ee{\mbox{$e^+$-$e^-$}}
\def\ppbar{\mbox{$p$-$\bar p$}}

\def\pt{$p_t$}
\def\mt{$m_t$}
\def\yt{$y_t$}
\def\nch{$n_{ch}$}

\begin{document} 

\setlength{\pdfpagewidth}{8.5in}
\setlength{\pdfpageheight}{11in}

\setpagewiselinenumbers
\modulolinenumbers[5]

\preprint{Version 2.6}

\title{
Charge-multiplicity and collision-energy dependence of $\bf p_t$ spectra from $p$-$p$ collisions\\ at the relativistic heavy-ion collider and large hadron collider
}

\author{Thomas A.\ Trainor}\affiliation{CENPA 354290, University of Washington, Seattle, Washington 98195}


\date{\today}

\begin{abstract}
A two-component (soft + hard) model (TCM) of hadron production in yields and spectra derived from  the charge-multiplicity dependence of 200 GeV {\it p-p} collisions at the relativistic heavy ion collider (RHIC) is extended to describe {\it p-p} spectrum data from the large hadron collider (LHC) up to 13 TeV. The LHC data include spectrum ratios that provide only partial information on the TCM. The LHC ratio method is applied to well-understood 200 GeV spectrum data to derive an algebraic link between spectrum ratios and the full TCM. Some aspects of the form of the hard component on transverse momentum are found to be $n_{ch}$ dependent. LHC spectrum ratios are then analyzed to obtain $n_{ch}$ and collision-energy (over three orders of magnitude) dependence of isolated soft and hard TCM spectrum components. The energy dependence of the spectrum soft component is a new result suggesting a relation to Gribov diffusion. The spectrum hard component varies simply and smoothly with $n_{ch}$ suggesting bias of the underlying jet spectrum and linearly with QCD parameter $\log(s/s_0)$, its properties consistent with minimum-bias reconstructed-jet spectrum measurements. 
\end{abstract}


\maketitle

 \section{Introduction} \label{intro}

A two-component (soft + hard) model (TCM) of hadron production near mid-rapidity was derived from the charge-multiplicity \nch\ dependence of \pt\ spectra from 200 GeV \pp\ collisions~\cite{ppprd}. The TCM, interpreted to represent longitudinal projectile-nucleon dissociation and minimum-bias (MB) transverse dijet production near midrapidity, has been extended recently to consider the \nch\ dependence of \pt-integral angular correlations~\cite{ppquad}. Analysis of new high-statistics  \pt\ spectra in the latter study gave results quantitatively consistent with the earlier spectrum analysis. The \nch\ dependence of 2D angular correlations revealed a significant nonjet quadrupole ($v_2$) component as a novel aspect of \pp\ collisions. The \nch\ dependence of yields, spectra and correlations has played a key role in establishing (a) the nature of hadron production mechanisms in \pp\ collisions and (b) that the dijet contribution to \pt\ spectra is quantitatively consistent with predictions based on reconstructed jets~\cite{fragevo,jetspec,jetspec2}. 

The phenomenology of \pp\ collision data serves as an essential reference for high-energy \pa\ and \aa\ collisions, specifically regarding claims of novel physical mechanisms such as formation of a quark-gluon plasma~\cite{perfect} or possible manifestations of hydrodynamic flows (``collectivity'') even in small collision systems~\cite{ppflow,moreppflow}. It is important therefore to extend the TCM for \pp\ collisions formulated with data from the relativistic heavy ion collider (RHIC) to higher energies with analysis of \pp\ \pt\ spectra from the large hadron collider (LHC). The question of recently-claimed collectivity or flows in small (\pp\ and \pa) systems may be addressed in terms of the resolved TCM soft and hard components and evidence (or not) for radial flow in differential study of \pt\ spectra~\cite{hardspec}. 

In Ref.~\cite{alicespec} analysis of the \nch\ dependence of 13 TeV \pt\ spectra was presented with indirect reference (via a spectrum ratio) to a 7 TeV spectrum. The relation to previous TCM results  is not clear due to the spectrum-ratio strategy adopted. Comparisons with Monte Carlo (MC) models seem inconclusive. Those results raise significant questions about spectrum analysis: what information is conveyed by \pt\ spectra, what is the best method to extract all significant information, how should spectrum information be used to test theoretical models?

An initial study of the \nch\ dependence of \pt\ spectra from 200 GeV \pp\ collisions established that based on their \nch\ dependence \pt\ spectra can be decomposed into two distinct components subsequently identified as ``soft'' (associated with projectile-nucleon dissociation) and ``hard'' (associated with minimum-bias large-angle parton scattering and jet formation)~\cite{ppprd}. The decomposition does not rely on imposed model functions. The shapes of the isolated data components were found to be approximately independent of multiplicity and could be modeled by simple functions if the soft component was expressed on transverse mass \mt\ and the hard component on transverse rapidity \yt. The 200 GeV TCM decomposition has been confirmed with higher-statistics data~\cite{ppquad}. The present study shows that the TCM is {required} by data from a broad array of \pp\ collision systems.

The present study begins with a review of TCM methods as applied to \pt\ spectra in Refs.~\cite{ppprd,ppquad}. Spectrum data from the RHIC and LHC as used in this analysis are introduced, and the analysis strategy to be employed is described. In App.~\ref{200gevspec} spectrum-ratio methods from Ref.~\cite{alicespec} are applied to 200 GeV \pp\ spectra from Ref.~\cite{ppquad} to illustrate the consequences for a data system with established properties.  Results reveal that the TCM hard component has a substantial \nch\ dependence not previously accommodated. The hard-component \nch\ dependence is determined first for 200 GeV data and then 13 TeV data.  Combining those results with  spectra at 17.2 GeV and 0.9 TeV an accurate description of the energy dependence of \pp\ \pt\ spectra extending over all presently-accessible \pp\ collision energies is established. 
The resulting TCM  arguably represents all information carried by the \pt\ spectra of unidentified hadrons. The hard component is quantitatively related to the properties of isolated jets, and Monte-Carlo-simulated spectrum ratios are interpreted physically in relation to data systematics.

This article is arranged as follows:
Section~\ref{pptcm} introduces a TCM description of spectra and yields from \pp\ collisions.
Section~\ref{data} summarizes \pt\ spectra from RHIC \pp\ collisions based on the TCM in Refs.~\cite{ppprd,ppquad} and from LHC \pp\ collisions featuring spectrum ratios in Ref.~\cite{alicespec}.
Section~\ref{hardev} presents a revised hard-component model that describes the full \nch\ dependence of 200 GeV spectrum data.
Section~\ref{13tevspecc} applies the TCM to LHC  spectrum-ratio data to determine the \nch\ dependence of those spectra.
Section~\ref{edepp} describes the collision-energy evolution of \pp\ \pt\ spectra via the TCM.
Section~\ref{syserr} discusses systematic uncertainties.
Sections~\ref{disc} and~\ref{summ}  present discussion and summary.
Appendix~\ref{200gevspec} introduces TCM analysis of 200 GeV \pp\ spectra applying spectrum-ratio methods as in Ref.~\cite{alicespec}.
Appendix~\ref{900gev} compares a TCM-predicted 0.9 TeV \pt\ spectrum with data as a quantitative test of TCM energy dependence.
Appendix~\ref{lhcmult} reviews 13 TeV probability distributions on event multiplicity and multiplicity collision-energy dependence relevant to Ref.~\cite{alicespec}.

\section{A TCM for $\bf p$-$\bf p$ collisions}  \label{pptcm}

Final-state hadrons from high-energy nuclear collisions are distributed within a momentum space approximated near midrapidity ($\eta = 0$) by the cylindrical space $(p_t,\eta,\phi)$, where $p_t$ is transverse momentum, $\eta$ is pseudorapidity and $\phi$ is azimuth angle. Transverse mass is $m_t = \sqrt{p_t^2 + m_h^2}$ with hadron mass $m_h$. Pseudorapidity is  $\eta = -\ln[\tan(\theta/2)] $ ($\theta$ is polar angle relative to collision axis $z$), and $\eta \approx \cos(\theta)$ near $\eta = 0$. To improve visual access to low-$p_t$ structure and simplify description of the \pt\ spectrum hard component (defined below) spectra may be presented on transverse rapidity $y_t = \ln[(m_t + p_t) / m_h]$ with $p_t = m_h \sinh(y_t)$ and $m_t = m_h \cosh(y_t)$.  For unidentified hadrons $y_t$ with pion mass assumed (80\% of hadrons) serves as a regularized logarithmic $p_t$ measure $y_t \approx \ln(2p_t / m_h)$.
A typical acceptance limit $p_t > 0.15$ GeV/c corresponds to $y_t > 1$.

The \pp\ spectrum soft component is most efficiently described on transverse mass \mt\ whereas the spectrum hard component is most efficiently described on transverse rapidity \yt. The spectrum TCM thus requires a heterogeneous set of variables for its simplest definition. The components can be easily transformed from one variable to the other by a Jacobian factor defined below.

\subsection{TCM context}

The two-component model of hadron production in high-energy nuclear collisions has been reviewed in Refs.~\cite{ppprd,pptheory} for \pp\ collisions and Refs.~\cite{hardspec,fragevo,anomalous} for \aa\ collisions. The TCM represents both a mathematical data model and a system of inferred data components isolated via their systematic properties. The TCM model functions may serve as a reference for interpretation of data properties based on comparisons with theory (e.g.\ Ref.~\cite{fragevo}). Differences between data and TCM reference functions may reveal novel physical mechanisms.  

The TCM applied to elementary collisions has been interpreted to represent two principal sources of final-state hadrons near midrapidity: longitudinal projectile-nucleon dissociation (soft) and large-angle-scattered (transverse) parton fragmentation (hard). 
In \aa\ collisions the two processes scale respectively proportional to $N_{part}$ (participant nucleons $N$) and $N_{bin}$ (\nn\ binary encounters) as determined by a Glauber Monte Carlo model. Analogous scalings for \pp\ collisions in terms of charge multiplicities, as described in Sec.~\ref{ppdijet}, have been described in Refs.~\cite{pptheory,ppquad}. 
The TCM accurately represents hadron yield and spectrum systematics~\cite{ppprd,hardspec} and related aspects of  angular correlations~\cite{porter2,porter3,ppquad,anomalous}.

\subsection{TCM for p-p single-particle $\bf p_t$ or $\bf y_t$ spectra} \label{ppspec2}

The joint single-charged-particle (SP) 2D (azimuth integral) density on $y_t$ and $\eta$ is denoted by  $\rho_0(y_t,\eta) = d^2 n_{ch}/ y_t dy_t d\eta$. The $\eta$-averaged (over $\Delta \eta$) \yt\ spectrum is $\bar \rho_0(y_t;\Delta \eta)$. The $y_t$-integral mean angular density is $\bar \rho_0(\Delta \eta) = \int dy_t y_t \bar \rho_0(y_t;\Delta \eta) = n_{ch} / \Delta \eta$  averaged over acceptance $\Delta \eta$. 
According to the \pt\ spectrum TCM hadron density $\bar \rho_0$ has soft $\bar \rho_s$ and hard $\bar \rho_h$ components related by $\bar \rho_h = \alpha \bar \rho_s^2$ for $\alpha = O(0.01)$ and  $\bar \rho_0 = \bar \rho_s + \bar \rho_h$~\cite{ppprd}. Given some hypothesis $\alpha$ determined by spectrum analysis (see below) $\bar \rho_s$ and $\bar \rho_h$ can be obtained for any $\bar \rho_0$ as the solutions to a quadratic equation. For 200 GeV the value determined by spectrum analysis is $\alpha \approx 0.006$~\cite{ppprd,ppquad}.

The $n_{ch}$ dependence of \pp\ SP $p_t$ spectra over a large \nch\ interval (10-fold increase) was established  in Ref.~\cite{ppprd}.  Systematic variation of spectrum shapes leads to identification of two {approximately} fixed forms whose relative amplitudes vary smoothly with $n_{ch}$ (approximately linear and quadratic respectively for soft and hard). The TCM was not assumed initially, emerged instead from an inductive study. The relation of the hard component to isolated-jet properties was established in Ref.~\cite{fragevo}. The soft component in \auau\ collisions retains its fixed form but the hard-component form changes substantially with centrality, reflecting {\em quantitative} jet modification~\cite{fragevo}.

A TCM for \pp\ $y_t$ spectra conditional on uncorrected $n_{ch}'$ integrated over $2\pi$ azimuth and averaged over some $\eta$ acceptance $\Delta \eta$ was defined in Ref.~\cite{ppprd} by
\bea \label{ppspec}
\bar \rho_0(y_t;n_{ch}') &=&  S(y_t;n_{ch}') + H(y_t;n_{ch}')
\\ \nonumber
&\approx& \bar \rho_s(n_{ch}') \hat S_0(y_t)  +  \bar \rho_{h}(n_{ch}') \hat H_0(y_t),
\eea
where  $\bar \rho_s = n_s / \Delta \eta$ and $\bar \rho_h = n_h / \Delta \eta$ are corresponding $\eta$-averaged soft and hard hadron densities. The soft and hard $y_t$ spectrum shapes [unit normal $\hat S_0(y_t)$ and $\hat H_0(y_t)$] inferred from data were assumed to be independent of $n_{ch}'$, with parametrized forms defined in Refs.~\cite{ppprd,hardspec,fragevo}.  
Conversion from densities on \pt\ or $m_t$ to densities on \yt\ is easily accomplished via the Jacobian factor $p_t m_t / y_t$.

The fixed unit-normal soft component is most efficiently approximated by a L\'evy distribution on $m_t$
\bea \label{s0}
\hat S_0(m_t) &\equiv& \frac{A(T,n)}{[1 + (m_t - m_h) / (n T)]^n},
\eea
with hadron mass $m_h$, slope parameter $T$ and L\'evy exponent $n$, that goes to a Maxwell-Boltzmann exponential on \mt\ in the limit $1/n \rightarrow 0$. $\hat S_0(m_t)$ describes spectra normalized as ratios $\bar \rho_0(y_t;n_{ch}')/\bar \rho_s$ in the limit $n_{ch}' \rightarrow 0$. The L\'evy distribution may represent a system near equilibrium represented by parameter $T$, with additional parameter $1/n$ measuring deviations from full equilibrium~\cite{wilk}. 

The fixed unit-normal hard-component model $\hat H_0(y_t)$ is most efficiently approximated by a Gaussian plus exponential tail on \yt\ determined by Gaussian centroid $\bar y_t$, Gaussian width $\sigma_{y_t}$ and ``power-law'' parameter $q$. The slope is required to be continuous at the transition point from Gaussian to exponential. An algorithm for computing $\hat H_0(y_t)$ is provided in Ref.~\cite{hardspec} (App.\ A). Whereas $\hat H_0(y_t)$ on \yt\ has an exponential form $\propto e^{ - q y_t}$ at larger \yt\ the corresponding $\hat H_0(p_t)$ approximates the power-law form $\propto 1/p_t^{q + 2}$ at larger \pt\ (hence ``power-law'' tail). Hard component $H(y_t) = \alpha \bar \rho_s^2 \hat H_0(y_t)$ is well-approximated  by measured \pp\ fragmentation functions convoluted with a measured 200 GeV minimum-bias (MB) jet spectrum~\cite{fragevo,jetspec2}. Note that the two TCM ``power-law'' exponents $n$ and $q$ represent distinct soft and hard hadron production mechanisms near $\eta = 0$.

The fixed hard-component model developed in previous studies as described above is revised twice in the present study in response to higher-statistics 200 GeV spectrum data, first to accommodate data above the hard-component mode in Sec.~\ref{hcev} and then  to accommodate data below the mode in Sec.~\ref{hcbelow}. Hard-component parameters  $\sigma_{y_t}$ and $q$ are allowed to vary smoothly with \nch\ as  required by the spectrum data. Centroid $\bar y_t$ does not vary significantly according to data.




\subsection{p-p hard components and dijet production} \label{ppdijet}

Based on a dijet interpretation for the spectrum hard component~\cite{fragevo,hardspec,jetspec} the corresponding yield should be  $\bar \rho_h  \equiv \epsilon(\Delta \eta) f(n_{ch}') 2\bar  n_{ch,j}$, where $f(n_{ch}')$ is the dijet frequency per collision and per unit $\eta$, $\epsilon(\Delta \eta) \in [0.5,1]$ is the average fraction of a dijet appearing in acceptance $\Delta \eta$ and $2\bar  n_{ch,j}$ is the mean dijet fragment multiplicity. 
For 200 GeV non-single-diffractive (NSD) \pp\ collisions with $ \bar \rho_s \approx 2.5$ and mean fragment multiplicity $2\bar n_{ch,j} \approx 2.5 \pm 0.5$ inferred from measured jet systematics~\cite{ppprd} frequency $f_{NSD} = 0.006\times  2.5^2 /(0.55 \times 2.5) \approx 0.027$ is inferred from \pp\ spectra integrated within $\Delta \eta = 2$~\cite{ppquad}. 

That value can be compared with results from isolated-jet measurements in the form  $f_{NSD} = (1/\sigma_\text{NSD}) d\sigma_\text{dijet} /  d\eta  \approx   (1/36.5~\text{mb}) \times 1~\text{mb}  \approx 0.028$ for 200 GeV \pp\ collisions~\cite{jetspec2} based on a measured jet spectrum~\cite{ua1} and NSD \pp\ cross section~\cite{ua5}. Measured NSD  hard-component density $\bar \rho_h$~\cite{ppprd} is thus quantitatively consistent with dijet systematics derived from eventwise-reconstructed jets~\cite{ua1,cdfjets,jetspec2}.
If a non-NSD \pp\ event sample with arbitrary mean $n_{ch}'$ is selected the dijet frequency should vary with soft hadron density $\bar \rho_s$ as
\bea \label{nj1}
f(n_{ch}') &\approx& 0.027  \left[\frac{\bar \rho_{s}(n_{ch}')}{\bar \rho_{s,NSD}}\right]^2
\eea
with $\bar \rho_{s,NSD} = 2.5$ for 200 GeV \pp\ collisions according to spectrum results from Ref.~\cite{ppprd}. The same quadratic production trend is observed for dijet manifestations in 2D angular correlations~\cite{ppquad}, further supporting the relation.


\section{$\bf p$-$\bf p$ $\bf p_t$ spectrum data} \label{data}

\pt\ spectra for the present study are obtained from SPS data at 17.2 GeV, RHIC data at 200 GeV and LHC data at several energies. The RHIC data are in the form of isolated spectra over an extended range of collision multiplicities whereas the LHC data are in the form of a few isolated spectra and spectrum ratios over a limited \nch\ range. The goal of the study is to obtain an accurate and self-consistent TCM parametrization for a broad range of event multiplicities and collision energies.

\subsection{RHIC $\bf p_t$ spectrum analysis}

The multiplicity dependence of \pt\ spectrum structure from 200 GeV \pp\ collisions was reported in Ref.~\cite{ppprd}, and the trend for angular correlation structure was reported more recently in Ref.~\cite{ppquad}. The latter also included an updated TCM spectrum analysis of high-statistics data.

\begin{table}[h]
  \caption{Multiplicity classes based on observed (uncorrected) multiplicity $n_{ch}'$ falling within acceptance $|\eta| < 1$ or $\Delta \eta = 2$. The efficiency-corrected density is $\bar \rho_0 = n_{ch} / \Delta \eta$. Event numbers are in millions (M = $1 \times 10^6$).  The table entries are based on $\alpha = 0.006$ and tracking efficiency $\xi = 0.66$.  
}
  \label{multclass}
\begin{center}
\begin{tabular}{|c|c|c|c|c|c|c|c|} \hline
 Class $n$ &1 & 2 & 3 & 4 & 5 & 6 & 7 \\ \hline
 $n_{ch}'$  &2-3& 4-6 & 7-9 & 10-12 & 13-17 & 18-24 & 25-50 \\ \hline
$\langle n_{ch}' \rangle $ & 2.52 & 4.87 & 7.81 & 10.8 & 14.3  & 19.6 & 26.8 \\ \hline
 $\bar \rho_0(n_{ch}')$ & 1.90 & 3.65 & 5.82& 8.00 &  10.6 & 14.3 & 19.3 \\ \hline 
 $\bar \rho_s(n_{ch}')$ & 1.88  & 3.57 & 5.63 & 7.65 & 9.96 & 13.3 & 17.5 \\ \hline 
Events (M) & 2.31 & 2.21 & 0.91 & 0.33  & 0.14 & 0.02 & 0.001  \\ \hline
\end{tabular}
\end{center}
\end{table}

Table~\ref{multclass} defines seven multiplicity classes for the study in Ref.~\cite{ppquad} that apply to the spectrum data  considered below. $n_{ch}'$ is an uncorrected multiplicity within $\Delta \eta = 2$ related to corrected multiplicity \nch\ by $n_{ch}' = \xi n_{ch}$. The  $\langle n_{ch}' \rangle$ are distribution-weighted bin mean values. $\bar \rho_0$ and $\bar \rho_s$ are corrected for efficiencies and \pt\ acceptance. $n_{ch}'$ in this text replaces symbol $\hat n_{ch}$ from Ref.~\cite{ppprd}.

\begin{figure}[h]
  \includegraphics[width=1.65in,height=1.6in]{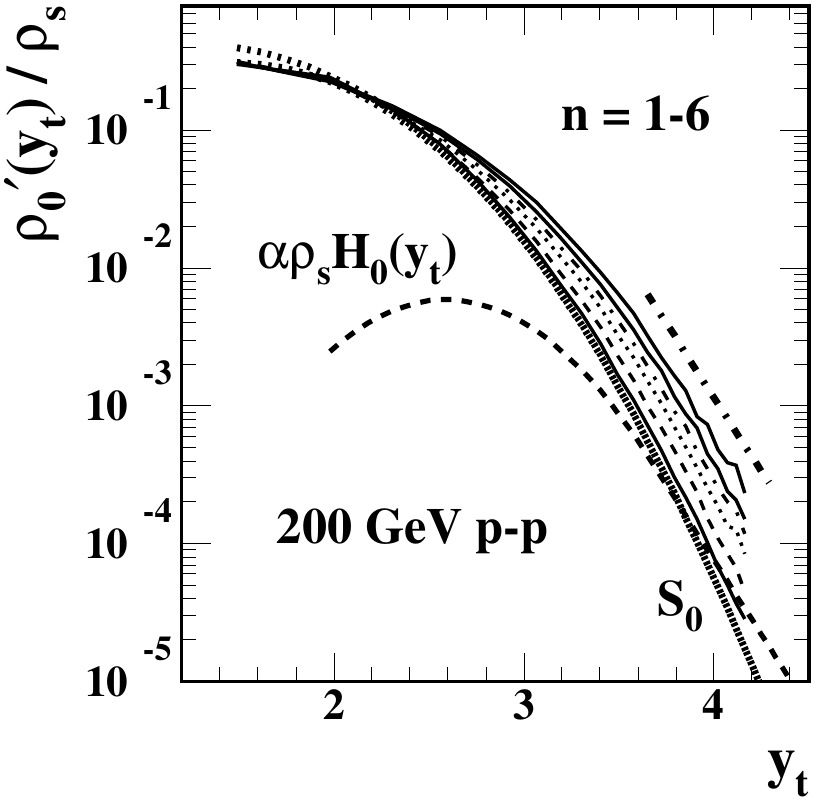}
  \includegraphics[width=1.65in,height=1.6in]{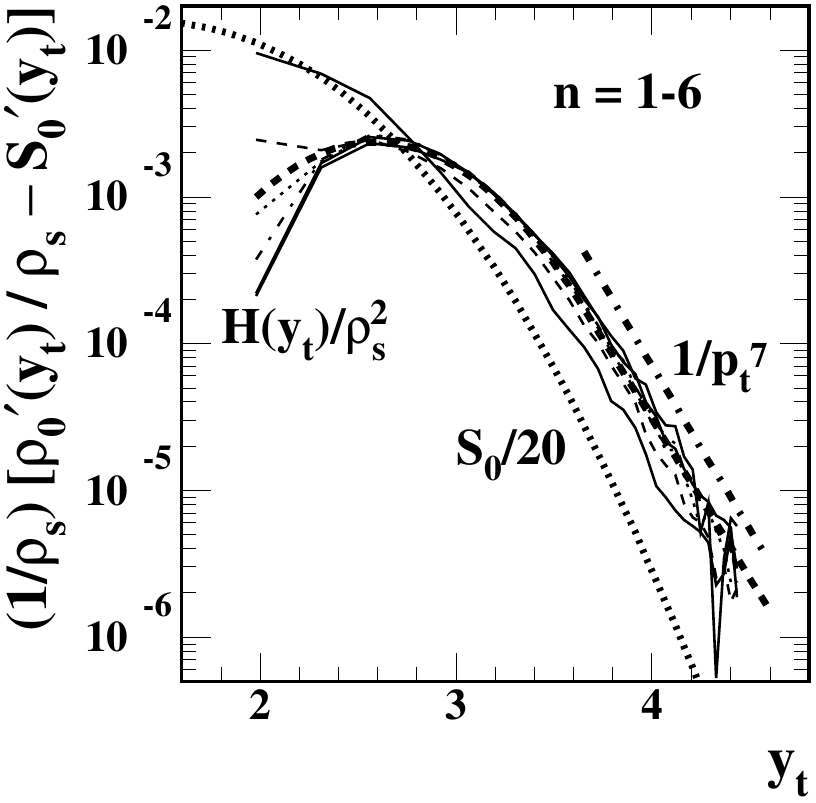}
\caption{\label{ppspec1}
Left: Normalized \yt\ spectra for six multiplicity classes of 200 GeV \pp\ collisions increasing near $y_t = 4$ with multiplicity class $n=1,\ldots,6$ ( see Table~\ref{multclass}). $\hat S_0(y_t)$ is the soft-component model function for corrected (upper dotted) and uncorrected (lower dotted) data. $\bar \rho_s$ is the corrected soft-component multiplicity assuming $\alpha = 0.006$ (see text), and the spectra are averaged over acceptance $\Delta \eta = 2$. The bold dashed curve is a fixed hard-component model.
Right: Spectrum-data hard components for $n = 1$-6 in the form $H(y_t) / \bar \rho_s^2$ with $H(y_t)$ defined by Eq.~(\ref{ppspec}) (several line styles) compared to hard-component model function $\alpha \hat H_0(y_t)$ (bold dashed curve). Bars and carets are omitted from figure labels.
 }   
\end{figure}

Figure \ref{ppspec1} (left) shows $y_t$ spectra for six multiplicity classes. The spectra (uncorrected for tracking inefficiencies) are normalized by corrected soft component $\bar \rho_s(n_{ch}')$. A common $y_t$-dependent inefficiency function is introduced for comparison of this analysis with corrected spectra in Ref.~\cite{ppprd}, indicated below $y_t = 2$ by the ratio of two bold dotted curves representing uncorrected $S_0'(y_t)$ and corrected (unit-normal) $\hat S_0(y_t)$ model functions. Data spectra are represented by spline curves rather than individual points to emphasize systematic variation with $n_{ch}'$.
The bold dashed curve labeled $\alpha \bar \rho_s \hat H_0(y_t)$ (for $\bar \rho_s = 2.5$) estimates a {\em fixed} hard component for 200 GeV \pp\ collisions in relation to corresponding soft component $\hat S_0(y_t)$ (bold dotted curve). The NSD curves cross near $y_t = 3.75$ ($p_t \approx 3$ GeV/c) where $S(y_t) = H(y_t)$ in Eq.~(\ref{ppspec}).

Figure \ref{ppspec1} (right) shows normalized spectra from the left panel for six multiplicity classes  in  the form $[\bar \rho_0'(y_t) / \bar \rho_s - S_0'(y_t)] / \bar \rho_s \approx H(y_t) / \bar \rho_s^2$ based on TCM Eq.~(\ref{ppspec}). The bold dashed curve is fixed hard-component model  $\alpha \hat H_0(y_t)$ with $\alpha = 0.006$. Those results are consistent with Ref.~\cite{ppprd}.
The bold dash-dotted line at right in each panel represents power-law trend $\hat H_0(p_t) \propto 1/p_t^{7}$ reflecting (by hypothesis) the underlying jet spectrum over a {\em limited} jet energy interval~\cite{fragevo,jetspec2}.  
A UA1 200 GeV jet spectrum~\cite{ua1} is approximately $d\sigma_j / dp_{jet} \propto 1/p_{jet}^{6}$ near 10 GeV/c, as in Fig.~8 of Ref.~\cite{jetspec2}. The trend $\hat H_0(p_t) \propto dn_h / p_t dp_t \propto 1/p_t^7$ is then fully consistent with convoluting a fixed fragmentation-function (FF) ensemble with the underlying {\em measured} jet spectrum~\cite{fragevo}.

In App.~\ref{200gevspec} a revised  TCM defined in terms of {\em spectrum ratios} is applied to these 200 GeV data in preparation for analysis of LHC spectrum ratios. A fixed hard-component model is assumed in App.~\ref{200gevspec} but variations to accommodate spectrum data are explored in Sec.~\ref{hardev}.
 
\subsection{LHC $\bf p_t$ spectrum analysis} \label{lhc}

Reference~\cite{alicespec} (ALICE collaboration) reports \pt\ spectra from 13 TeV \pp\ collisions for several event-multiplicity classes in the form of spectrum ratios. It is acknowledged that both hard and soft QCD processes may play a role in hadron production (the two elements of the TCM) but they are not isolated in that study.
The study is based on less than 1.5 million (M) events (vs 3M accepted events for Ref.~\cite{ppprd} and 6M events for Ref.~\cite{ppquad}). Reference~\cite{ppprd} is cited but its detailed spectrum analysis  is not acknowledged.

The evolution of \pp\ \pt\ spectra with \nch\ and collision energy is studied via spectrum ratios that discard some of the information in the individual spectra as demonstrated below. \pp\ data are said to provide a reference for \aa\ data and particularly for spectrum ratio $R_{AA}$ intended to study jet modification in \aa\ collisions. But as a ratio $R_{AA}$ also discards essential information: in particular it conceals {\em most of the jet contribution} (whatever appears below 4 GeV/c) that is essential to understand  QCD processes in high-energy nuclear collisions~\cite{fragevo,nohydro}.

Measurements include extension of charge $\eta$ density to 13 TeV (Fig.~2), a \pt\ spectrum extending to 20 GeV/c (Fig~3), a spectrum ratio comparing 13 and 7 TeV spectra (Fig.~4) and spectrum ratios comparing 13 TeV \pt\ spectra from three multiplicity classes to a common reference (Fig.~5). The spectrum-ratio data are said to show ``...rich features when correlated with the charged-particle multiplicity....'' It is concluded that spectrum ratios in Fig.~5 demonstrate stronger correlation of spectra with \nch\ at higher \pt, but the structure of {\em individual} spectra varies most rapidly with \nch\ at the spectrum hard-component mode near 1 GeV/c as shown in Ref.~\cite{ppprd}. It is acknowledged that jets may play a role in \pt\ spectra, but no previous  analysis addressing that subject is considered (e.g. Refs.~~\cite{ppprd,fragevo,hardspec}).  Qualitative comments are offered about Monte Carlo comparisons with data.

\begin{figure}[h]
  \includegraphics[width=1.65in]{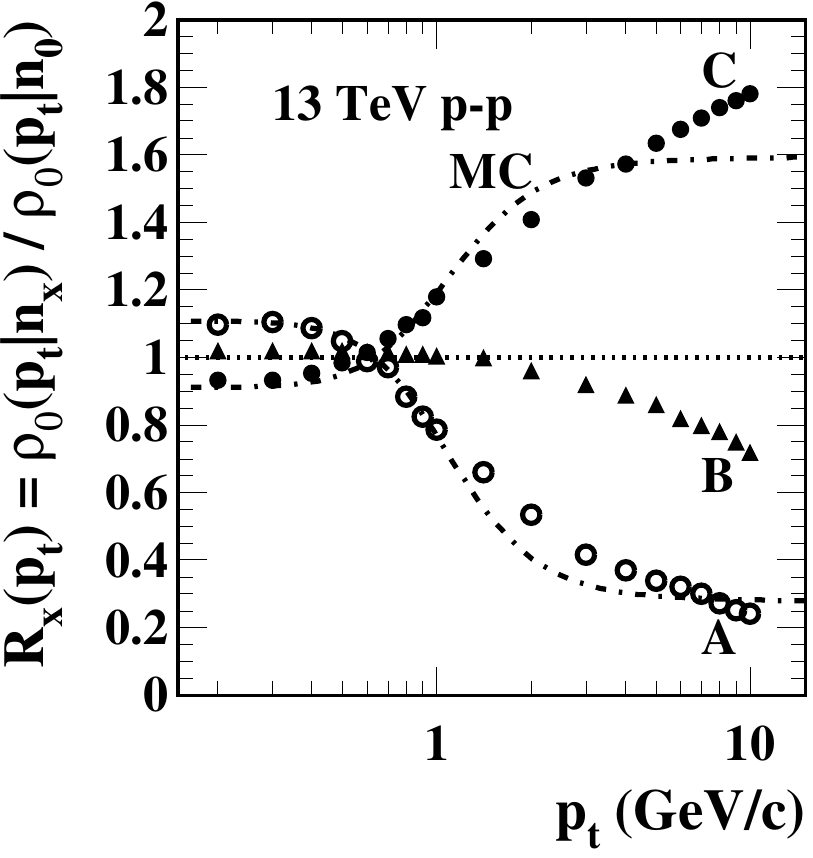}
  \includegraphics[width=1.65in]{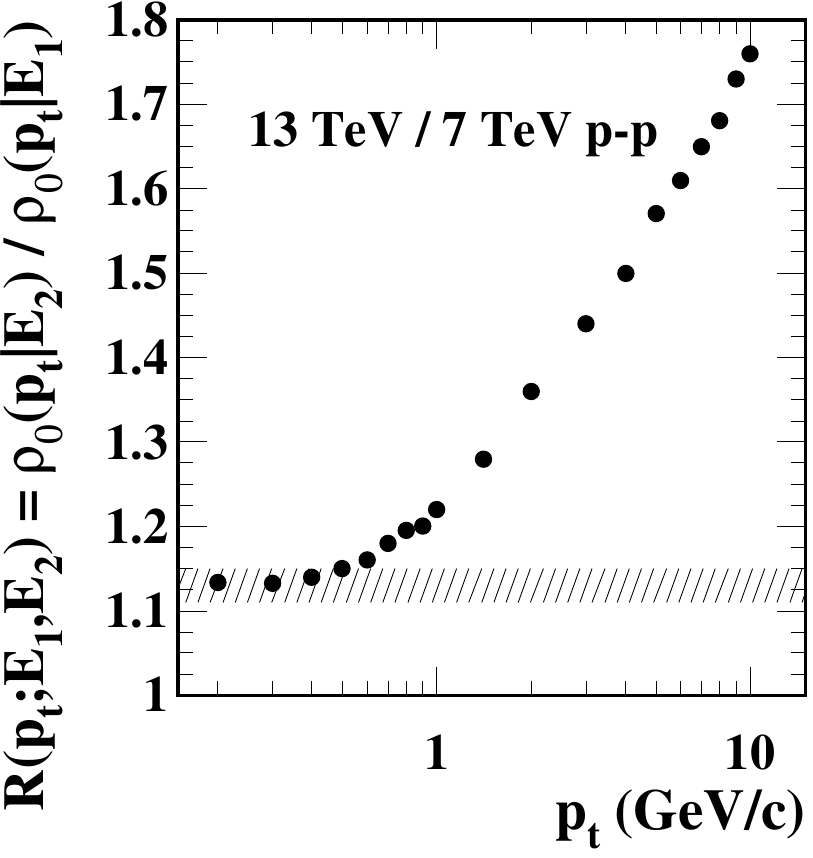}
\caption{\label{ratdata}
Left: Spectrum-ratio data obtained from Fig.~5 of Ref.~\cite{alicespec} (points) for three event multiplicity classes. 
The dash-dotted curves approximate MC results (see Sec.~\ref{13tevspecc}).
Right:  Data derived from a spectrum ratio comparing 13 and 7 TeV INEL $> 0$ spectra, from Fig.~4 of Ref.~\cite{alicespec} (points).
 } 
\end{figure}

Figure~\ref{ratdata} (left) shows ratio data from Ref.~\cite{alicespec} (points) for three $n_{ch}'$ conditions (multiplicity bins A, B and C in App.~\ref{lhcmult})  relative to an INEL $> 0$ (inelastic collisions with at least one detected particle in $\Delta \eta$) reference and ensemble-mean accepted $\bar n_{ch}' \rightarrow \bar \rho_{00}' = \bar \rho_{s0}' + \bar \rho_{h0}$. The statistical uncertainties are typically smaller than the data points. Tracking efficiencies are assumed to cancel in ratios, and  the \pt\ lower limit affects $\bar \rho_s$ but not $\bar \rho_h$ because the latter is localized on \pt. Limiting values at small \pt\ are $R \approx$ 1.12 for bin A, 1.02 for bin B and 0.91 for bin C. The dash-dotted curves approximate MC results (e.g.\ PYTHIA~\cite{pythia}) in Fig.~5 of Ref.~\cite{alicespec} as discussed in Sec.~\ref{13tevspecc}. 

Figure~\ref{ratdata} (right) shows a spectrum ratio comparing 13 and 7 TeV data (points) derived from  Fig.~4 of Ref.~\cite{alicespec}. The low-\pt\ limit of $R(p_t;E_1,E_2)$ is density ratio $\bar \rho_{s2} / \bar \rho_{s1} \approx 1.125$ (hatched band). Ratios of full spectra confuse soft and hard TCM components, are insensitive to energy-dependent jet physics obscured at lower \pt\ by the spectrum soft component, and the changes between 7 and 13 TeV are relatively small as demonstrated below. Improved sensitivity to jet physics could be obtained by analyzing spectra over a larger \nch\ interval as in Refs.~\cite{ppprd} and \cite{ppquad} and  energy interval as in the present study.

\subsection{Spectrum analysis strategy}

The TCM for \nch\ dependence of 200 GeV yields, spectra and correlations was established and confirmed in several papers~\cite{ppprd,ppquad,pptheory}. Those results were obtained from up to ten {\em isolated} event classes distributed over a broad \nch\ interval. The fixed soft component appears to be universal as confirmed in the present study. Previously the spectrum hard component,  a peaked distribution with exponential tail on \yt, was assumed fixed to preserve model simplicity. However, results from the present study reveal significant variation of hard-component parameters that should be accommodated in an updated TCM. Recent LHC data at higher energies include spectrum ratios over a more-limited \nch\ interval. A revised strategy is required to extract all information about \nch\ and energy dependence from the available spectrum data.

{\bf $\bf n_{ch}$ dependence:} The need for an \nch-dependent hard-component model is revealed by the 200 GeV spectrum-ratio study in App.~\ref{200gevspec}. Fortunately, the availability of individual high-statistics 200 GeV spectra (not ratios) over a broad \nch\ interval permits direct examination of TCM data trends in Sec.~\ref{hardev} to produce a revised hard-component model. To establish similar results for 13 TeV spectrum ratios requires a more intricate algebraic exercise as described in Sec.~\ref{13tevspecc}. The published ratios are transformed in several steps to isolate hard/soft data ratios $T(p_t;n_{ch}')\equiv \hat H_0(p_t;n_{ch}') / \hat S_0(p_t)$. The soft-component model $\hat S_0(p_t)$ for the higher energy is then determined from a single 13 TeV spectrum fit, and the  multiplicity-dependent TCM hard components $\hat H_0(p_t;n_{ch}')$ are finally isolated from inferred data ratios $T(p_t;n_{ch}')$.

{\bf Energy dependence:} With \nch-dependent TCMs defined at 200 GeV and 13 TeV it is possible to establish an energy-dependent TCM covering the full available range of \pp\ collision energies relevant to dijet production near midrapidity. In Sec.~\ref{esoft} an energy parametrization of soft-component L\'evy exponent $n$ is defined with the aid of SPS spectrum data at 17.2 GeV. In Sec.~\ref{ehard} an energy parametrization of hard-component exponent $q$ is defined based on the 200 GeV and 13 TeV results and a simple $\log(s/s_0)$ QCD trend for $1/q$. The energy dependence of $\bar y_t$ and $\sigma_{y_t}$ are supplemented by analysis of a 13 vs 7 TeV spectrum ratio. A key element is the energy dependence of TCM parameter $\alpha$ that relates soft and hard yields.  Its energy dependence is predicted in Sec.~\ref{parsum} based on measured jet-related QCD quantities and a revised trend for jet fragment multiplicities $2\bar n_{ch,j}$.

A major point of this exercise is a demonstration that the \nch\ and energy dependence of available spectrum data {\em require} a specific model inferred via an inductive study of data properties. The resulting TCM is not an arbitrary model based on hypotheses, is intimately related to measured jet properties and QCD expectations.

\section{200 G$\bf e$V $\bf p_t$ spectrum TCM $\bf vs$ $\bf n_{ch}$} \label{hardev}

A main achievement of Ref.~\cite{ppprd} was isolation of two spectrum components based on dramatically different scalings with \nch\ variation. The ``hard'' component was later interpreted as jet-related after comparison with measured jet properties (FFs and jet spectrum). It was noted that significant deviations from the TCM hard component occurred for the two lowest \nch\ classes.

In App.~\ref{200gevspec} TCM analysis of 200 GeV spectrum ratios reveals a substantial systematic \nch\ dependence of the hard component at higher \yt. In this section variations of hard-component parameters $\sigma_{y_t}$ and $q$ are found to accommodate spectrum data {\em above} the hard-component mode. But significant deviations from the fixed model {\em below} the mode are closely related: the entire hard-component shape is biased by a changing \nch\ condition. 
It is  desirable to develop a complete TCM description accommodating all aspects of \nch\ evolution.
The term ``data'' below refers to normalized spectra $\bar \rho_0(y_t;n_{ch}') / \bar \rho_s$.

\subsection{Hard-component $\bf n_{ch}$ evolution above the mode} \label{hcev}

Figure.~\ref{specrat2} (right) of App.~\ref{200gevspec} shows ratios of spectrum data from Fig.~\ref{ppspec1} (left) to the corresponding TCM expression in Eq.~(\ref{ppspec}) with fixed hard component.  Above $y_t = 3$ significant systematic variation (10\% increase per multiplicity class at 4 GeV/c) suggests that power-law exponent $q$ should decrease with increasing \nch, as might be expected if demand for larger multiplicities biases to more jet fragments by distorting the jet spectrum.

\begin{figure}[h]
  \includegraphics[width=1.65in]{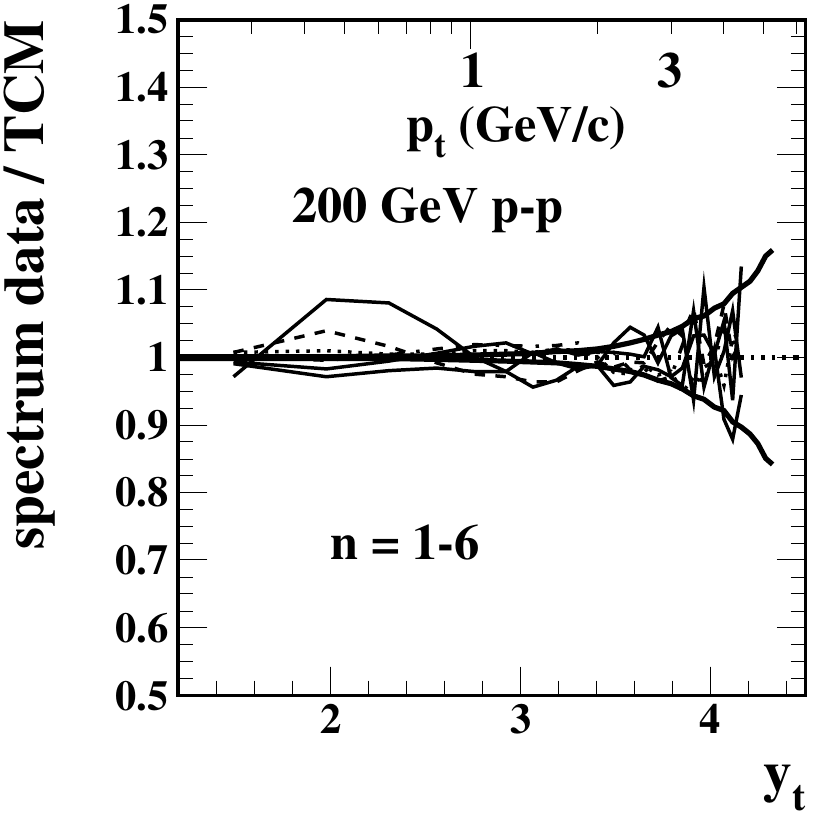}
  \includegraphics[width=1.65in]{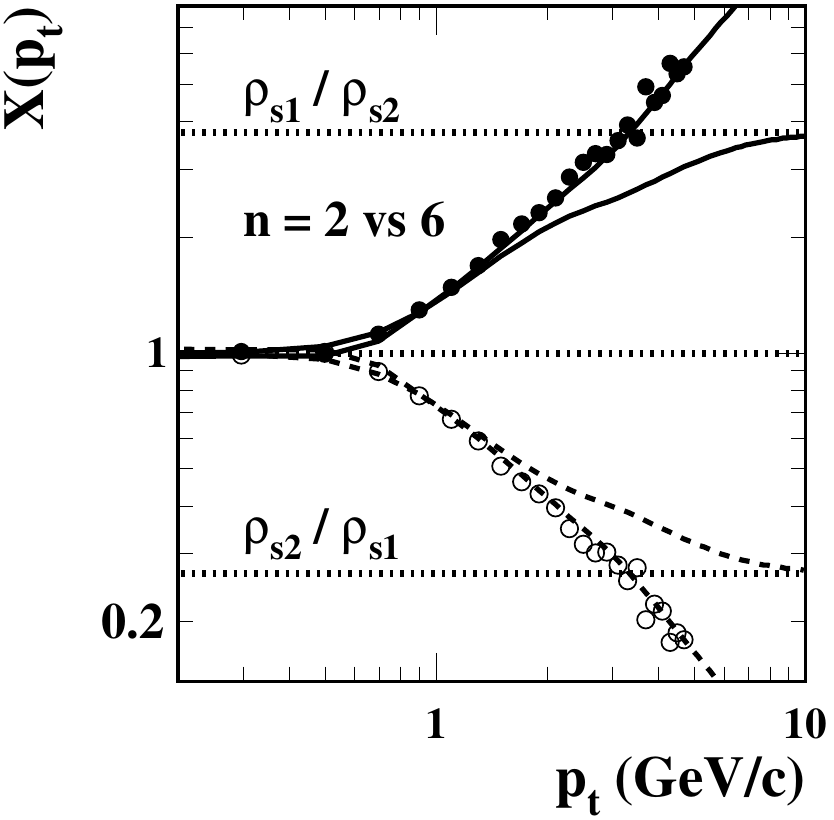}
\caption{\label{corrrat}
Left:  Ratios of data \yt\ spectra to TCM equivalents with a {\em varying} hard component for six multiplicity classes.  The data-model deviations at larger \yt\ are consistent with statistical uncertainties (bold solid curves symmetric about unity) and tracking errors. The overall unweighted r.m.s.\ deviation for curves 2-6 is 3\%.
Right: Figure~\ref{specrat1} (left) replotted with revised TCM including varying hard-component model (curves through points) and log-log plotting format.
 }  
\end{figure}

Figure~\ref{corrrat} (left) shows data/TCM ratios based on a revised TCM with two hard-component $\hat H_0(p_t)$ parameters  varying. Whereas $\hat H_0(y_t)$ was previously held fixed with power-law index $q = 5$ and width $\sigma_{y_t} = 0.465$ those parameters are now varied to accommodate individual spectra as described below. The modified TCM describes data {above the mode} within statistics (bold solid curves). 
Substantial spectrum deviations from the fixed TCM correlated with \nch\ appear in Fig.~\ref{ppspec1} (right) below the mode, but the corresponding manifestations in Fig.~\ref{corrrat} (left) below $y_t = 3$ are strongly suppressed by the ratio format. Further modification could also accommodate those deviations (with the exception of the $n = 1$ high solid curve).

Figure~\ref{corrrat} (right) repeats  Figure~\ref{specrat1} (left) with a revised TCM including varying $\hat H_0(y_t;q,\sigma_{y_t})$. The updated TCM ratios (new solid and dashed curves) pass through all data points (modulo statistical fluctuations), but the asymptotic limit in Eq.~(\ref{eqnx}) is no longer $\bar \rho_{s1} / \bar \rho_{s2}$ (dotted lines), is instead $\bar \rho_{s1}\hat H_0(p_t;q_1,\sigma_{y_t1}) / \bar \rho_{s2}\hat H_0(p_t;q_2,\sigma_{y_t2})$ confirming that dijet production changes significantly with increasing event multiplicity. The log-log format reveals the reciprocal relation of the two ratios whereas the linear format in Fig.~\ref{specrat1} gives the misleading impression that a ratio and its reciprocal carry different information.

\begin{figure}[h]
\includegraphics[width=1.62in]{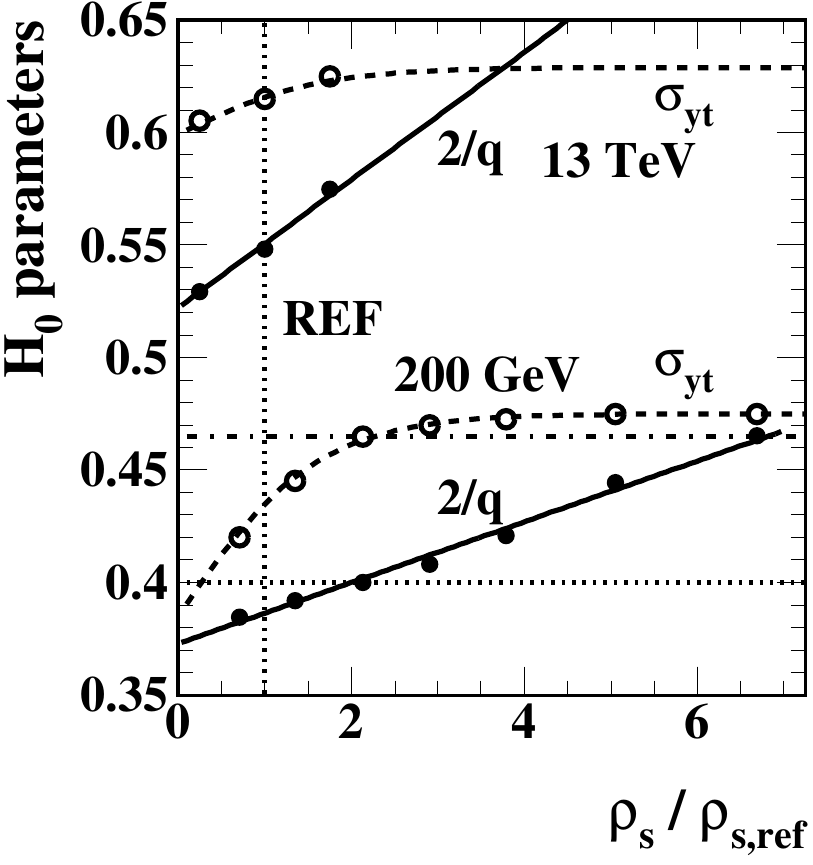}
    \includegraphics[width=1.7in]{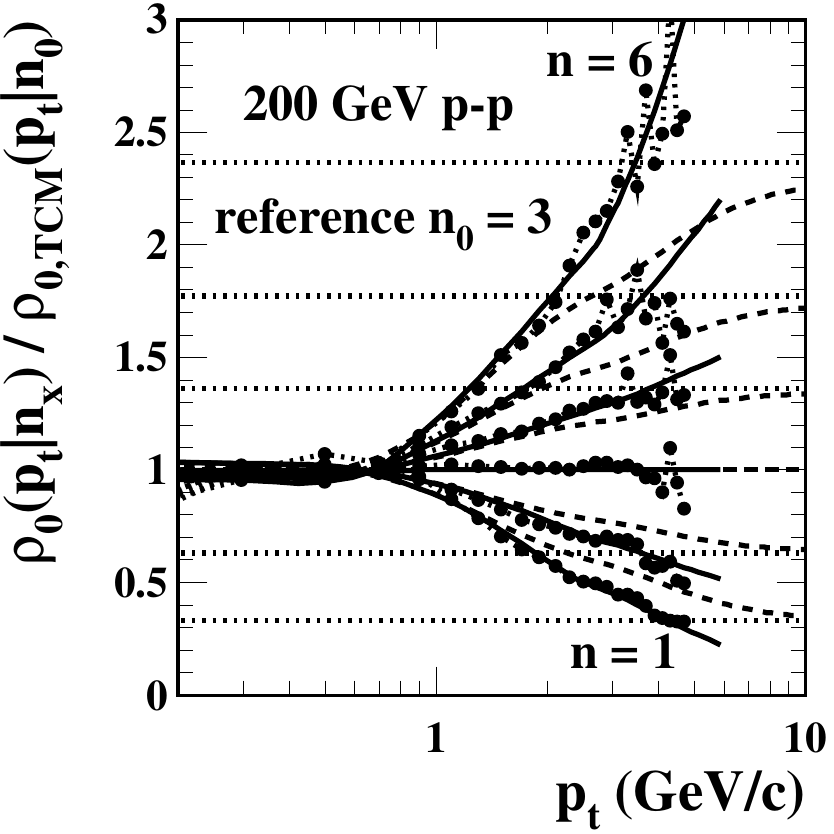}
\caption{\label{check}
Left: Hard-component parameters varying with $n_{ch}'$ or $\bar \rho_s$ for a revised TCM. 200 GeV solid and dashed curves through parameter data are defined by Eqs.~(\ref{parameq}). The 13 TeV points and curves are discussed in Sec.~\ref{13tevspec}. $\bar \rho_{s,ref} = 2.5$ for 200 GeV NSD \pp\ collisions and 6 for 13 TeV INEL $> 0$ (inelastic events with at least one charged particle accepted) collisions. The factor 2 in $2/q$ permits greater plot sensitivity.  The mean-value energy trend for $\sigma_{y_t}$ is shown in Fig.~\ref{params} (left) and for $1/q$ is shown in Fig.~\ref{enrat1} (left).
Right: Spectrum ratios for data (points) and TCM with varying hard component (solid curves) relative to an $n_0 = 3$ TCM reference, emulating Fig.~5 of Ref.~\cite{alicespec} and demonstrating the accuracy of the 200 GeV TCM. The dashed curves represent the fixed hard-component TCM of Refs.~\cite{ppprd,ppquad}.
 } 
\end{figure}

Figure~\ref{check} (left) shows the variation of two $\hat H_0(y_t)$ parameters with $\bar \rho_s$ (lower solid and dashed curves) that provides accurate description of spectrum ratios above the hard-component mode for all multiplicity classes. Optimized 200 GeV parameters follow simple $\bar \rho_s$ trends
\bea \label{parameq}
2/q &=& 0.373 + 0.0054 \bar \rho_s~~~\text{(solid)}
\\ \nonumber
\sigma_{y_t} &=& 0.385 + 0.09 \tanh(\bar \rho_s/4)~~~\text{(dashed)}.
\eea
The nominal parameter values for the 200 GeV fixed $\hat H_0(y_t)$ model  (bold dashed curves in Fig.~\ref{ppspec1}) are represented by the dotted and dash-dotted lines (corresponding to parameter values for $\bar \rho_s / \bar \rho_{s,ref} \approx 2$). Variation of two parameters in combination serves to broaden the hard-component model above the mode toward higher \pt. The saturation of $\sigma_{y_t}$ at larger $\bar \rho_s$ is a consequence of increasing $2/q$. The transition point on $\hat H_0(y_t)$ from Gaussian to exponential form then moves back toward the mode and the exponential/power-law tail increasingly dominates the higher-\pt\ structure. $\hat H_0(p_t)$ broadening could represent hardening of the underlying jet spectrum and/or modified jet formation. The corresponding parameter trends for 13 TeV are discussed in Sec.~\ref{13tevspec}.

Figure~\ref{check} (right)  provides a check on the overall consistency of the TCM description and corresponds in format to Fig.~5 of Ref.~\cite{alicespec} for comparison. The reference spectrum in this case is the TCM spectrum for $n_0 = 3$, with $\hat H_0(y_t)$ parameters that happen to coincide with the fixed model from Refs.~\cite{ppprd,ppquad}. Following Ref.~\cite{alicespec} each spectrum is normalized by its integral $\bar \rho_0'(n_{ch}')$. The solid curves represent TCM spectra with varying hard component. The revised TCM is accurate at the percent level for all multiplicity classes.  The dashed curves represent TCM spectra with fixed hard component and are retained for comparison with previous results. Dotted lines represent asymptotic limits $(\bar \rho_{sn}^2 / \bar \rho_{0n})(\bar \rho_{03}/\bar \rho_{s3}^2)$ for a spectrum-ratio TCM with fixed hard component. 
Note that in this case spectrum ratios compare six spectra to a single $n_0 = 3$ TCM reference whereas ratios in Figs.~\ref{specrat2} (right) and \ref{corrrat} (left) compare each data spectrum to a corresponding TCM spectrum for the same conditions. Also note that whereas comparison of classes 2 and 6 provides an {\em example} spectrum ratio the TCM for class 3 above serves as a {\em reference} to which all data are compared.

The present study demonstrates that high-statistics data from Ref.~\cite{ppquad} convey substantial new information about the dijet contribution to \pp\ hadron spectra that is simply represented by smooth variation of existing $\hat H_0(y_t)$ parameters. It also  illustrates the utility of the TCM as a reference relative to which novel data properties can be detected and interpreted. Data/TCM spectrum ratios are apparently more easily described and interpreted than data/data ratios presented without a reference.

\subsection{Optimized hard component below the mode} \label{hcbelow}

Figure~\ref{corrrat} (left)  shows high-statistics 200 GeV \pp\ spectra for six multiplicity classes from Ref.~\cite{ppquad} compared {in ratio} to the TCM with hard component varying as in Fig.~\ref{check} (left). One could conclude that the TCM data description is good except for the lowest $(n = 1)$ multiplicity class (high solid curve). However, two bold solid curves symmetric about unity indicate one-sigma statistical errors that become {\em very small in ratio} at lower \yt. Spectrum ratios tend to strongly suppress statistically-significant  information below the hard-component mode. To progress further requires a {\em differential} data-TCM comparison relative to bin-wise statistical errors.

\begin{figure}[h]
  \includegraphics[width=1.65in]{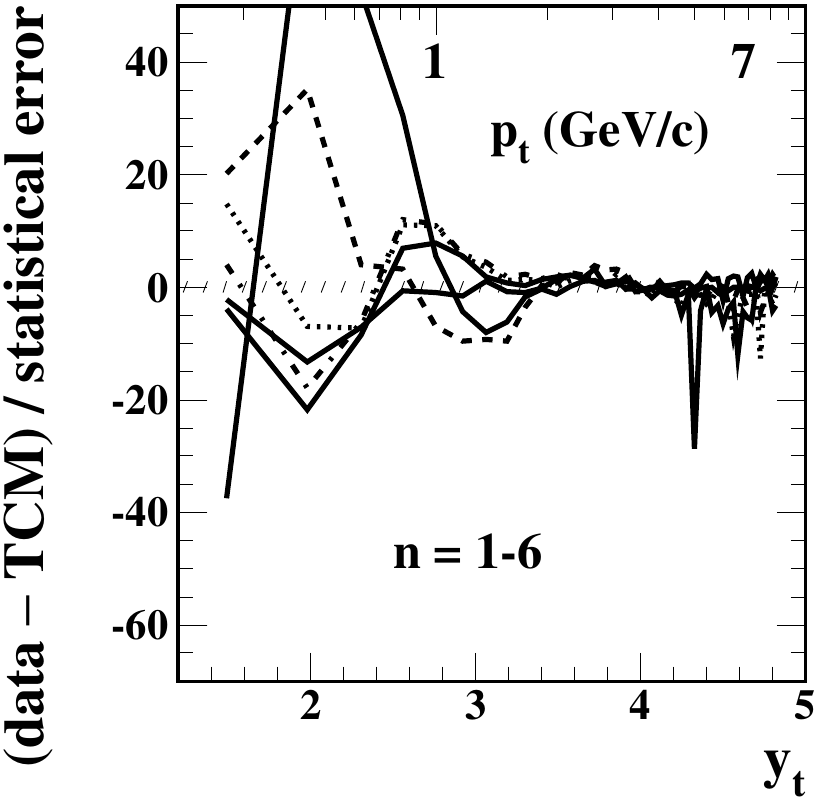}
  \includegraphics[width=1.65in]{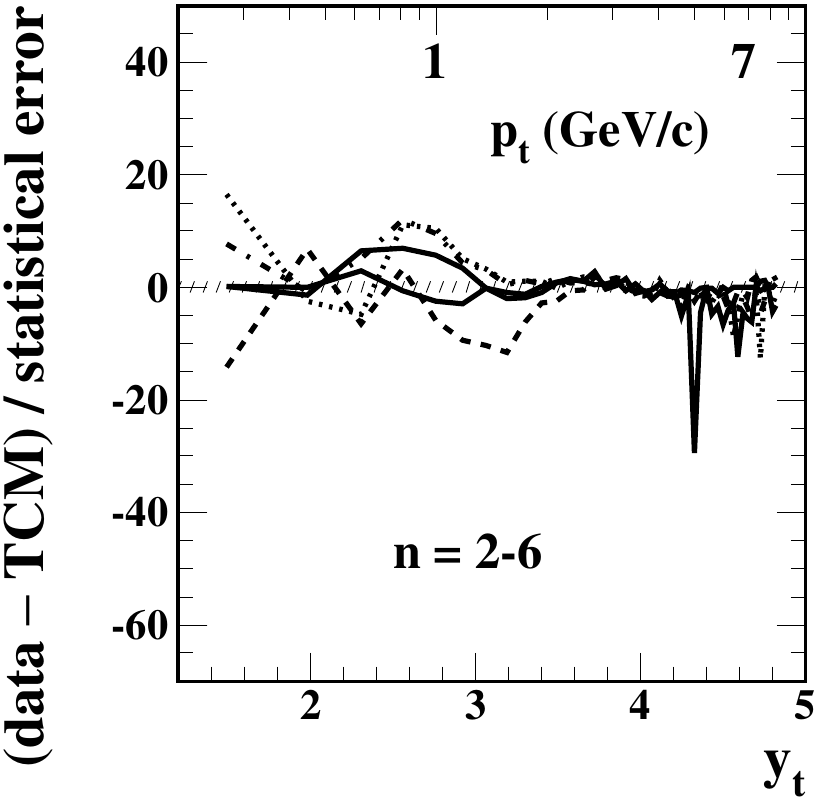}
\caption{\label{varyno}
Left:  The difference between spectrum data and the TCM from Sec.~\ref{hcev} compared to statistical errors as in Eq.~(\ref{poisson2}). Above the hard-component mode near $p_t = 1$ GeV/c residuals are consistent with statistical errors (hatched band). Below the mode there are large systematic excursions.
Right: Same format as the left panel but hard-component parameters have been further adjusted to accommodate data below the  mode as described in the text. The remaining low-\yt\ residuals are consistent with point-to-point systematic errors.
}   
\end{figure}

Figure~\ref{varyno} (left) is equivalent in principle to the ratios in Fig.~\ref{corrrat} (left)  but instead of the $\Delta \rho / \rho_{ref} + 1$ form of the latter this comparison has the {\em per-particle} form ${\Delta \rho}/{\sqrt{\rho_{ref}}}$ discussed in Sec.~\ref{significance}, comparable to Fig.~6 of Ref.~\cite{ppprd}. The hatched band about zero indicates one-sigma statistical errors {\em uniform on \yt\ in this plot format}. The variable hard component reduces residuals above the mode to the statistical level, but residuals below the mode remain {\em very large compared to statistical errors} and represent substantial spectrum information not accommodated within the revised TCM derived in Sec.~\ref{hcev}.

Figure~\ref{varyno} (right) shows the result of further modification of the hard-component model. Gaussian widths $\sigma_{y_t+}$ above and $\sigma_{y_t-}$ below the mode are varied separately. The values given in Fig.~\ref{check} (left) and Eq.~(\ref{parameq}) (lower) are retained for  $\sigma_{y_t+}$ above the mode, but the values for $\sigma_{y_t-}$ are varied independently (in the form $1/\sigma_{y_t-}^2$) to accommodate  data below the mode (except for $n = 1$). The resulting residuals for $n = 2$-6 are consistent with point-to-point systematic errors (about 1 {\em permil} of data values).

\begin{figure}[h]
 \includegraphics[width=1.66in]{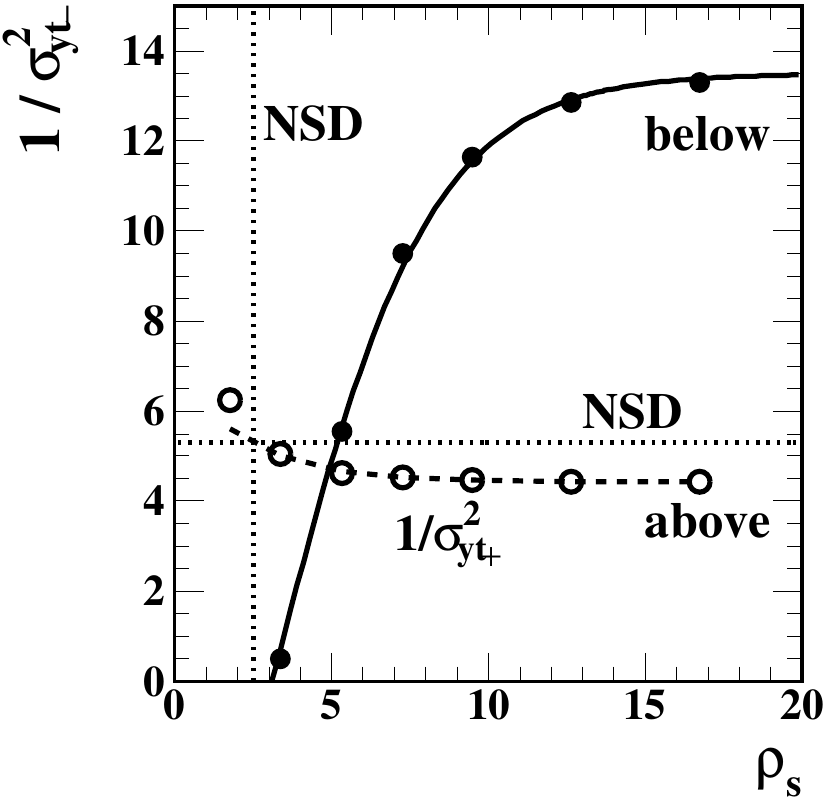}
  \includegraphics[width=1.64in]{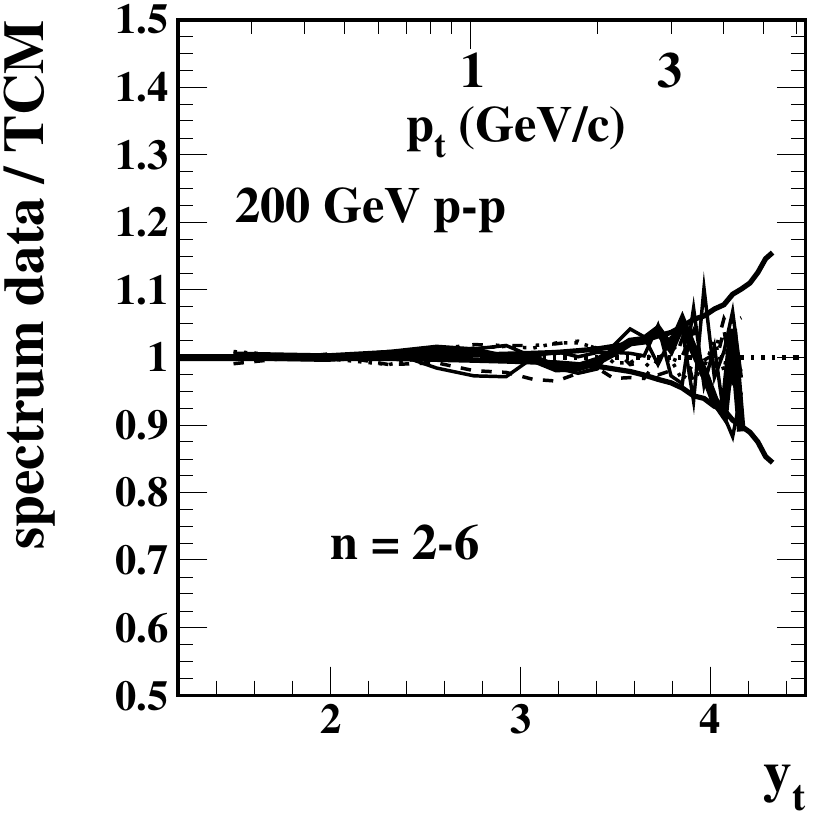}
\caption{\label{varyyes}
Left:  Variation of the Gaussian width below the hard-component mode $\sigma_{y_t-}$ (solid points) for $n = 2$-7 that accommodates data in that \yt\ interval, as in Fig.~\ref{varyno} (right). The Gaussian width above the mode $\sigma_{y_t+}$ (open points) as in Fig.~\ref{check} (left) is included for comparison. The curves are defined in the text. The correlation of two trends is notable.
Right:  Data in Fig.~\ref{varyno} (right) plotted as conventional spectrum ratios demonstrating that important information below 1 GeV/c tends to be concealed by such ratios.
}  
\end{figure}

Figure~\ref{varyyes} (left)  shows variation of the hard-component width required to accommodate data {\em below} the mode in the form $1/\sigma_{y_t-}^2$ (solid points). The solid curve through points is $13.5 \tanh[(\bar \rho_s - 3.1)/5]$.  Also included is the trend for the width above the mode from Fig.~\ref{check} (left) and Eq.~(\ref{parameq}) (lower) as $1/\sigma_{y_t+}^2$ (open points and dashed curve respectively) demonstrating correlation of the two trends. The  two widths become equal near $\bar \rho_s \approx 5$, or  $\bar \rho_s / \bar \rho_{s,ref} \approx 2$ in Fig.~\ref{check} (left), where the hard-component peak model is then approximately symmetric as in Refs.~\cite{ppprd,ppquad}.

Figure~\ref{varyyes} (right) shows final residuals in a conventional ratio format. Residuals below the mode are no longer visible (less than 1 {\em permil}) in the ratio format, illustrating the extent to which significant residuals structure at smaller \yt\ may be concealed by that format whereas relatively minor effects at larger \yt\ may be exaggerated.

\subsection{200 GeV TCM $\bf n_{ch}$ dependence summary}

\begin{figure}[h]
  \includegraphics[width=1.64in]{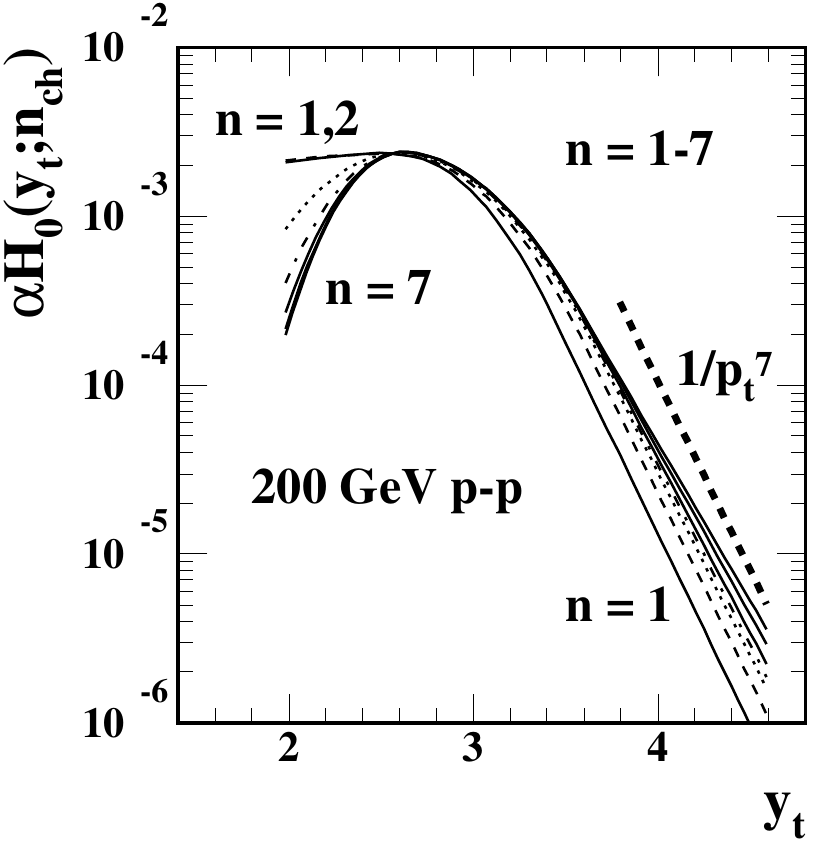}
  \includegraphics[width=1.65in]{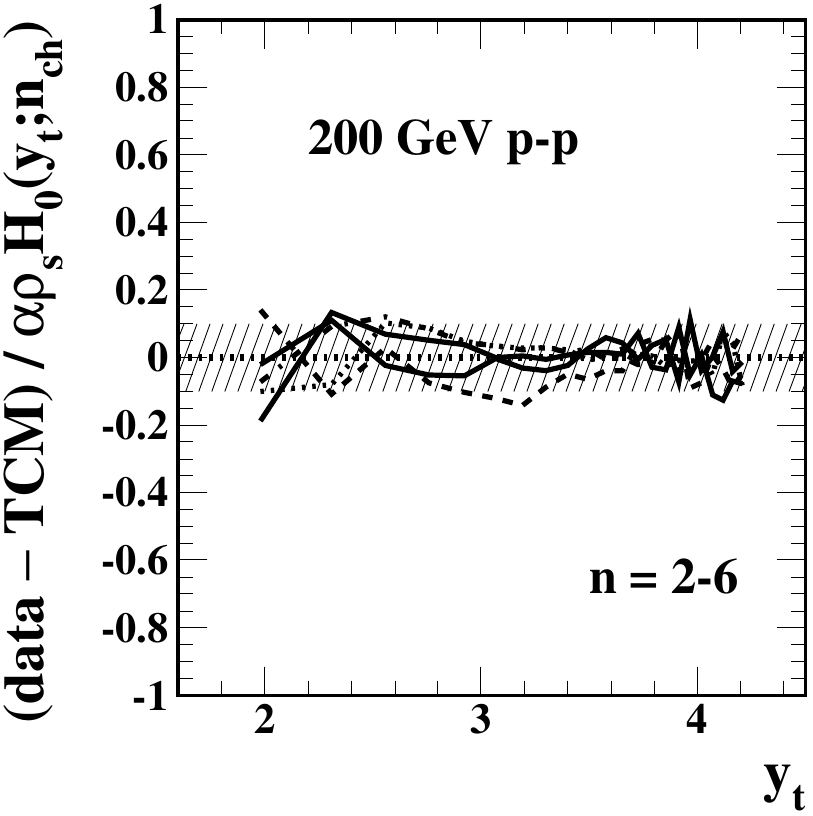}
\caption{\label{chi2x}
Left: Evolution of the hard-component model over seven multiplicity classes that exhausts all information in high-statistics spectrum data from Ref.~\cite{ppquad}. The model for $n = 1$ is undefined below the hard-component mode $y_t \approx 2.6$.
Right:  Difference between spectrum data and the full TCM relative to the hard-component model demonstrating that the latter is accurate to a few percent over the relevant \yt\ interval.
 } 
\end{figure}

Figure~\ref{chi2x}  (left)  summarizes the revised 200 GeV TCM hard-component model for seven multiplicity classes. The model for class $n = 1$ cannot accommodate data below the mode: the trend for $1/\sigma_{y_t-}^2$ in Fig.~\ref{varyyes} (left)  requires a negative entry for $n = 1$, and the shape of the model function would remain very different from the data. The value of $\sigma_{y_t-}$ for $n = 2$ is retained for $n = 1$. This model can be compared with data in Fig.~\ref{ppspec1} (right).

Figure~\ref{chi2x} (right)  shows the scaled differential $\Delta \bar \rho_0(y_t) / \bar \rho_s$ denoted  by expression  (data $-$ TCM) divided by the TCM hard component in the form $\alpha \bar \rho_s \hat H_0(y_t;n_{ch}')$. Whereas full-spectrum ratios [e.g.\ Fig.~\ref{corrrat} (left)] may be misleading this ratio to the hard-component model alone is informative, revealing  that residuals between full (soft + hard) data spectra and a full revised TCM are less than 10\% of hard-component values for $n = 2$-6 over the entire \yt\ range relevant to jet-related spectrum structure.

\section{13 T$\bf e$V $\bf p_t$ spectrum ratios $\bf vs$ $\bf n_{ch}$} \label{13tevspecc}

In this section the spectrum TCM is applied to 13 TeV LHC  \pt\  spectrum ratios from Ref.~\cite{alicespec} to infer separate soft- and hard-component spectrum models from those data. Relations derived for 200 GeV data in App.~\ref{200gevspec} are based on normalized spectra as defined in Ref.~\cite{ppprd} over a large \nch\ interval and include a known soft component and fixed hard-component model. Relations defined in this section are based on ratios of spectra as normalized in Ref.~\cite{alicespec} over a more-limited \nch\ interval with no soft-component model previously established but include an \nch-varying hard component as in the previous section. 
Expressions below including an ``$\equiv$'' symbol that define data quantities (points)  are followed by  corresponding TCM expressions including an ``$\approx$'' symbol (curves).

\subsection{13 TeV spectrum-analysis strategy}

The 13 TeV spectrum study in Ref.~\cite{alicespec} defined three multiplicity classes A, B and C as described in App.~\ref{multclass1}. Ratios were formed in which the \pt\ spectrum from each class is divided by the INEL $> 0$ ensemble-mean spectrum adopted as a reference. The results are presented in Fig.~5 of Ref.~\cite{alicespec} repeated in Fig.~\ref{ratdata} (left) of the present study. It is concluded that ``correlation of the spectrum with multiplicity...is stronger at high $p_T$....with jets presumably dominating the high-multiplicity domain.'' It is further noted that ``The general features...are similar to those first seen at $\sqrt{s} = 0.9$ TeV~\cite{alice9}.'' There is no reference to previous spectrum studies e.g.\ in Refs.~\cite{ppprd,fragevo,hardspec,jetspec} where {\em differential} analysis of \pp\ \pt\ spectrum \nch\ dependence and quantitative theoretical interpretations were first established, as reviewed  in Sec.~\ref{pptcm} and App.~\ref{200gevspec}.

Appendix~\ref{200gevspec}  demonstrates that for high-statistics 200 GeV \pp\ spectra 
a TCM with fixed hard component deviates significantly from ratio data at higher \pt.  A hard-component model with two parameters varying with \nch\ is {\em required} by those data.  For analysis of spectrum-ratio data from Ref.~\cite{alicespec} a varying hard component is assumed from the beginning to establish parameter \nch\ trends.

The spectrum ratios in  Fig.~\ref{ratdata} (left) provide only indirect information on TCM elements in the form of ratios $T(p_t;n_{ch}') \equiv \hat H_0(p_t;n_{ch}') / \hat S_0(p_t)$ per Eq.~(\ref{eqnr}) below. By suitable transformation of ratio data  ($R \rightarrow X \rightarrow Y \rightarrow T$ below) ratios $T(p_t;n_{ch}')$ for multiplicity bins A and C can be isolated as in Fig.~\ref{blog} (left) below (the bin-B ratio data provide no significant information). 

An intermediate parametrization to describe 13 TeV $T(p_t;n_{ch}')$ ratio data is defined in terms of 200 GeV $\hat S_0(p_t;T,n)$ by adjusting $\hat H_0(p_t;n_{ch}')$ model parameters to fit the inferred ratio data. That intermediate parametrization is not intended as a final description of 13 TeV spectra. From that parametrization a model $T_0(p_t)$ for the 13 TeV INEL $> 0$ reference spectrum is determined by interpolation.
To obtain the correct soft-component model for 13 TeV data the full reference spectrum in Fig.~3 of Ref.~\cite{alicespec} is then fitted by varying only parameter $n$ of $\hat S_0(p_t,n)$ in the modified TCM expression
\bea
 \bar \rho_0(p_t;n_{ch}') &=& \bar \rho_s(n_{ch}') \hat S_0(p_t,n)\left[1  + \alpha \bar \rho_{s}(n_{ch}')  T_0(p_t)\right],~~~
\eea
where $n_{ch}'$ here corresponds to INEL $> 0$. 

Given a correct 13 TeV $\hat S_0(p_t,n)$ model  the 13 TeV hard-component model is recovered from 13 TeV spectrum-ratio data by adjusting hard-component parameters to fit ratios $T(p_t;n_{ch}') =\hat H_0(p_t;n_{ch}') / \hat S_0(p_t,n)$ for bins A and C and the INEL $> 0$ spectrum, thus establishing a full \nch-dependent TCM for 13 TeV \pt\ spectra.

\subsection{Spectrum ratios vs $\bf n_{ch}$ and hard/soft ratio $\bf T(p_t)$}

Spectrum ratios $R(p_t;n_{ch}')$ are formed relative to a reference spectrum $\bar \rho_{00}'(p_t;\Delta \eta)$ (e.g.\ INEL $>0$) for each of three multiplicity bins indexed by $n_{ch}'$.
Spectra averaged over acceptance $\Delta \eta = 1.6$ are first normalized by their \pt\ integrals as $\bar \rho_0'(p_t;n_{ch}',\Delta \eta) / \bar \rho_0'(n_{ch}',\Delta \eta)$. The spectrum-ratio data are then represented by the first line of
\bea \label{eqnr}
R(p_t;n_{ch}') &\equiv&  \frac{\bar \rho_{00}'(\Delta \eta)}{\bar \rho_{0}'(n_{ch}',\Delta \eta)} \frac{\bar \rho_{0}'(p_t;n_{ch}',\Delta \eta)}{\bar \rho_{00}'(p_t;\Delta \eta)}
\\ \nonumber
&\approx &\left(\frac{\bar \rho_{s0}' + \bar \rho_{h0}}{\bar \rho_s' + \bar \rho_h}\right) \frac{\bar \rho_s \hat S_0(p_t) + \bar \rho_h \hat H_0(p_t;n_{ch}')}{\bar \rho_{s0} \hat S_0(p_t) + \bar \rho_{h0} \hat H_{00}(p_t)}
\\ \nonumber 
&=&  \left(\frac{1 + \alpha' \bar \rho_{s0}'}{1 + \alpha' \bar \rho_s'}\right) \left(\frac{\bar \rho_{s0}' \bar \rho_s}{\bar \rho_s' \bar \rho_{s0}}\right)   \frac{1 + \alpha \bar \rho_{s} T(p_t;n_{ch}')}{1 + \alpha \bar \rho_{s0} T_0(p_t)}
\\ \nonumber 
&\rightarrow& \left(\frac{1 + \alpha' \bar \rho_{s0}'}{1 + \alpha'\bar  \rho_s'}\right)~~~\text{at small \pt}.
\eea
The TCM for unit-normal spectrum ratios based on Eq.~(\ref{ppspec}) is described by the  last three lines, where the $\bar \rho_{x0}$ refer to the reference INEL $>0$ event class and primes refer to incomplete \pt\ acceptance. The middle factor in the third line is assumed to be unity based on efficiency cancellations. The ratio limiting value at small \pt\ (first factor) has the approximate form $[1 + \alpha' \bar \rho_{s0}'(1 - \bar \rho_{s} / \bar \rho_{s0})]$.

\begin{figure}[h]
  \includegraphics[width=1.65in]{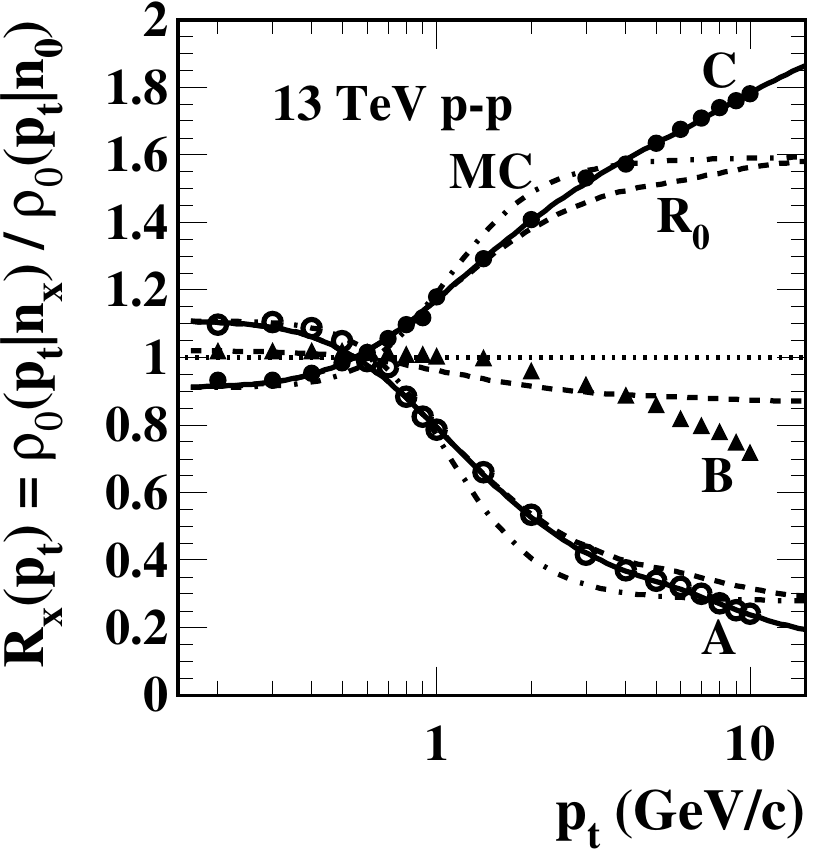}
  \includegraphics[width=1.65in]{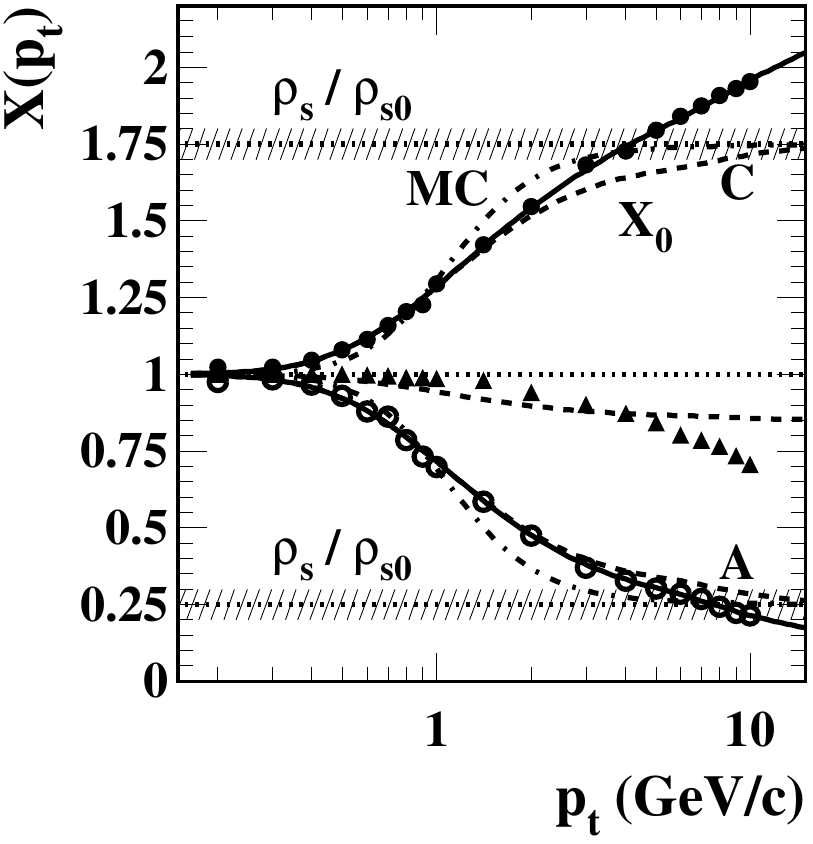}
\caption{\label{basicrat}
Left: Spectrum-ratio data obtained from Fig.~5 of Ref.~\cite{alicespec} (points) for three event multiplicity classes. The solid curves are generated by Eq.~(\ref{eqnr}) (third line) with varying TCM hard component as described in Sec.~\ref{13tevspec}.  The dashed curves are generated by Eq.~(\ref{eqnr}) with fixed TCM hard component $T(p_t) \rightarrow T_0(p_t)$ in the numerator. The dash-dotted curves approximate MC results (see text).
Right: Data from the left panel transformed as in the first line of Eq.~(\ref{eqnx2}) (points). The solid curves are generated by Eq.~(\ref{eqnx2}) (second line). The dashed curves correspond to a TCM with fixed hard component  $T(p_t) \rightarrow T_0(p_t)$ in the numerator. The dotted lines correspond to asymptotic limits $\bar \rho_s / \bar \rho_{s0}$.
 }  
\end{figure}

Figure~\ref{basicrat}  (left) shows ratio data from Fig.~\ref{ratdata} (left) (points) for three $n_{ch}'$ conditions (multiplicity bins A, B and C)  relative to the INEL $> 0$ reference and ensemble-mean accepted $\bar n_{ch}' \rightarrow \bar \rho_{00}' = \bar \rho_{s0}' + \bar \rho_{h0}$. 
Tracking efficiencies are assumed to cancel in ratios, and  the \pt\ lower limit affects $\bar \rho_s \rightarrow \bar \rho_s'$ but not $\bar \rho_h$ because of its localization on \pt. Limiting values at small \pt\ are $R \approx$ 1.12 for bin A, 1.02 for B and 0.91 for C. TCM curves are defined by Eq.~(\ref{eqnr}) with reference ratio $T_0(p_t)$ (dashed) and \nch-dependent $T(p_t;n_{ch}')$ (solid) defined below. The solid curve for case B does not appear (in this and later plots) because the $\bar \rho_s  /\bar \rho_{s0}$ ratio is close to unity and spectrum ratio B is relatively insensitive to hard-component structure. The dash-dotted curves represent MC results (e.g.\ PYTHIA~\cite{pythia}) in Fig.~5 of Ref.~\cite{alicespec}.

The general form of the ratio data is a clear manifestation of the spectrum TCM, as anticipated by Fig.~3 (left) of Ref.~\cite{ppprd} with spectra normalized by $\bar \rho_0$ ($n_{ch}$ within $\Delta \eta = 1$) rather than $\bar \rho_s$ as in Fig.~\ref{ppspec1} (left) of the present study. 
The excursions about unity at lower \pt, defined by $\alpha' \bar \rho_{s0}'(\bar \rho_s  /\bar \rho_{s0} - 1) \approx 0.1$ with $\alpha' \bar \rho_{s0}' \approx 0.15$, can be compared with similar excursions at 200 GeV in Fig.~\ref{check} (right) where $\alpha' \bar \rho_{s0}' \approx 0.025$ but the  $\bar \rho_s  /\bar \rho_{s0}$ range is larger.

Figure~\ref{basicrat} (right) shows the first intermediate quantity $X(p_t;n_{ch}')$ extracted from $R(p_t;n_{ch}')$ data (points) for conditions $n_{ch}'$ represented in Eq.~(\ref{eqnx2}) for data (first line) and defined for the TCM (second line)
\bea \label{eqnx2}
X(p_t;n_{ch}') &\equiv& R(p_t;n_{ch}') \left(\frac{1 + \alpha'\bar  \rho_s'}{1 + \alpha' \bar \rho_{s0}'}\right) \\ \nonumber
&&  \hspace{-.7in}   \approx \frac{1 + \alpha \bar \rho_s T(p_t;n_{ch}')}{1 + \alpha \bar \rho_{s0} T_0(p_t)}  \rightarrow \frac{\bar \rho_s}{ \bar \rho_{s0}}\frac{T(p_t;n_{ch}')}{T_0(p_t)}~~\text{(at large \pt)}
\eea
In App.~\ref{ppprdspecrat} ratio $X(p_t)$ emerges directly from  the spectrum ratio in Eq.~(\ref{eqnx}), thus bypassing quantity $R(p_t)$ because of the choice of normalization in Ref.~\cite{ppprd} and Fig.~\ref{ppspec}. Reference $X_0(p_t)$ (dashed) results if $T(p_t;n_{ch}') \rightarrow T_0(p_t)$ in the numerator.
The asymptotic limits at right (dotted lines) are $\bar \rho_s / \bar \rho_{s0} \approx$ 0.25 for bin A, 0.85 for bin B and 1.75 for bin C. Given corrected $\bar \rho_{s0} \approx 6$ for 13 GeV INEL	$> 0$ collisions the mean values 1.5, 5 and 10.5 for  three bins do not match values 3, 9 and 15 estimated in App.~\ref{multclass1}. 

From the combination of limiting values for $R(p_t;n_{ch}')$ and $X(p_t;n_{ch}')$ at small and large \pt\ respectively the product $\alpha \bar \rho_{s0}$ for NSD collisions can be inferred uniquely from the spectrum-ratio data. $\bar \rho_s / \bar \rho_{s0}$ is estimated from limiting cases of Eq.~(\ref{eqnx2}) at larger \pt\ and combined with Eq.~(\ref{eqnr}) limiting cases at smaller \pt\ to provide the estimate $\alpha' \bar \rho_{s0}' \approx 0.15 \approx \alpha \bar \rho_{s0} / \xi$. Since $\xi \approx 0.6$ for these data  $\alpha \bar \rho_{s0} \approx 0.09$ and (for $\bar \rho_{s0} \approx 6$) $\alpha \approx 0.015\pm0.0015$ at 13 TeV vs $\alpha \approx 0.006\pm0.001$ at 200 GeV~\cite{ppprd,ppquad}. The jet-fragment yield per participant ($\sim \alpha \bar \rho_{so}$) is 6 times larger at 13 TeV than at 200 GeV and the total jet-fragment yield ($\sim \bar \rho_h \approx \alpha \bar \rho_{s0}^2$) is 14 times larger. The energy variation of TCM soft-hard parameter $\alpha$ is discussed in Sec.~\ref{parsum} in connection with Fig.~\ref{params}.

\begin{figure}[h]
  \includegraphics[width=1.61in]{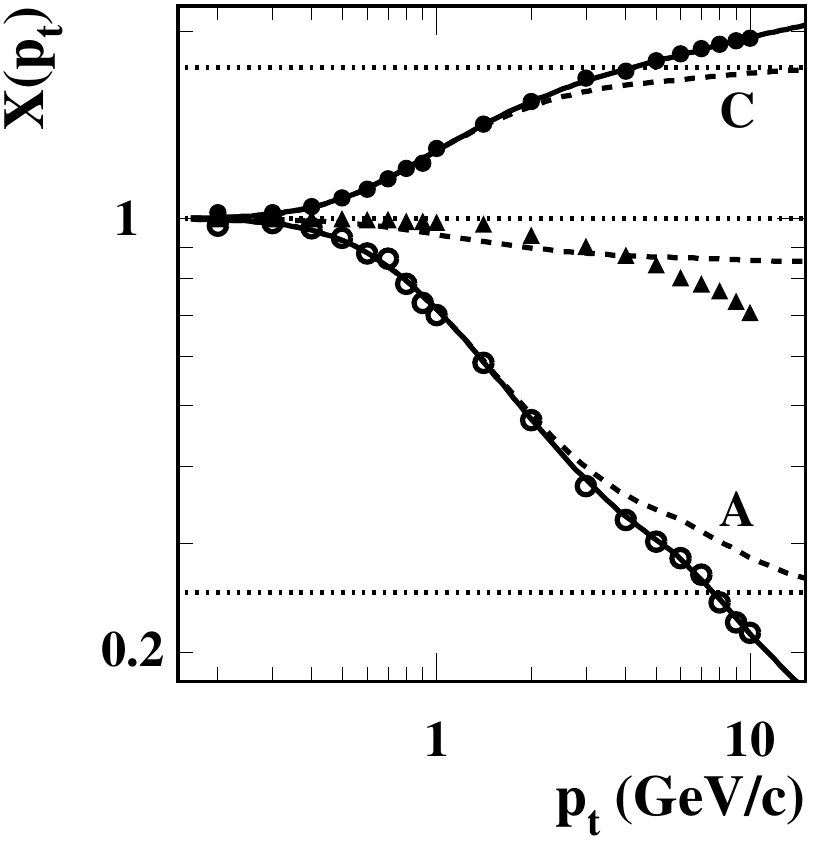}
  \includegraphics[width=1.65in]{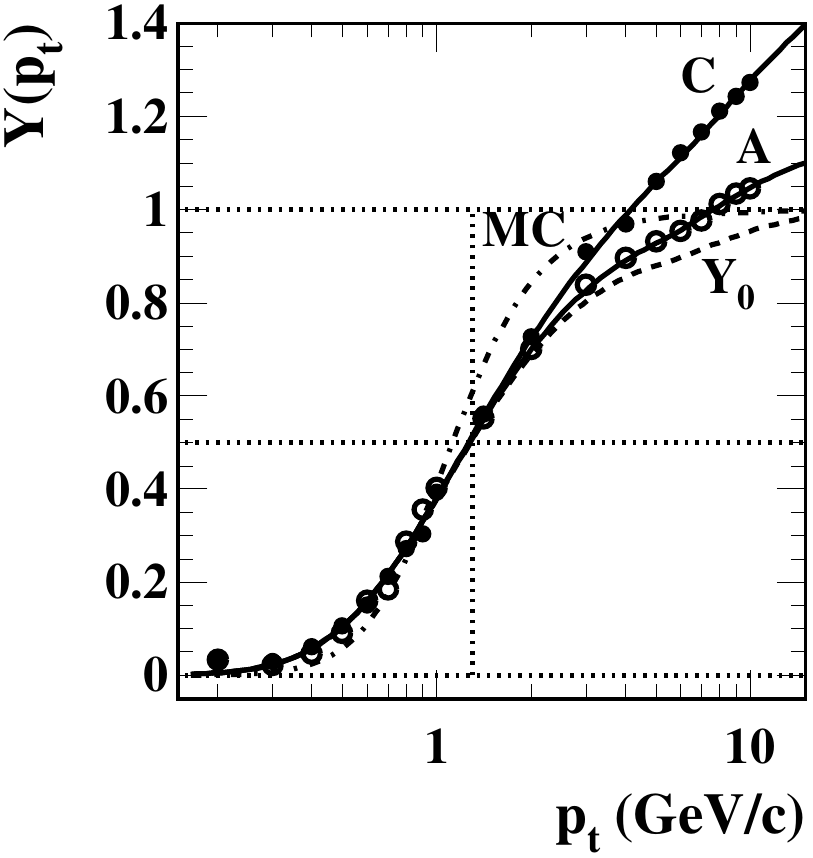}
\caption{\label{morerats}
Left: Figure~\ref{basicrat} (right) replotted in log-log format to demonstrate the relation between fixed (dashed) and variable (solid) hard-component TCM trends for bins A and C. Compare also with Fig.~\ref{corrrat} (right).
Right: Data from Fig.~\ref{basicrat} (right) transformed according to Eq.~(\ref{ypt}) (first line) for two multiplicity bins (points). The dashed curve is $Y_0(p_t)$ [factor in parentheses in Eq.~(\ref{ypt}) (second line)] with a fixed hard component. The solid curves are $Y(p_t;n_{ch}')$ generated by complete  Eq.~(\ref{ypt}) (second line) with varying hard component as described in Sec.~\ref{13tevspec}.
 } 
\end{figure}

\begin{figure}[h]
  \includegraphics[width=1.65in]{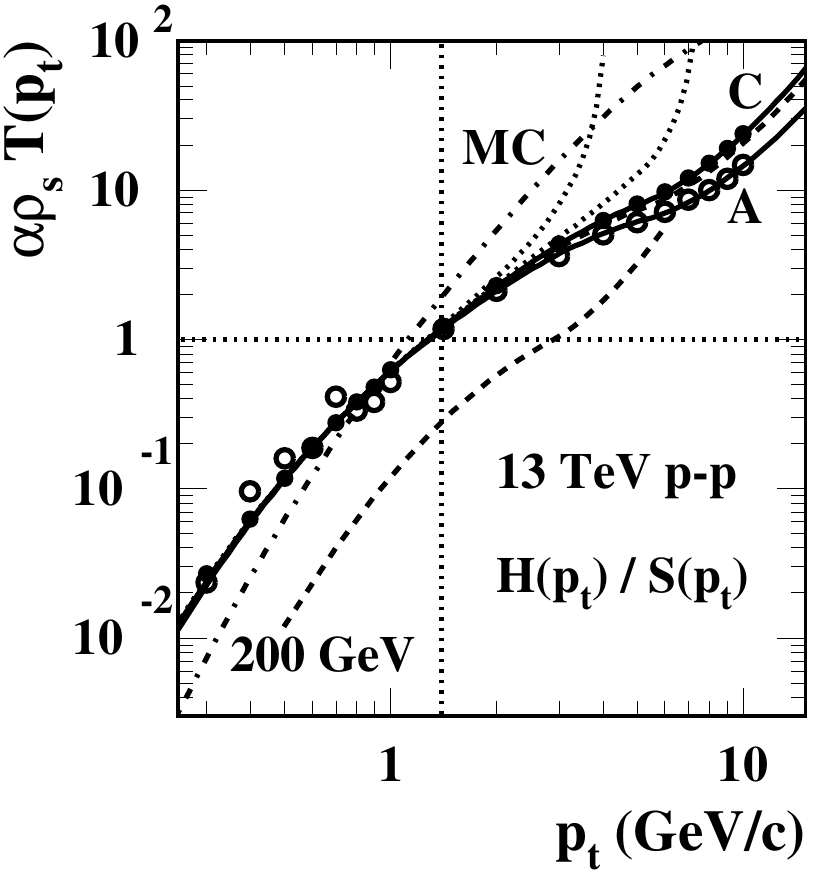}
 \includegraphics[width=1.67in]{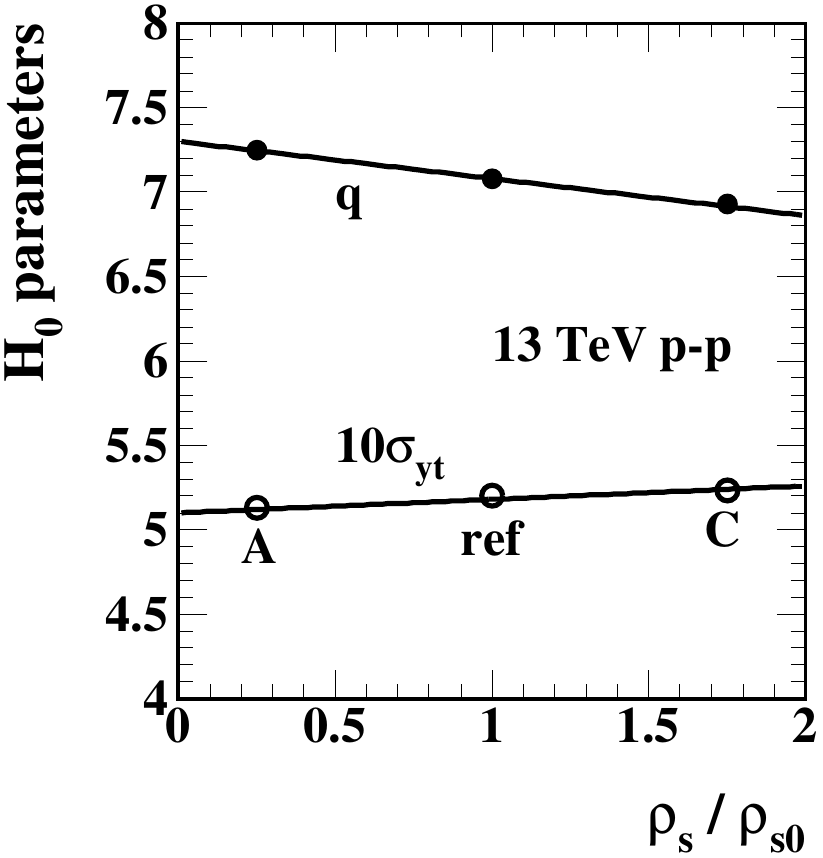}
 \caption{\label{blog}
Left: Data from Fig.~\ref{morerats} (right) transformed according to Eq.~(\ref{tpt}) (points) with $Y_0(p_t)$ and $T_0(p_t)$ defined above. The solid curves are solid curves in the previous figure transformed to $T(p_t;n_{ch}')$ in the same way. The upper dashed curve $T_0(p_t)$ is determined by fitted model parameters. The lower dashed curve is $T_0(p_t)$ for 200 GeV for comparison. The dotted curves indicate the result if data were transformed assuming a fixed hard component. The dash-dotted curve follows from the MC curves in Fig.~\ref{basicrat} (left).
Right: 
Hard-component $\hat H_0(p_t)$ parameter variations required to describe previous data from bins A and C relative to the 200 GeV $\hat S_0(p_t)$. The parameter values relative to a 13 TeV  $\hat S_0(p_t)$ are shown in Fig.~\ref{check} (left).
 } 
\end{figure}

Figure~\ref{morerats} (left) converts Fig.~\ref{basicrat} (right) to a log-log format to demonstrate that the change in $\hat H_0(p_t;n_{ch}')$ with \nch\ {\em relative to the fixed reference in} $T_0(p_t)$ is larger for bin A data (open points) than for bin C data (solid points) explaining why bin A data in Fig.~\ref{blog} (left) are further from the reference (upper dashed curve) than bin C data. Compare that panel with the symmetry of Fig.~\ref{corrrat} (right).

Figure~\ref{morerats} (right) shows the second intermediate quantity
\bea \label{ypt}
Y(p_t;n_{ch}') &\equiv&  \frac{1}{\bar \rho_s / \bar \rho_{s0} - 1}\left[X(p_t;n_{ch}') - 1\right] 
\\ \nonumber
&& \hspace{-.6in} \approx \frac{(\bar \rho_s / \bar \rho_{s0})T(p_t;n_{ch}')/ T_0(p_t) - 1}{\bar \rho_s / \bar \rho_{s0} - 1}
\left(\frac{\alpha \bar \rho_{s0} T_0(p_t)}{1 + \alpha \bar \rho_{s0} T_0(p_t)}\right)
\eea
defined in the first line (points) as derived from $X(p_t)$ data in the previous figure using $\bar \rho_s / \bar \rho_{s0}$ values inferred from that panel. The TCM is described by the second line. The first factor contains information on $T(p_t;n_{ch}')$ variation with \nch\  class and leads to the solid curves representing the revised TCM. The factor in parentheses defines $Y_0(p_t)$ (dashed) that would result if $T(p_t;n_{ch}') \rightarrow T_0(p_t)$. $Y_0(p_t) = 0.5$ when $H(p_t) = S(p_t)$ or $\alpha \bar \rho_{s0} T_0(p_t) = 1$. Compare with Fig.~\ref{specrat1} (right) for 200 GeV data.

Figure~\ref{blog} (left) shows the hard/soft ratio $T(p_t;n_{ch}')$ (solid and open points) obtained from $Y(p_t;n_{ch}')$ data as
\bea \label{tpt}
\alpha \bar \rho_{s}T(p_t;n_{ch}') \hspace{-.05in} &=& \hspace{-.05in} \alpha \bar \rho_{s0}T_0(p_t) \hspace{-.05in} \left[ \left( \frac{\bar \rho_{s}}{\bar \rho_{s0}}  \hspace{-.02in}- \hspace{-.02in}1\right)\frac{Y(p_t;n_{ch}')}{Y_0(p_t)}   \hspace{-.02in}+ \hspace{-.02in}1\right]~~~
\eea
with $T_0(p_t)$ (upper dashed) obtained from a 13 TeV spectrum fit as described below.  Data and TCM transform equivalently. Dotted curves show the result if data were transformed as in Eq.~(\ref{wrongy}) assuming a fixed $T_0(p_t)$. The lower dashed curve is $T_0(p_t)$ for 200 GeV from Fig.~\ref{specrat2} (left). The change in ratio $H(p_t) / S(p_t)$ indicates a larger role played by jets at {\em lower} \pt\ for 13 TeV \pp\ collisions. 

The dash-dotted curve is the MC curve(s) in  Fig.~\ref{basicrat} (left) transformed as fixed $T(p_t) \rightarrow T_0(p_t)$ indicating that MCs assume a fixed hard component, larger in amplitude and skewed to lower $p_t$  than data, similar to results presented in Sec.~IX of Ref.~\cite{ppprd} and consistent with an assumed parton/jet spectrum extending well below 3 GeV, therefore predicting a large excess of low-energy jets.\footnote{The conclusion in Ref.~\cite{ppprd} that the hard-component yield from PYTHIA is less than that inferred from data is incorrect. Plotted PYTHIA curves are comparable to $\alpha \bar \rho_s \hat H_0(y_t)$ with $\alpha \bar \rho_s \approx 0.015$ for 200 GeV NSD \pp\ collisions. The PYTHIA hard-component density $\bar \rho_h$ is thus 3-5 times {\em larger} than that inferred from data.}

A temporary parametrization for $T_0(p_t)$ (upper dashed curve) can be constructed by retaining the 200 GeV $\hat S_0(p_t)$ model and adjusting  $\hat H_0(p_t;n_{ch}')$ parameters to fit the 13 TeV $T(p_t;n_{ch}')$ data in Figure~\ref{blog} (left). Resulting parameter multiplicity trends can be interpolated to define $T_0(p_t)$ describing the INEL $> 0$ reference spectrum.

Figure~\ref{blog} (right) shows variation with $\bar \rho_s$ of two hard-component parameters required to match $T(p_t;n_{ch}')$ ratio data in the left panel. Those parameter values relative to a 200 GeV soft-component model are used only to parametrize the $T(p_t;n_{ch}')$ data temporarily. They do not describe a proper 13 TeV TCM hard-component model. Interpolated parameter values $\sigma_{y_t} = 0.52$ and $q = 7$ define reference $T_0(p_t)$ that appears as the upper dashed curve in Fig.~\ref{blog} (left). In the next subsection the full TCM for 13 TeV \pp\ collisions is derived by combining ratio and spectrum data. Trends are similar in form to those for 200 GeV in Sec.~\ref{hardev} but with reduced relative variation because of the limited \nch\ excursion at 13 TeV.

\subsection{Full 13 TeV TCM derived from fit to spectrum} \label{13tevspec}

Spectrum-ratio data provide information only about hard/soft ratio $T(p_t;n_{ch}') = \hat H_0(p_t;n_{ch}') / \hat S_0(p_t)$. A fit to at least one \pt\ spectrum is required to isolate individual data components and define TCM model functions. Given the parametrization of 13 TeV reference $T_0(p_t)$ defined in the previous subsection the 13 TeV $\hat S_0(p_t)$ model is derived from a spectrum fit, and a 13 TeV $\hat H_0(p_t;n_{ch}')$ model is inferred from $\hat S_0(p_t)$ and the $T(p_t;n_{ch}')$ data.

\begin{figure}[h]
  \includegraphics[width=1.65in]{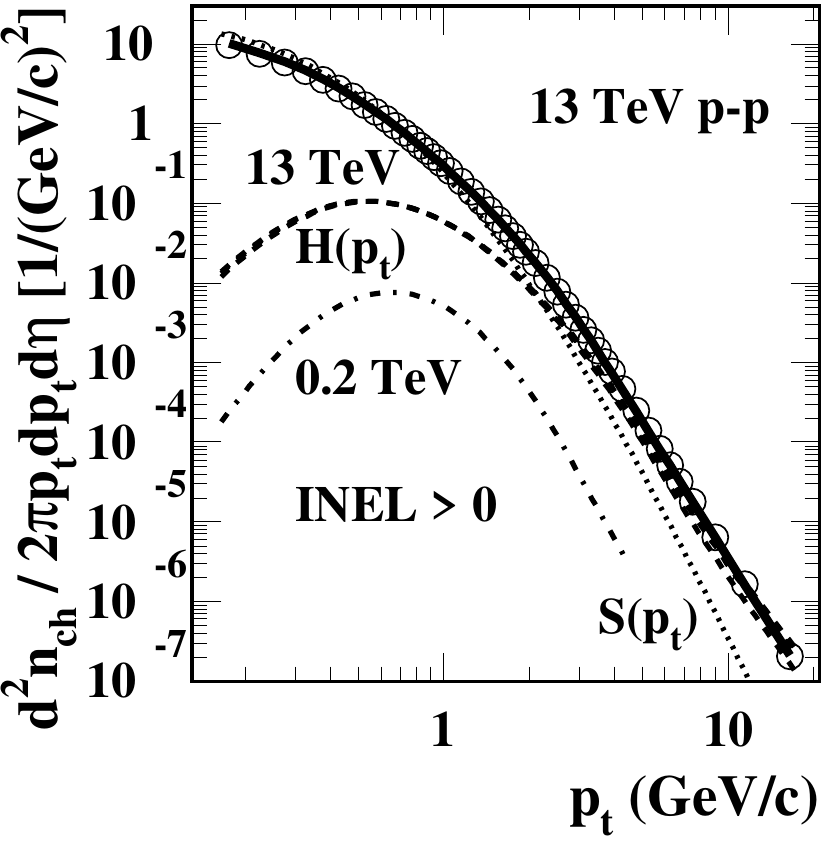}
 \includegraphics[width=1.67in]{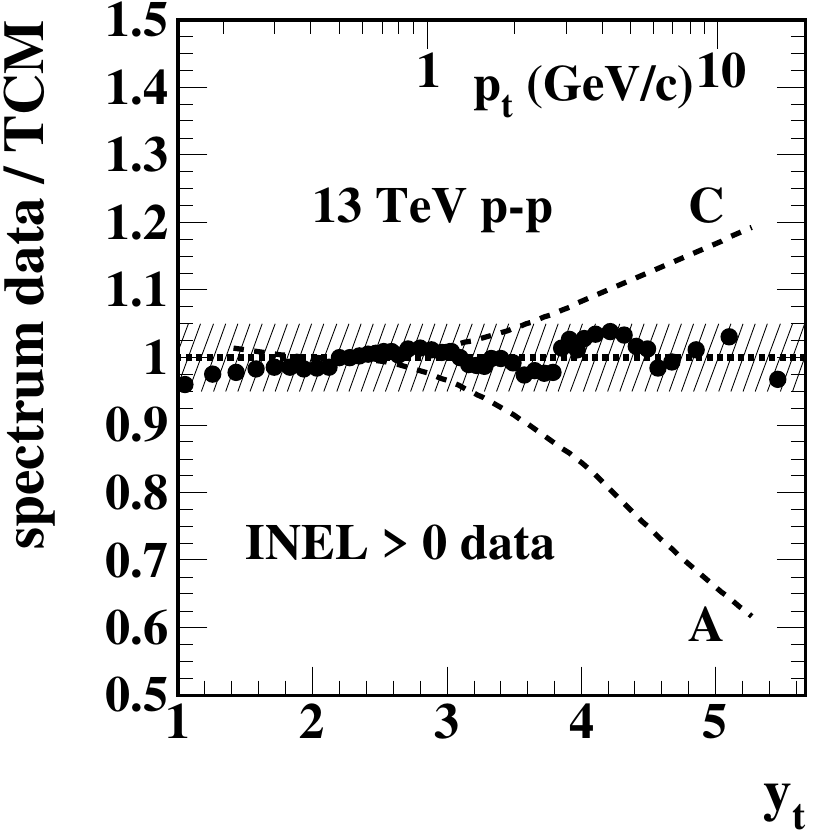}
\caption{\label{specfit}
Left: Data obtained from the INEL $> 0$ 13 TeV \pt\ spectrum in Fig.~3 of Ref.~\cite{alicespec} (points). The solid curve is the model described by Eq.~(\ref{13teveqn}) with $n$ optimized. The three dashed curves represent $H(p_t;n_{ch}')$ for $T_0(p_t)$ (middle) and for $T(p_t;n_{ch}')$ from bins A and C (lower and upper respectively). The 13 TeV soft component is represented by the dotted curve. The dash-dotted curve represents $H(p_t)$ for 200 GeV.
Right: Fit residuals for the left panel (points). 
Spectrum ratios with varying hard component for bins A and C (dashed) can be compared with 200 GeV ratio data in Fig.~\ref{specrat2} (right). The hatched band indicates $\pm5$\% deviations.
 } 
\end{figure}

Figure~\ref{specfit} (left)) shows the INEL $> 0$ spectrum (points from Fig.~3 of Ref.~\cite{alicespec}. The fitted TCM (solid) is 
\bea \label{13teveqn}
\bar \rho_{00}(p_t)    &=& \bar \rho_{s0}\hat S_0(p_t;T,n) \left[ 1 + \alpha \bar \rho_{s0} T_0(p_t)   \right]
\eea
with $\alpha \bar \rho_{s0} T_0(p_t)$ represented by the upper dashed curve in Fig.~\ref{blog} (left). The only TCM adjustment is variation of exponent $n$ in the L\'evy form of $\hat S_0(m_t;T,n)$ in Eq.~(\ref{13teveqn}) to fit the spectrum data, with slope parameter $T = 145$ MeV held fixed. The fitted L\'evy  exponent $n \approx 7.8$ at 13 TeV can be compared with $n \approx 12.5$ at 200 GeV~\cite{ppquad}.

Given optimized soft-component model $\hat S_0(p_t;T,n)$ the 13 TeV hard-component model parameters $\bar y_t$, $\sigma_{y_t}$ and $q$ are determined by fitting the TCM for $T(p_t;n_{ch}') = \hat H_0(p_t;n_{ch}') / \hat S_0(p_t;T,n)$ to data from Fig.~\ref{blog} (left) (bins A and C) and the reference spectrum in Fig.~\ref{specfit} (left) (INEL $> 0$ reference). That parametrization back transformed generates solid model curves that pass through data for bins A and C in all previous figures of Sec.~\ref{13tevspecc}. 

Figure~\ref{specfit} (right) demonstrates the quality of the TCM reference description with  fit residuals (points). There is no systematic deviation of data from the optimized TCM. Deviations are substantially less than 5\% (hatched band) over a momentum interval from 0.15 GeV/c to 16 GeV/c. That result can be compared with claims of ``log-periodic'' oscillations~\cite{wilkosc} in deviations of a so-called Tsallis distribution~\cite{tsallis,tsallis1} [equivalent to Eq.~(\ref{s0other}) of the present study] from a spectrum for  7 TeV \pp\ collisions that are similar to the residuals at 0.9 TeV in Fig.~\ref{tcm9} (b).

The soft and hard spectrum components in Fig.~\ref{specfit} (left) are $S(p_t) =  \bar \rho_{s0} \hat S_0(p_t) $ (dotted curve) and $H(p_t) = \alpha  \bar \rho_{s0}^2 \hat H_{00}(p_t)$ (dashed curve so labeled). The 200 GeV hard component (dash-dotted curve) is shown for comparison. Two other dashed curves in Figure~\ref{specfit} (left) correspond to Eq.~(\ref{13teveqn}) with $T_0(p_t) \rightarrow T(p_t;n_{ch}')$ represented by solid curves A and C in Fig.~\ref{blog} (left). The dashed curves in Figure~\ref{specfit} (right) are Eq.~(\ref{13teveqn}) including $\hat H_0(p_t;n_{ch}')$ (for bins A and C) in ratio to Eq.~(\ref{13teveqn}) including fixed $\hat H_{00}(p_t)$ (for the INEL $> 0$ reference) that can be compared with ratio data in Fig.~\ref{specrat2} (right). 

The fitted $\hat H_0(p_t;n_{ch}')$ parameter values for bins A and C and the INEL $> 0$ reference are included in Fig.~\ref{check} (left) (upper points, labeled 13 TeV) along with simple parametrizations (upper solid and dashed curves)  that show variations with $\bar \rho_s$ similar to 200 GeV, although over a reduced $\bar \rho_s / \bar \rho_{s0}$ interval. The 13 TeV INEL $> 0$ reference values are included in Table~\ref{engparam}.


These results plus previous figures in this section indicate that the 13 TeV TCM with \nch-varying hard component provides an accurate \pt\ spectrum description over a significant range of event multiplicities. It also buttresses results from Ref.~\cite{ppprd} that revealed a 200 GeV spectrum hard component with mode near $p_t = 1$ GeV/c and approximate power-law tail at higher \pt\ compatible with QCD predictions derived from reconstructed jets.

\section{$\bf p_t$ spectrum energy evolution}  \label{edepp}

Given \nch-dependent TCMs inferred from 200 GeV and 13 TeV spectrum data an energy-dependent TCM continuously covering the interval from 17 GeV to 13 TeV is derived using supplementary spectrum data.

\subsection{Energy evolution of spectrum soft exponent $\bf n$}  \label{esoft}

The spectrum soft component at 13 TeV is substantially different from that at 200 GeV from Ref.~\cite{ppprd}. The L\'evy exponent changes from $n \approx 12.5$ at 200 GeV to $n \approx 7.8$ at 13 TeV (harder spectrum). Results from CERN super proton synchrotron (SPS) \pp\ spectra extend the energy trend over a larger energy interval.

\begin{figure}[h]
  \includegraphics[width=1.65in,height=1.63in]{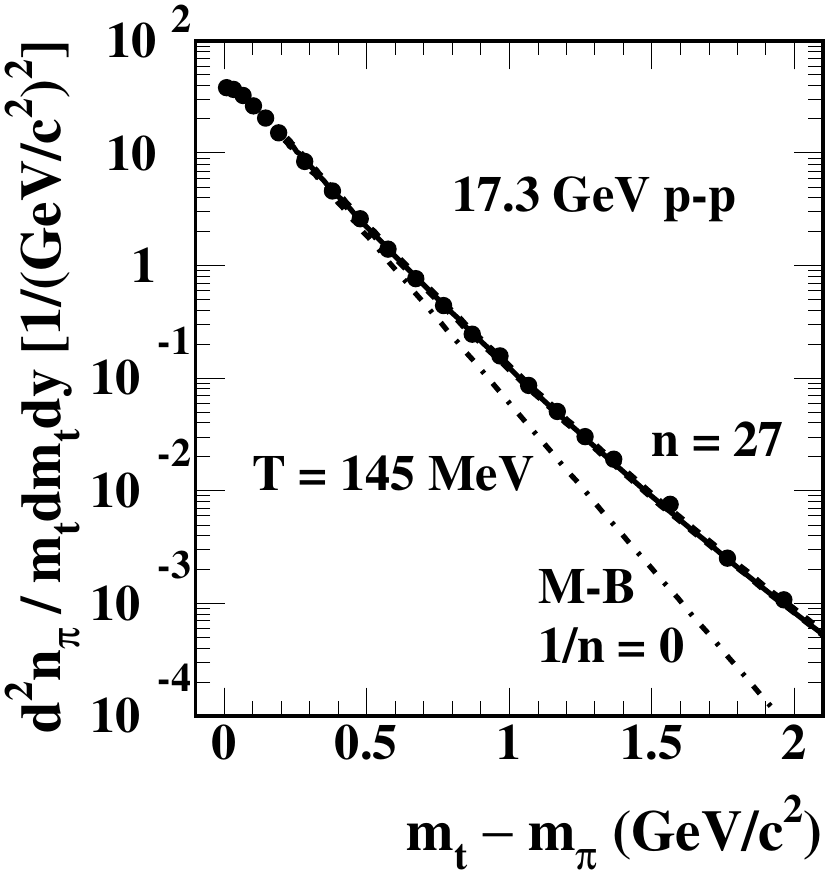}
  \includegraphics[width=1.65in,height=1.6in]{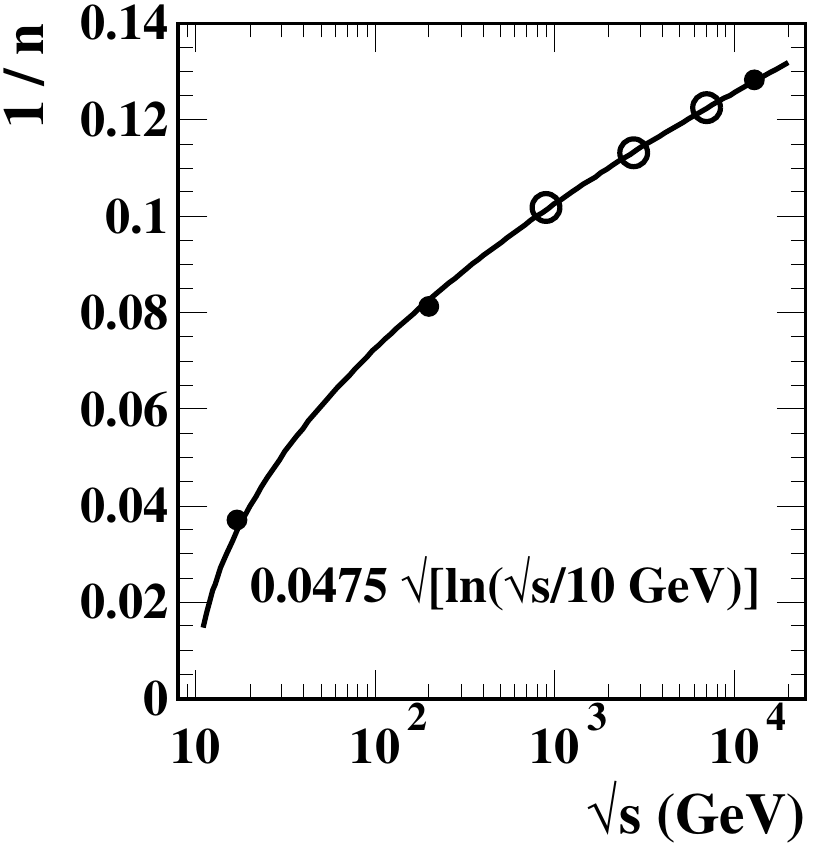}
\caption{\label{softcomp}
Left: \mt\ spectrum for identified charged pions from 17.2 GeV inelastic \pp\ collisions (points)~\cite{na49spec}. The solid curve is a fit of Eq.~(\ref{s0}) with $T = 145$ MeV held fixed to determine L\'evy exponent  $n = 27$. The dashed curve is a full TCM determined by the parameters for that energy in Table~\ref{engparam}. The dash-dotted curve is the corresponding Maxwell-Boltzmann exponential.
Right: Measured L\'evy exponents for three collision energies (solid points). The curve is a fit by eye of the function $A \sqrt{\ln(\sqrt{s} / \text{10 GeV})}$ (with $A = 0.0475$) motivated by the possibility of Gribov diffusion controlling the growth of transverse momentum for low-$x$ partons (gluons)~\cite{gribov}.  Open symbols are interpolations at energies relevant to this study.
 } 
\end{figure}

Figure~\ref{softcomp} (left) shows an \mt\ spectrum for identified pions (points) from 17.2 GeV inelastic \pp\ collisions~\cite{na49spec} well described by a L\'evy distribution (solid curve) with exponent $n \approx 27$ and the universal slope parameter $T = 145$ MeV. Also shown is the corresponding Maxwell-Boltzmann (M-B) distribution (dash-dotted curve) with $1/n \rightarrow 0$ and the same slope parameter. Given the known energy dependence of the jet contribution~\cite{jetspec2} the spectrum hard component should produce at most a slight deviation from the soft component in that \pt\ interval, especially for inelastic \pp\ collisions (the dashed curve represents a TCM predicted sum of soft + hard components --  see Sec.~\ref{hardedep}). The spectrum data at 17.2 GeV then constrain only soft component $\hat S_0(m_t;T,n)$.

Figure~\ref{softcomp} (right) shows soft-component exponents in the form $1/n$ inferred from spectrum data for three collision energies (solid points) at the SPS, RHIC and LHC. The solid curve is an algebraic hypothesis based on variation of the soft component due to Gribov diffusion~\cite{gribov}. Low-$x$ gluons result from a virtual parton splitting cascade within projectile nucleons whose mean depth on $x$ is determined by the collision energy. Each step of the cascade adds transverse-momentum components in a random-walk process. The depth of the cascade is proportional to $\ln(s/s_0)$, and $\sqrt{s_0} \approx 10$ GeV is inferred from dijet systematics~\cite{anomalous,jetspec2}. Given the properties of a random walk and with $1/n$ as a measure of transverse-momentum excursions~\cite{wilk} its trend is estimated as $\propto \sqrt{\ln(\sqrt{s} / \text{10 GeV})}$ (solid curve).  The open circles at 0.9, 2.76 and 7 TeV are interpolations of the L\'evy exponent to $n = 9.82$, 8.83 and $8.16$ respectively.

\subsection{Energy evolution of spectrum hard exponent $\bf q$} \label{ehard}

\begin{figure}[h]
  \includegraphics[width=1.67in]{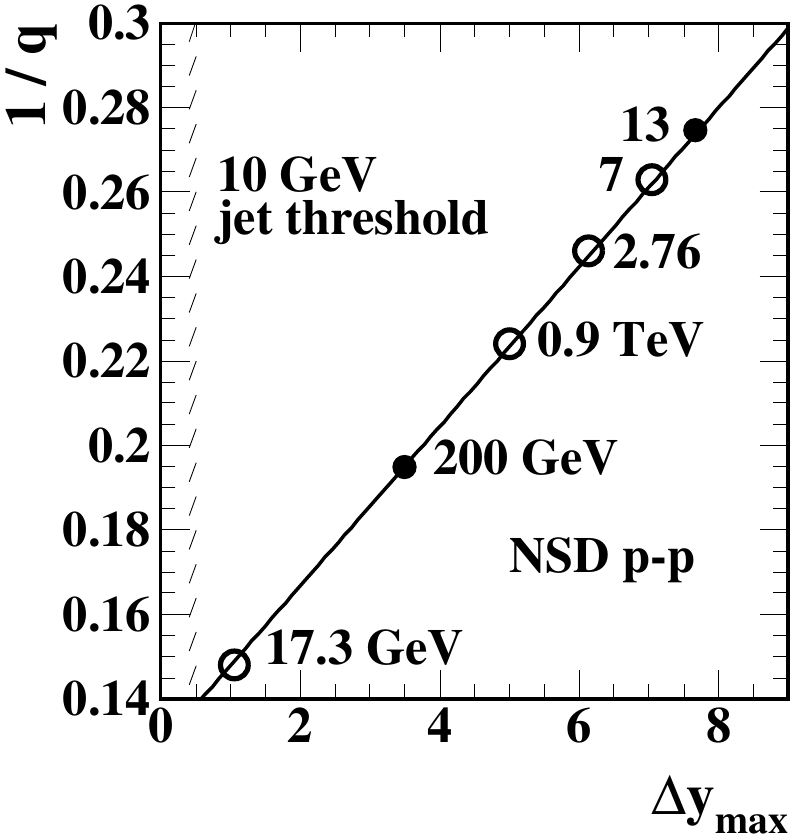}
  \includegraphics[width=1.65in]{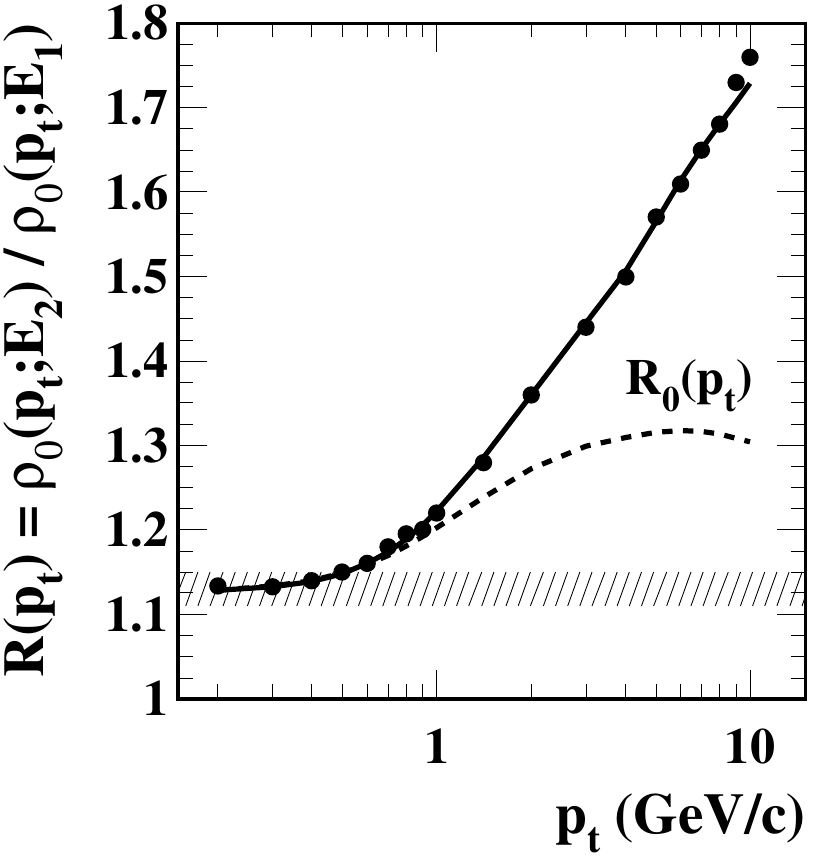}
\caption{\label{enrat1}
Left: Hard-component exponents $q$ determined by analysis of spectrum data (solid points) from Ref.~\cite{ppquad} and the present study. The solid curve is based on a jet-spectrum parametrization in Ref.~\cite{jetspec2} that also describes ensemble-mean-\pt\ hard-component energy variation~\cite{tomalicempt}. The open points are interpolations and an extrapolation relevant to this study.
Right: Data derived from a spectrum ratio comparing 13 and 7 TeV INEL $> 0$ spectra, from Fig.~4 of Ref.~\cite{alicespec} (points). The dashed curve is reference $R_0(p_t)$ assuming the same hard component for two energies in Eq.~(\ref{re1e2}). The solid curve is Eq.~(\ref{re1e2}) with 7 TeV $\hat H_0(p_t;E_1)$ adjusted to accommodate data.
 }  
\end{figure}

Figure~\ref{enrat1} (left) shows inverse values (solid points) of exponents $q = 5.15$ for 200 GeV as in Fig.~\ref{ppspec1} and $q = 3.65$ for 13 TeV as in Fig.~\ref{specfit} (left) plotted vs quantity $\Delta y_{max} \equiv \ln(\sqrt{s} / \text{6 GeV})$ observed to describe the energy trend for jet spectrum widths $\propto \Delta y_{max}$ from NSD \pp\ collisions assuming a jet spectrum low-energy cutoff near 3 GeV~\cite{jetspec2} [see Fig. 5 of Ref.~\cite{jetspec2} for a direct comparison with measured jet spectra]. The inverse $1/q$ effectively measures the hard-component peak width at larger \yt. The relation $1/q \propto  \Delta y_{max}$ (solid line) is expected given that the \pp\ \pt-spectrum hard component can be expressed as the convolution of a fixed \pp\ FF ensemble with a collision-energy-dependent jet spectrum~\cite{fragevo}, and the jet-spectrum width trend has the same dependence~\cite{jetspec2}. The vertical hatched band indicates an inferred cutoff to dijet production from low-$x$ gluon collisions near 10 GeV. That the same relation applies to the ensemble-mean-\pt\ hard component has been established in a separate study~\cite{tomalicempt}. The inverse values of $q = 3.80$ for 7 TeV, $q = 4.05$ for 2.76 TeV and $q = 4.45$ for 0.9 TeV (open circles) are obtained by interpolation. 

\subsection{$\bf p_t$ spectrum ratio for two LHC energies}

Figure 4 of Ref.~\cite{alicespec} [Fig.~\ref{ratdata} (right) of this paper] provides partial information on the energy variation of \pt\ spectra between 7 and 13 TeV. Supplementary information must be introduced to obtain a full spectrum description. Soft and hard spectrum components are considered separately.

The data spectrum ratio in Fig.~\ref{ratdata} (right) is represented by the first line of
\bea \label{re1e2}
R(p_t;E_1,E_2) &\equiv&   \frac{\bar \rho_0'(p_t;E_2)}{\bar \rho_0'(p_t;E_1)}
\\ \nonumber
&\approx& \frac{\bar \rho_{s2} \hat S_0(p_t;E_2) + \bar \rho_{h2} \hat H_0(p_t;E_2)}{\bar \rho_{s1} \hat S_0(p_t;E_1) + \bar \rho_{h1} \hat H_0(p_t;E_1)}
\\ \nonumber
&=& \hspace{-.05in}  \left(\frac{\bar \rho_{s2}}{\bar \rho_{s1}}\right) \left[\frac{\hat S_0(p_t;E_2)}{\hat S_0(p_t;E_1)}\right] \frac{1 + \alpha \bar \rho_{s2} T(p_t;E_2)}{1 + \alpha  \bar \rho_{s1} T(p_t;E_1)}.
\eea
The second line defines the TCM for this case and the third line indicates a factorization similar to that in Eq.~(\ref{eqnr}). However, the $\hat S_0$  ratio does not cancel, is determined instead by the exponent-$n$ trend in Fig.~\ref{softcomp} (right).

Figure~\ref{enrat1} (right) shows spectrum-ratio data (points) as in  Fig.~\ref{ratdata} (right). The low-\pt\ limit of $R(p_t)$ is density ratio $\bar \rho_{s2} / \bar \rho_{s1} \approx 1.12$ (hatched band) compared with expected ratio 1.10 derived from the soft-component trend (dotted curve) in Fig.~\ref{edep} of App.~\ref{lhcmult}. This ratio of full spectra confuses soft and hard TCM components, is insensitive to energy-dependent jet physics obscured at lower \pt\ by the spectrum soft component, and the changes between 7 and 13 TeV are small as shown below. Improved sensitivity to jet physics could be obtained by analyzing spectra from higher-multiplicity \pp\ collisions.

Given energy trends for the soft- and hard-component exponents derived above, the 7 TeV spectrum hard component can be isolated. Quantity $R_0(p_t)$ (dashed curve) is Eq.~(\ref{re1e2}) with both hard-component forms $\hat H_0(p_t;E)$ fixed at $E_2 = 13$ TeV so that $T(p_t;E_1) \rightarrow T_0(p_t;E_1) \equiv \hat H_0(p_t;E_2) / \hat S_0(p_t;E_1)$ with L\'evy $\hat S_0$ index $n = 8.16$ at 7 TeV derived from Fig.~\ref{softcomp} (right). $R_0(p_t;E_1,E_2)$  then represents only the soft-component contribution to $R(p_t;E_1,E_2)$ variation with energy, which is known.

%

\begin{figure}[h]
  \includegraphics[width=1.65in]{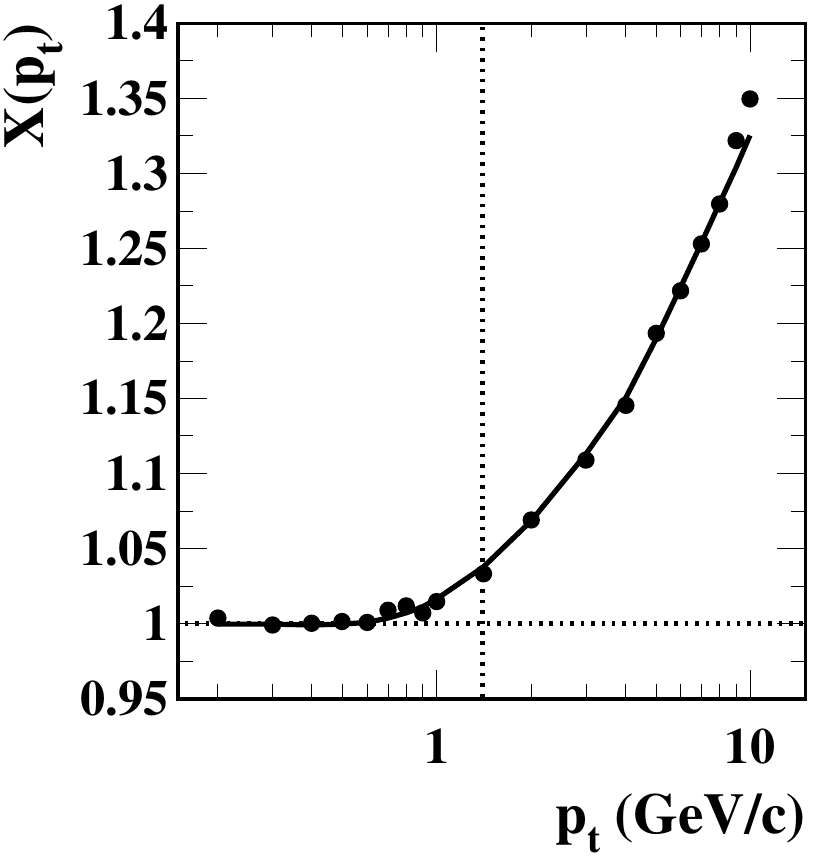}
  \includegraphics[width=1.63in]{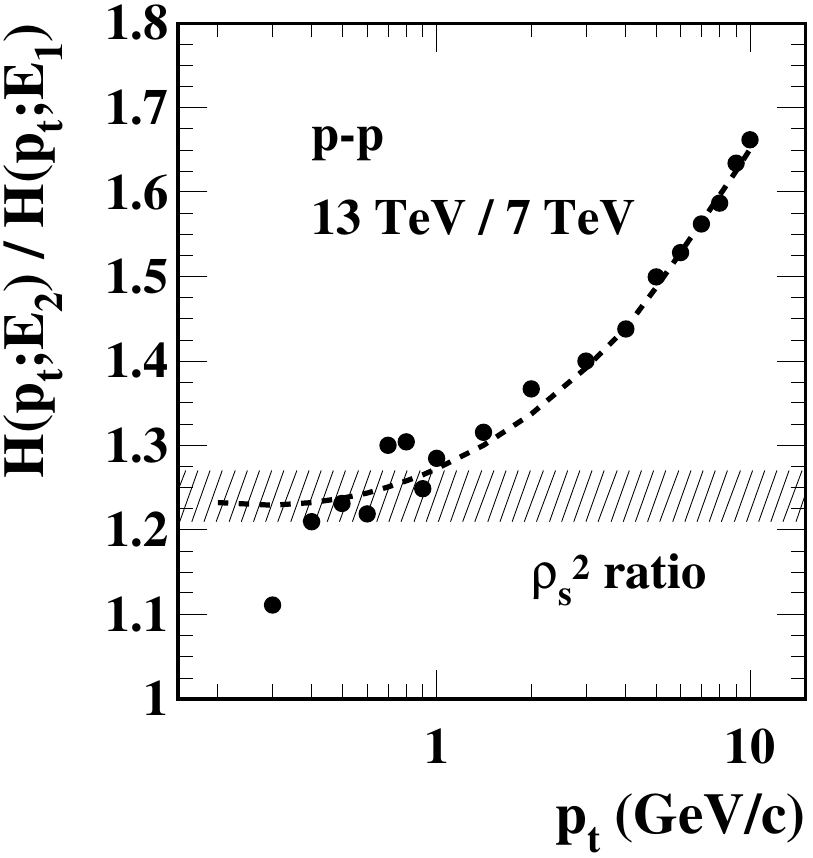}
\caption{\label{enrat2}
Left: Ratio $X(p_t)$ derived from data in the left panel with Eq.~(\ref{xpt}) (first line) (points). The solid curve is the second line with  7 TeV $\hat H_0(p_t)$ adjusted to accommodate the data by varying $\bar y_t$ and $\sigma_{y_t}$ to obtain the values appearing in Table~\ref{engparam}. TCM parameters $n$ and $q$ are interpolated.
Right: Hard-component ratio comparing 13 and 7 TeV (points) derived from points in the left panel via Eq.~(\ref{ent}) and $H(p_t;E) = \alpha(E) \bar \rho_s^2(E) \hat S[p_t;n(E)] T(p_t;E)$. The dashed curve is determined by 7 and 13 TeV TCM parameters in Table~\ref{engparam}.
 } 
\end{figure}

Figure~\ref{enrat2} (left) shows data for quantity $X(p_t;E_1,E_2)$ (points) defined by the first line of
\bea \label{xpt}
X(p_t;E_1,E_2)  &=& \frac{R(p_t;E_1,E_2)}{R_0(p_t;E_1,E_2)}  
\\ \nonumber
&\approx&   \frac{1 + \alpha \bar \rho_{s1} T_0(p_t;E_1)}{1 + \alpha  \bar \rho_{s1} T(p_t;E_1)}.
\eea
 The second line defines a TCM expression having asymptotic form $\hat H_0(p_t;E_2) / \hat H_0(p_t;E_1)$ at larger \pt. From Fig.~\ref{blog} (left) $S(p_t) = H(p_t)$ near $p_t \approx 1.4$ GeV/c (vertical dotted line). To the right of that point ratio $X(p_t)$ is dominated by the spectrum hard components.
With $T_0(p_t;E_1) \equiv T(p_t;E_2) \times \hat S_0(p_t;E_2) / \hat S_0(p_t;E_1)$ and $\alpha \bar \rho_{s1}$ already determined by data, either directly or by interpolation, data ratio $T(p_t;E_1)$ can be obtained via 
\bea \label{ent}
T(p_t;E_1) \hspace{-.07in} &=& \hspace{-.07in} \frac{1}{\alpha \bar \rho_{s1}}
\left\{ \frac{1 + \alpha \bar \rho_{s1} T_0(p_t;E_1)}{X(p_t)} -1\right\},
\eea
yielding $H(p_t;E_1) = \alpha \bar \rho_{s1}^2 \hat S_0(p_t;E_1) T(p_t;E_1)$.

Figure~\ref{enrat2} (right) shows data hard-component ratio $H(p_t;E_2) / H(p_t;E_1)$ (points)  derived from $X(p_t)$ data in the left panel via Eq.~(\ref{ent}). The TCM ratio (dashed) is obtained by combining $H(p_t;E_2)$ from the 13 TeV TCM in Fig.~\ref{specfit} (left) with 7 TeV $H(p_t;E_1)$ obtained by interpolating $\hat H_0(p_t)$ TCM parameters between 0.2 and 13 TeV (Table~\ref{engparam}). Small adjustments to $\hat H_0(y_t)$ Gaussian width $\sigma_{y_t}$ and centroid $\bar y_t$ were made to accommodate the data: $2.66 \rightarrow 2.64$ for the centroid and $0.60 \rightarrow 0.595$ for the width (see Table~\ref{engparam}). The asymptotic low-\pt\ limit is $(\bar \rho_{s2}/\bar \rho_{s1})^2 \approx 1.2^2 \approx 1.25$ (hatched band). 
 
\subsection{Energy evolution of spectrum hard component} \label{hardedep}

\begin{figure}[h]
  \includegraphics[width=1.65in]{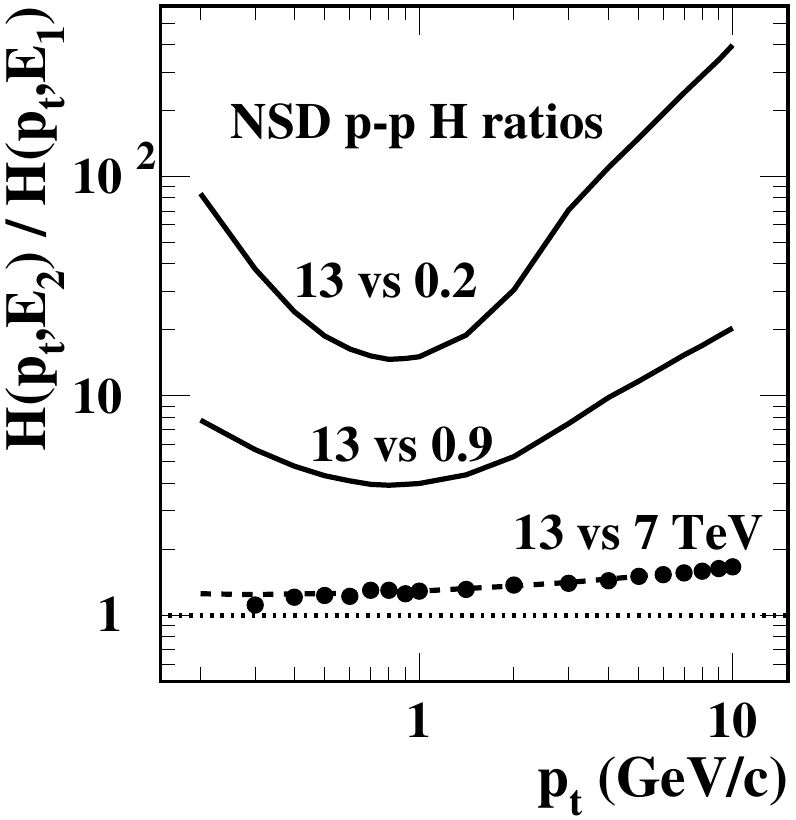}
  \includegraphics[width=1.65in]{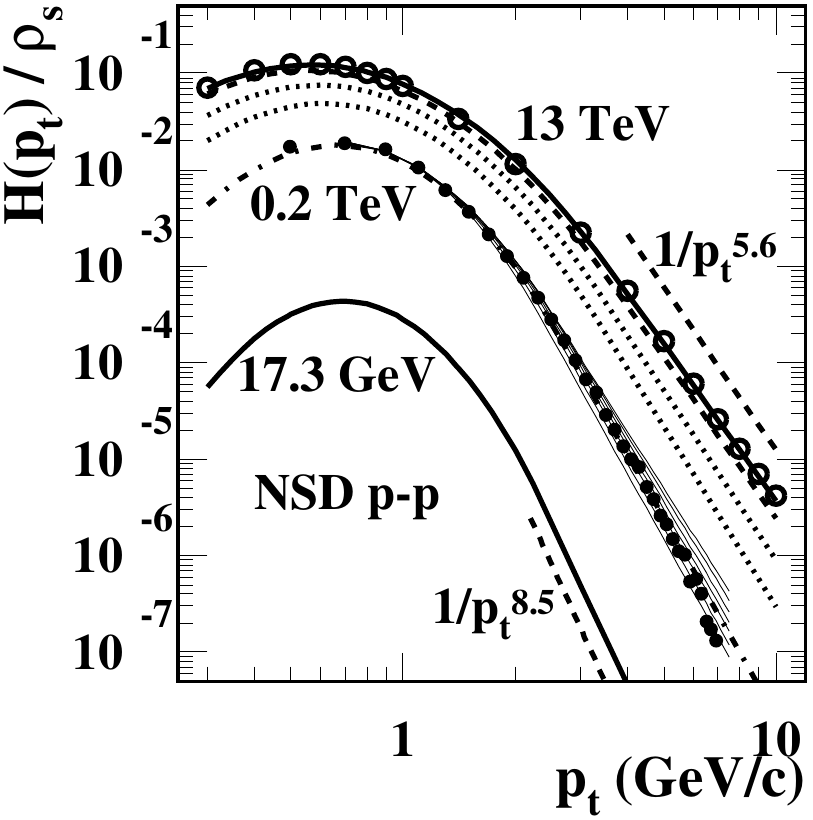}
\caption{\label{enrat3}
Left: Hard-component ratios for three energy combinations. The points are from Fig.~\ref{enrat2} (right). The curves are determined by parameters in Table~\ref{engparam}.
Right: A survey of spectrum hard components over the currently accessible energy range from threshold of dijet production (10 GeV) to LHC top energy (13 TeV). The curves are determined by parameters in Table~\ref{engparam} except for the 200 GeV fine solid curves determined also by the $\sigma_{y_t}$ and $q$ trends in Fig.~\ref{check} (left). The points are from Refs.~\cite{ppquad} (200 GeV) and \cite{alicespec} (13 TeV).
 }  
\end{figure}

Figure~\ref{enrat3} (left) shows TCM hard-component ratios for three energy pairs (curves), where $H(p_t;E) \approx \alpha(E) \bar \rho_s^2(E) \hat H_0(p_t;E)$ relates hard-component yield data to unit-normal model functions. The curves are determined by the parameters in Table~\ref{engparam}. The 13 vs 7 TeV comparison (points) reveals little about the energy evolution of $H(p_t;E)$, but larger energy intervals demonstrate that the hard-component width near its mode broadens significantly, and the high-\pt\ power-law tail falls much less rapidly at higher collision energies.  Those trends are quantitatively consistent with the measured \pp\ collision-energy dependence of underlying jet energy spectra~\cite{jetspec2}. 

Figure~\ref{enrat3} (right) shows the TCM for quantity $H(p_t;E) / \bar \rho_s(E) \approx \alpha(E) \bar \rho_s(E) \hat H_0(p_t;E)$ measuring the spectrum hard component  {\em per soft-component yield} corresponding to dijet production per participant low-$x$ gluon. The two dotted curves are for 0.9 and 2.76 TeV and the dashed curve is for 7 TeV. Isolated hard components rather than spectrum ratios clarify spectrum energy evolution and its relation to dijet production.

The predictions for six collision energies (curves) derived from parameter values in Table~\ref{engparam} are compared to data from four energies (13, 7 and 0.2 TeV above and 0.9  TeV in App.~\ref{900gev}).
The 17.2 GeV extrapolation described below indicates  no {\em significant} jet contribution to yields and spectra at that energy (dashed curve in Fig.~\ref{softcomp} -- left) and explains why no excess \pt\ fluctuations were observed at that energy~\cite{na49fluct,tomaliceptfluct}. However, evidence for SPS jets {\em is}  visible in 17 GeV azimuth correlations as a more sensitive detection method~\cite{ceres}.
The 200 GeV summary includes parametric variation of $\hat H_0(y_t;q,\sigma_{y_t})$ for seven multiplicity classes (thin solid curves) as described in Sec.~\ref{hcev}. Corresponding data (solid points) represent NSD \pp\ collisions. The overall result is a comprehensive and accurate description of dijet contributions to \pt\ spectra vs \pp\ collision energy over three orders of magnitude.

\subsection{Spectrum TCM parameter summary} \label{parsum}

Table~\ref{engparam} summarizes NSD \pp\ TCM parameters for a broad range of energies. The entries are grouped as soft-component parameters $(T,n)$, hard-component parameters $(\bar y_t,\sigma_{y_t},q)$, hard-soft relation parameter $\alpha$ and soft density $\bar \rho_s$.
Slope parameter $T = 145$ MeV is held fixed for all cases consistent with observations.  Its value is determined solely by a lowest-\yt\ interval where the hard component is negligible. The interpolated L\'evy exponent $n$ values are derived from Fig.~\ref{softcomp} (right) (open circles). Interpolated hard-component $q$ values are derived from Fig.~\ref{enrat1} (left) (open circles). $\bar \rho_s$ values are derived from the universal trend in Fig.~\ref{edep} (dotted curve) inferred from correlation and yield data. All 0.9 and 2.76 TeV values are predicted via interpolations. 
All remaining (unstarred) numbers are obtained from fits to data.

\begin{table}[h]
  \caption{Spectrum TCM parameters for NSD \pp\ collisions within $\Delta \eta \approx 2$ at several energies.
Starred entries are estimates by interpolation or extrapolation. Unstarred entries are derived from fits to yield, spectrum or spectrum-ratio data.
}
  \label{engparam}
\begin{center}
\begin{tabular}{|c|c|c|c|c|c|c|c|} \hline
 Eng.\. (TeV) & T\. (MeV) & $n$ & $\bar y_t$ & $\sigma_{y_t}$ & $q$ & $100\alpha$ & $\bar \rho_s$ \\ \hline
 0.0172  & 145  & 27 & 2.55$^*$ & 0.40$^*$  & 6.75$^*$  & 0.07$^*$ & 0.45 \\ \hline
 0.2  & 145  & 12.5 & 2.59 & 0.435  & 5.15  & 0.6 & 2.5 \\ \hline
 0.9  & 145  & 9.82$^*$ & 2.62$^*$ & 0.53$^*$  &  4.45$^*$  & 1.0$^*$ & 3.61 \\ \hline
 2.76  & 145  & 8.83$^*$ & 2.63$^*$ & 0.56$^*$  &  4.05$^*$  & 1.2$^*$ & 4.55 \\ \hline
7.0  & 145  & 8.16$^*$ & 2.64 & 0.595  & 3.80$^*$  & 1.4$^*$ & 5.35  \\ \hline
13.0  & 145  & 7.80 & 2.66 & 0.615  & 3.65  & 1.5 & 5.87 \\ \hline
\end{tabular}
\end{center}
\end{table}
 
\begin{figure}[h]
  \includegraphics[width=1.65in]{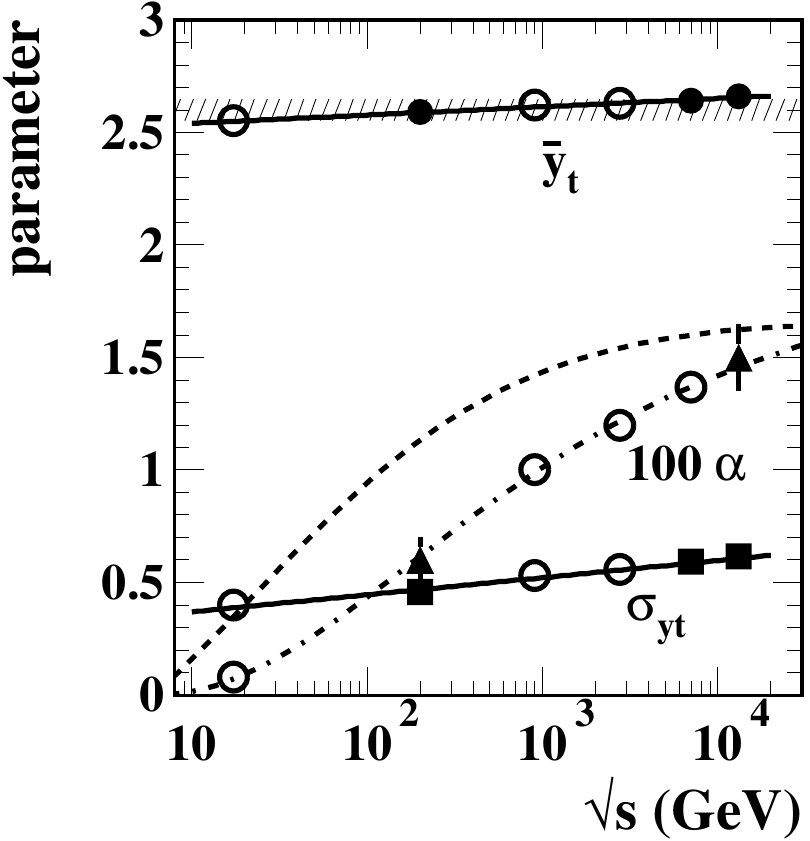}
  \includegraphics[width=1.65in]{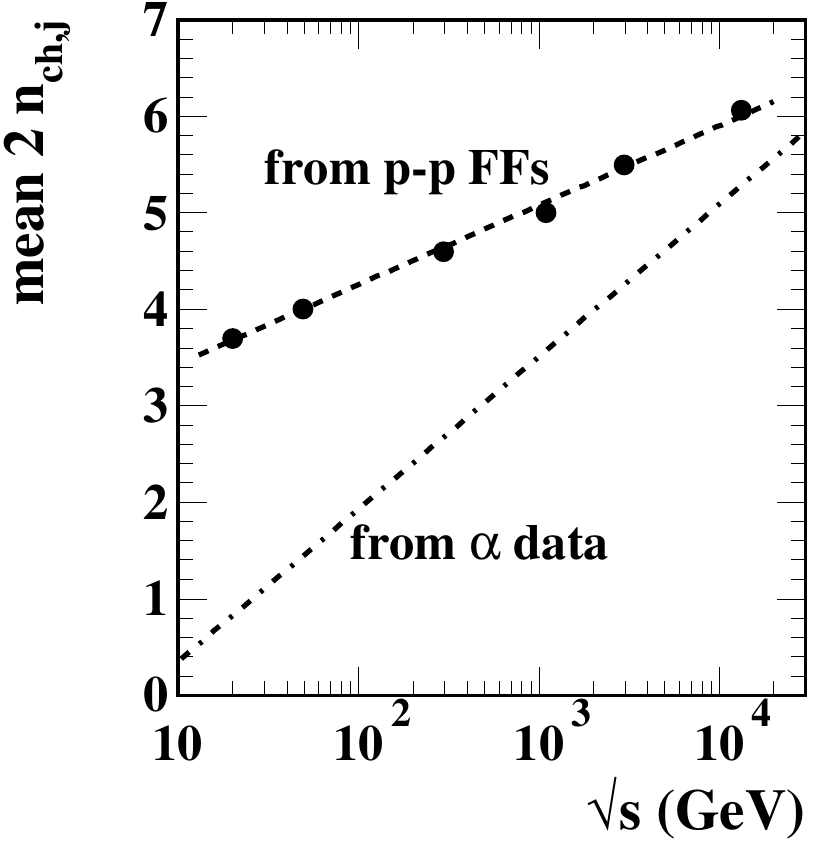}
\caption{\label{params}
Left: TCM NSD hard-component parameters determined by analysis of spectrum data (solid points). Open circles are interpolations or extrapolations relevant to this study. The solid lines are fits to data. The dashed and dash-dotted curves related to $\alpha(\sqrt{s})$ are described in the text.
Right: Two scenarios describing the energy dependence of mean dijet fragment multiplicity $2 \bar n_{ch,j}(\sqrt{s})$ based on Refs.~\cite{eeprd,fragevo,jetspec2}.
 }  
\end{figure}

Figure~\ref{params} (left) shows NSD TCM hard-component model parameters (points) vs collision energy. The solid points are derived from data. The open points are interpolations or extrapolations derived from the inferred or predicted trends in the figure (curves). The trends for $\bar y_t$ and $\sigma_{y_t}$ are consistent with straight lines. Whereas $\sigma_{y_t}$ increases by 50\% the upper limit on $\bar y_t$ variation is five percent (hatched band) and $\bar y_t$ may not actually vary significantly over the large energy interval. 

Soft-hard parameter $\alpha$ is defined by $\bar \rho_h = \alpha \bar \rho_s^2$. Values inferred from differential analysis of \pt\ spectra, as for 200 GeV in Ref.~\cite{ppprd} and for 13 TeV in the present study, are denoted by solid triangles in Fig.~\ref{params} (left). A model for $\alpha$ energy dependence can be determined by the following argument. $\alpha$ is related to jet systematics via $\bar \rho_h$ as
\bea
\bar \rho_{h,NSD} &=& \epsilon(\Delta \eta) f_{NSD}  2\bar n_{ch,j}
\eea
for a given collision energy, where $2\bar n_{ch,j}$ is the mean hadron fragment multiplicity per dijet averaged over a jet spectrum for that energy~\cite{eeprd} and $f_{NSD} = (1/\sigma_{NSD}) d\sigma_{jet} /  d\eta$~\cite{fragevo}. The energy trends for those quantities, inferred from reconstructed-jet data, and $\bar \rho_s$ from App.~\ref{nsedep} can be used to predict the energy trend for $\alpha(\sqrt{s})$. Certain defined kinematic quantities are useful: $y_{max} = \ln(2E_{jet}/m_\pi)$ is a logarithmic representation of jet energy, and $y_{b} = \ln(\sqrt{s}/m_\pi)$ similarly represents the \pp\ collision energy. $\Delta y_b = \ln(\sqrt{s}/\text{10 GeV})$ represents an observed cutoff of dijet production near $\sqrt{s} = 10$ GeV, and  $\Delta y_{max} = \ln(\sqrt{s}/\text{6 GeV})$ responds to an inferred infrared cutoff of jet spectra near $E_{jet} = 3$ GeV. Reference~\cite{jetspec2} reports  $d\sigma_{jet}/d\eta \approx 0.026 \Delta y_b^2 \Delta y_{max}$ and $\sigma_{NSD} \approx 0.83 (32 + \Delta y_b^2)$. From Fig.~\ref{edep} $\bar \rho_s \approx 0.81 \Delta y_b$. Dijet acceptance factor $\epsilon \approx 0.6$ is an estimate for  $\Delta \eta = 1.5$ - 2~\cite{jetspec}.  Combining various elements the $\alpha(\sqrt{s})$ trend is
\bea \label{alphas}
\alpha(\sqrt{s}) &\approx& \frac{\epsilon(\Delta \eta)}{\sigma_{NSD}} \frac{d\sigma_{jet}}{d\eta} \frac{2\bar n_{ch,j}}{\bar \rho_s^2}
\\ \nonumber
&\approx& \frac{ 0.6 \times 0.026\Delta y_b^2 \Delta y_{max} \times 2\bar n_{ch,j}}{0.83(32 + \Delta y_b^2)  \times (0.81 \Delta y_b)^2}
\\ \nonumber
 &\approx& \frac{0.03 \Delta y_{max}}{32 + \Delta y_b^2} \times 2\bar n_{ch,j}(\sqrt{s})
\eea
It remains to determine the systematic variation of dijet fragment multiplicity $2\bar n_{ch,j}$ with \pp\ collision energy.

Figure~\ref{params} (right) shows the energy trend for factor $2\bar n_{ch,j}(\sqrt{s})$ in Eq.~(\ref{alphas}) estimated in two ways. The first estimate is based on published FFs and jet spectra. For each collision energy a parametrization of the jet spectrum for that energy from Ref.~\cite{jetspec2} is used to obtain the weighted mean $2\bar n_{ch,j} (\sqrt{s})$ of dijet fragment yields $2n_{ch,j} (E_{jet})$ for \pp\ collisions from Fig.~6 of Ref.~\cite{fragevo}. The weighted means $2\bar n_{ch,j}(\sqrt{s})$ for six collision energies (points) are parametrized by $2\bar n_{ch,j}(\sqrt{s}) \approx 3(1+\Delta y_{max}/10)$ (dashed). When inserted into Eq.~(\ref{alphas}) that expression produces the dashed curve in the left panel which deviates substantially from the $\alpha(\sqrt{s})$ trend inferred from spectrum analysis (solid triangles).

An alternative estimate is the simple proportionality $2\bar n_{ch,j}(\sqrt{s}) \approx 0.7 \Delta y_{max}$ (dash-dotted) in the right panel.  When inserted into Eq.~(\ref{alphas}) that result produces the dash-dotted curve in the left panel that describes well the $\alpha(\sqrt{s})$ trend inferred from spectrum analysis. The same expression is used to generate the solid curve in Fig.~\ref{edep}. This comparison establishes an absolute relation between jet fragments within reconstructed jets and jet fragments manifesting as spectrum hard components. The comparison suggests that FFs from \ppbar\ collisions, already substantially modified (sharply reduced at lower fragment momentum) as compared to \ee\ collisions~\cite{fragevo}, may still overestimate total fragment yields at lower \pp\ collision energies by 50-100\%. Those issues may be related to factor-2 disagreements between NLO pQCD theory predictions and measured \pt\ spectra~\cite{alicerats}.

\section{Systematic uncertainties} \label{syserr}

The main purpose of this study is to extend the \pp\ \pt\ spectrum TCM established at the RHIC to the highest available collision energies. Previous studies of TCM \nch\ dependence have been based on  isolated \pt\ spectra~\cite{ppprd,ppquad} whereas spectrum ratios are presently available at LHC energies as in Ref.~\cite{alicespec}. Spectrum ratios do not clearly distinguish two dominant hadron production mechanisms and are thus difficult to parametrize accurately or to interpret. However, within the TCM framework spectrum ratios can be processed to isolate soft and hard components as demonstrated in this study. In this section the accuracy of the extended TCM is evaluated.

\subsection{200 GeV spectrum TCM and spectrum ratios}

Systematic uncertainties for 200 GeV \pp\ \pt\ spectra as in Fig.~\ref{ppspec1} are described in Sec. VIII of Ref.~\cite{ppprd}. The most uncertain aspect of the TCM is the shape of the low-\pt\ part of hard component $H(p_t)$ (below the mode) that results from subtraction of inferred model $\hat S_0(p_t)$, and the uncertainty is greatest for the lowest multiplicity class. While adjusting the amplitude of $\hat S_0(p_t)$ by a few percent may alter the structure below the mode the result does not match the shape variation with \nch\ observed there.

Given the imposed \nch\ bins Table~\ref{multclass} uncertainties for $\bar \rho_0$ and $\bar \rho_s$ depend on the estimates for $\xi = 0.66\pm0.02$ and $\alpha = 0.006\pm0.001$. The value of $\xi$ is estimated from the integrals of $\hat S_0(p_t)$ and $S_0'(p_t)$ (dotted curves in Fig.~\ref{ppspec1} -- left) that describe the data. The value of $\alpha$ is derived from an iterative process described in Refs.~\cite{ppprd,ppquad} where it was established empirically that $\bar \rho_h = \alpha \bar \rho_s^2$ with $\alpha \approx 0.006$ and $\bar \rho_h$ is the integral of spectrum hard component $H(p_t)$. The variations in $\hat H_0(p_t)$ introduced in Sec.~\ref{hcev} do not change that relation since small changes to the high-\pt\ tail do not significantly affect the integral. Figure~\ref{corrrat} (left) indicates that the resulting spectrum scaling is self consistent to a few percent.

A new aspect of this study is the \nch\ dependence of the TCM hard component. The 200 GeV parameter values in Fig.~\ref{check} (left) provide an indication of the precision. The points represent best-fits-by-eye, and the values were not adjusted after the fit procedure. Small deviations from the simple $\bar \rho_s$ trends in Eq.~(\ref{parameq}) suggest parameter precision at the few-percent level.
Fig.~\ref{check} (right)  appears to confirm  Fig.~\ref{corrrat} (left) as to the quality of the TCM description over a {\em 100-fold variation in dijet production rate and ten-fold variation in soft-hadron density}.

The 200 GeV spectrum ratios are in principle precise since \pt-dependent tracking inefficiencies cancel and $\bar \rho_s$ is well-defined. The relation of absolute spectra to the TCM with varying hard component is suggested by Fig.~\ref{corrrat} (left) with r.m.s.\ deviation 3\% for $n  =2$-6 spectra. However, the apparent precision can be misleading as discussed in the next subsection and Sec.~\ref{hardev}. Spectrum ratios alone representing only a fraction of the available information must be supplemented by isolated spectra as in Sec.~\ref{13tevspec}. And comparisons between models and data must demonstrate the degree of {\em statistical precision}. Qualitatively, the \nch\ dependence of spectrum ratios and the precision of inferred $T(p_t;n_{ch}',\sqrt{s})$ for several collision systems {\em do} confirm a peaked spectrum hard component with mode near 1 GeV/c in agreement with Ref.~\cite{ppprd}.

\subsection{Spectrum precision and statistical significance} \label{significance}

It is useful to specify the r.m.s.\ Poisson errors for spectrum $ \bar  \rho_0'(y_t;n_{ch}')$ on \yt\ (assuming uncorrected spectra)
\bea \label{poisson}
\delta \bar  \rho_0'(y_t;n_{ch}') &=& \frac{\sqrt{\bar  \rho_0'(y_t;n_{ch}') }}{\sqrt{y_t dy_t \Delta \eta N_{evt}(n_{ch}')}}
\eea
where $N_{evt}(n_{ch}')$ is the number of events for event class $n_{ch}'$. If spectra are normalized by soft charge density $\bar \rho_s$ the errors for $ \bar  \rho_0'(y_t;n_{ch}')/\bar \rho_s$ are
\bea \label{poisson2}
\delta \bar  \rho_0'(y_t;n_{ch}') / \bar \rho_s &=&  \frac{\sqrt{\bar  \rho_0'(y_t;n_{ch}') / \bar \rho_s }}{\sqrt{y_t dy_t \Delta \eta \bar \rho_s N_{evt}(n_{ch}')} }
\eea
The form ${\Delta \rho}/{\sqrt{\rho_{ref}}}$ equivalent to Pearson's normalized covariance~\cite{pearson} has been introduced previously as a {\em per-particle} measure of two-particle angular correlations~\cite{anomalous}. In Eq.~(\ref{info}) the equivalent quantity for SP spectra,  introduced with similar structure to assess the statistical significance of spectrum structure, is based on the Poisson error expression in Eq.~(\ref{poisson}) and measures fit residuals relative to statistical errors as in Fig.~\ref{tcm9} (b). 
Comparing spectra in the form of a ratio $\rho_{0,dat}(p_t) / \rho_{0,ref}(p_t) \sim \Delta \rho / \rho_{ref} + 1$ as in Fig.~\ref{corrrat} (left) introduces an additional factor $1/\sqrt{\rho_{ref}}$ compared to ${\Delta \rho}/{\sqrt{\rho_{ref}}}$ that strongly suppresses {\em apparent} residuals at lower \pt\ given typical spectrum variation with \pt\ over 6-8 orders of magnitude. 

In Ref.~\cite{ppprd} Figs.~2 (left) and  6 show residuals scaled by statistical errors in the form ${\Delta \rho}/{\sqrt{\rho_{ref}}}$, with ``data'' referring to normalized spectra as in Eq.~(\ref{poisson2}) above and $\Delta \eta = 1$. The assumed prefactor $\sqrt{y_t N_{evt}(n_{ch}')}$ in those figures then omits factors $dy_t  \bar \rho_s(n_{ch}')$. The main effect of the additional factors would be to reduce the scaled residuals for the lower event classes while increasing those for the higher, but the \pt\ or \yt\ structure is comparable. 

In Fig.~\ref{specrat2} (right) the \nch\ dependence of $\hat H_0(y_t,n_{ch}')$ at higher \pt\ appears substantial whereas no such structure is evident in Fig.~6 of Ref.~\cite{ppprd} describing the same collision system. There are two reasons: (a) The event number for Ref.~\cite{ppprd} was 3M distributed across eleven multiplicity classes whereas for Ref.~\cite{ppquad} the event number was 6M distributed across seven multiplicity classes, and the acceptance for the former was $\Delta \eta = 1$ whereas for the latter it was $\Delta \eta = 2$, the combination producing a 3-fold difference in the spectrum r.m.s.\ statistical error. (b) Figure~6 of Ref.~\cite{ppprd} is in the form ${\Delta \rho}/{\sqrt{\rho_{ref}}}$ whereas Fig.~\ref{specrat2} (right) of the present study is in the form $\Delta \rho / \rho_{ref} + 1$. The extra factor $1/\sqrt{\rho_{ref}}$ in the latter (varying over {\em several orders of magnitude}) strongly suppresses deviations at lower \pt\ and exaggerates deviations at higher \pt.
Spectrum comparisons in the form $\Delta \rho / \rho_{ref}$ may indicate a certain level of precision but provide no indication of statistical significance. Highly significant deviations at smaller \pt\ or \yt\ may be strongly suppressed leading to misconceptions.  Spectrum comparisons in the form $\Delta \rho /{\sqrt{\rho_{ref}}}$ relate data-model differences to statistical uncertainties presented uniformly over the full \pt\ acceptance. The differences in two descriptions of the same data can be dramatic as demonstrated in  Sec.~\ref{hardev}.

\subsection{13 TeV $\bf n_{ch}$ binning and $\bf n_{ch}$ energy dependence}

In Refs.~\cite{ppprd,ppquad} accurate independent determination of $\bar \rho_s$ for each multiplicity class facilitated the 200 GeV analysis. In order to provide equivalent information for analysis of Ref.~\cite{alicespec} data information on multiplicity distributions from Ref.~\cite{alicemult} was employed, specifically the NBD parameters in  Table~\ref{nbdtab}. The NBD fits to MD data are accurate to a few percent up to $n_{ch} / \Delta \eta = 20$ which includes almost all events. Based on multiplicity bins defined in Ref.~\cite{alicespec} (denoted here by A, B and C) the bin means were determined from the  NBD parameters in Table~\ref{nbdtab}.

Figure~\ref{edep} shows various charge-density measurements vs collision energy and can be used to estimate systematic uncertainties. The solid dots are mean values for several energies obtained from the NBD parameters in Table~\ref{nbdtab}. They are described by the TCM trend (solid curve) within 5\% and are consistent with results from Table 12 of  Ref.~\cite{alicemult} with 4\% uncertainties. The open squares summarize data for three event types (INEL, NSD, INEL $> 0$) from Table 7 of Ref.~\cite{alicemult}. The combined uncertainty is less than 3\% for each point. The summary suggests that the NBD parametrizations in Table~\ref{nbdtab} provide mean values with accuracy better than 5\%. 

Based on the bins defined in Ref.~\cite{alicespec} and indicated in Fig.~\ref{nbd} (right) the corrected bin means as $\bar \rho_0 = n_{ch} / \Delta \eta$ are 3, 9 and 15 and the ensemble mean is 6.2. Bin-mean {\em ratios} to the ensemble mean are 0.48, 1.45 and 2.4 for bins A, B and C respectively and should be independent of efficiency or acceptance. Assuming $\xi \approx 0.6$ for Ref.~\cite{alicespec} data the corresponding $N_{ch}^{acc}$ values within $\Delta \eta = 1.6$ should be approximately 3, 8.6 and 14.5 with ensemble mean $\langle N_{ch}^{acc} \rangle \approx  6$ (compared to 6.73 from Ref.~\cite{alicespec}). The bin means themselves are not reported in Ref.~\cite{alicespec}

Returning to the energy trends in Fig.~\ref{edep} the NBD averages (solid dots) appear to favor the TCM trend (solid curve) whereas  direct density measurements (open squares) seem to favor the ``power law'' trend (dashed curve). While the two data sets are consistent within their systematic uncertainties those estimates may not reflect point-to-point uncertainties that could be significantly smaller. MB dijet production continues to scale as $\bar \rho_h / \bar \rho_s \sim \log(\sqrt{s}/\text{10 GeV})$ as low as 62 GeV~\cite{anomalous} and presumably still depends on low-$x$ gluon participants at that point represented by $\bar \rho_s$. Differential spectrum analysis at lower energies might help resolve apparent conflicts.

\subsection{13 TeV spectrum ratios}

Figure~\ref{basicrat} (left) shows three spectrum ratios from Ref.~\cite{alicespec}. The ratio data should in principle be precise due to cancellation of instrumental effects but do present certain difficulties. In the simpler form of Eq.~(\ref{eqnx}) (fourth line) quantity $X(p_t) \rightarrow \bar \rho_{sx} / \bar \rho_{s0}$ at higher \pt\ for bin $x$, and that ratio should be accurately known to a few percent. But in Fig.~\ref{basicrat} (right) the approximate ratios are 0.25, 0.85 and 1.75 for bins A, B and C respectively whereas 0.48, 1.45 and 2.4 are expected based on results from the previous subsection. That the bin B  ratio is asymptotically {\em less} than 1 suggests that {\em inverses} of  described spectrum ratios might actually be plotted. As demonstrated in Fig.~\ref{corrrat} (right) the same information should emerge from a TCM analysis in either case. 

In Eq.~(\ref{eqnr}) (fourth line) the asymptotic limit at lower \pt\ is determined by the same ratio $\bar \rho_{sx} / \bar \rho_{s0}$ as at higher \pt, thus providing a consistency check on the data. To obtain consistency the ratio for bin  A was rescaled by factor 0.97 and that for bin C was rescaled by factor 1.03, both adjustments within the systematic uncertainties ($\pm5$\%) of the spectrum data in Fig.~3 of Ref.~\cite{alicespec}. Bin B data were not processed further because deviations from unity (the physically relevant structure) are not significant.

Those issues notwithstanding, the TCM description of bin A and C ratio data is quite accurate as indicated by  Fig.~\ref{specfit} (right): fit residuals well within $\pm5$\% up to 16 GeV/c and consistent with individual data uncertainties. The TCM descriptions of other quantities [e.g.\ $R(p_t)$, $X(p_t)$, $Y(p_t)$] are of similar quality. The 13 TeV parameters extracted for the varying $\hat H_0(p_t;n_{ch}')$ model as shown in Fig.~\ref{check} (left) seem consistent within a few percent.

\subsection{TCM parameter energy systematics} \label{tcmenergy}

Table~\ref{engparam} summarizes the TCM over a large energy interval. The model accuracy can be assessed both by the simplicity of parameter variations and by comparisons with data that did not contribute to parameter inference. 

The simplicity of TCM parameter variations with collision energy is demonstrated by Figs.~\ref{softcomp} (right), \ref{enrat1} (left), \ref{params} (left) and \ref{edep}. Except for L\'evy exponent $n$ the TCM parameters vary approximately linearly with $\log(s/s_0)$. In Fig.~\ref{softcomp} (right) the $1/n$ uncertainties are typically the size of the solid points (17.2 GeV 5\%, 200 GeV 2\%, 13 TeV 3\%). In Fig.~\ref{enrat1} (left) $1/q$ uncertainties are again the size of the points (about 2\%). Hard-component mode $\bar y_t$ is approximately constant at $\bar y_t \approx 2.6$ with at most 4\% variation across all energies. One should distinguish between 200 GeV fixed-$\hat H_0(p_t)$ parameters consistent with the  $n = 3$ multiplicity class (see Fig.~\ref{check} -- left) and the 200 GeV NSD values in Table~\ref{engparam}. The fixed reference for 13 TeV is based on the table values for that energy. The values in Table~\ref{engparam} apply only to spectrum data below 10 GeV/c. For any collision energy the effective power-law parameter $q$ is expected to increase at higher \pt\ consistent with the energy trend of underlying jet spectra~\cite{jetspec2}.

The TCM can also be evaluated as a predictive model by comparison with data not contributing to its definition, as in Fig.~\ref{spec9} where a 0.9 TeV spectrum and spectrum ratios are addressed. All TCM parameters for that energy are the result of TCM predictions with no adjustment. The data descriptions in Fig.~\ref{spec9} are of similar quality to those for 200 GeV and 13 TeV albeit with reduced statistics.

Figure~\ref{tcm9} provides more  information on spectrum \nch\ dependence at 0.9 TeV in the form of power-law Eq.~(\ref{s0other}) fit parameters for spectrum data [open points in panels (c) and (d)]. The solid points are obtained from power-law fits to the 0.9 TeV TCM as defined by Table~\ref{engparam}. They show good agreement except for the lowest \nch\ values. Figure~\ref{tcm9} (b) suggests that the power-law model describes spectrum data adequately since the fit residuals are consistent with statistical uncertainties, but the conclusion is misleading. The small number of collision events translates to large statistical uncertainties -- the 0.9 TeV data cannot test models effectively.

The 200 GeV study in Ref.~\cite{ppprd} was based on 3M events distributed over eleven multiplicity classes within $\Delta \eta = 1$ whereas the study in Ref.~\cite{ppquad} and the present analysis are based on 6M events distributed over seven multiplicity classes within $\Delta \eta = 2$.  The two lowest multiplicity classes in Ref.~\cite{ppquad}  include more than 2M events each as in Table~\ref{multclass}. The difference relative to  Ref.~\cite{ppprd} is a nearly 3-fold decrease in r.m.s.\ statistical errors  that reveals new details of hard-component evolution with \nch\ (Sec.~\ref{hardev}).  

In contrast, the 0.9 TeV analysis in Ref~\cite{alice9} is based on less than 0.3M events distributed over more than twenty multiplicity classes instead of seven.  The lowest multiplicity classes included only 40K events each (the equivalent for Ref.~\cite{ua1} that introduced the power-law fit model was 10K events). The r.m.s.\ statistical error is then more than 7-fold larger than for the present study, and fine details of TCM component evolution cannot be resolved.

\section{discussion} \label{disc}

Reference~\cite{alicespec} mentions ``rich features'' exhibited by the \nch\ dependence of 13 TeV \pp\ \pt\ spectra based on spectrum ratio data represented by Fig.~\ref{ratdata} (left) of the present study. There is indeed substantial new information conveyed by the \nch\ systematics of \pt\ spectra but a TCM context is required to fully access that information.

\subsection{Monte-Carlo comparisons with data}

Reference~\cite{alicespec} includes comparisons of spectrum data with several MC models. Given the increasing availability of high-statistics data, systematic deviations of many tens of statistical error bars may appear between MCs and data suggesting that such models should be rejected. Large deviations persist despite efforts to tune model parameters to accommodate data. Complex MC structure and the many parameters make interpretation problematic. Reference~\cite{alicespec} concludes that spectrum data are in ``fair agreement'' with MCs ``but not in all details.''

In contrast, the TCM provides an accurate description of most data to the limits of statistical precision. Separate components are individually comparable with QCD theory (as in Ref.~\cite{fragevo}), and the few parameters have simple physical interpretations. Systematic variation of model parameters over a broad range of charge multiplicity and collision energy is limited, smooth and controlled by the $\log(s/s_0)$ trends expected for QCD.  TCM analysis of MC output as in Ref.~\cite{ppprd} and the present study could allow isolation of MC soft and hard components for independent examination. And MCs offer the possibility to switch off and on specific collision mechanisms to investigate correspondence with TCM observables. 

Based on TCM results from the present study MC deviations from data can be interpreted physically. In Fig.~5 of Ref.~\cite{alicespec} [Fig.~\ref{basicrat} (left) of this study] MCs diverge more quickly than data from the common intercept at unity and then saturate at larger \pt\ whereas the data do not. The shape of ratio $R(p_t)$ is controlled by the hard/soft ratio $T(p_t) = \hat H_0(p_t) / \hat S_0(p_t)$. Figure~5 of Ref.~\cite{alicespec} then suggests that the MC hard/soft ratio rises more quickly than data in that \pt\ interval. 
The corresponding MC $T_0(p_t)$ in Fig.~\ref{blog} (left) confirms that the MC hard component is shifted to lower \pt\ and is substantially larger in amplitude than implied by spectrum data, a result established for PYTHIA in Fig.~9 of Ref.~\cite{ppprd}.  

Saturation of MC spectrum ratios at larger \pt\ appears consistent with a fixed hard-component model independent of \nch, whereas spectrum data reveal strong bias of the underlying jet spectrum with increasing event multiplicity. Similar MC-data conflicts have been encountered with angular correlations~\cite{anomalous} and ensemble-mean-\pt\ data~\cite{alicempt,tomalicempt}

\subsection{A universal TCM for p-p hadron production}

The TCM provides a unifying description of two distinct hadron production mechanisms in high-energy nuclear collisions and is required by a large body of data, of which the present study and Ref.~\cite{alicespec} provide examples. The two elements of the TCM, both directly related to the QCD structure of projectile nucleons (i.e.\ parton distribution functions), represent projectile dissociation and large-angle parton scattering followed by dijet formation near midrapidity. The TCM as invoked in this study applies to yields and spectra near midrapidity. As reported in Ref.~\cite{ppquad} accurate description of 2D angular correlations also requires a third (nonjet quadrupole) component.

The few TCM parameters have simple and interconnected energy and multiplicity dependences permitting accurate parameter prediction by interpolation and extrapolation, as in Figs.~\ref{check} (left), \ref{varyyes} (left), \ref{softcomp} (right), \ref{enrat1} (left) and \ref{params} (left) and as summarized in Table~\ref{engparam}. Energy dependence relates to $\log(s/s_0)$ with $\sqrt{s_0} = O(\text{10 GeV})$ representing a kinematic limit on dijet production from low-$x$ gluons.  Hard and soft components are directly connected by the empirical relation $\bar \rho_h \approx \alpha(\sqrt{s}) \bar \rho_s^2$, and there is quantitative correspondence between spectrum hard components and eventwise-reconstructed jets~\cite{fragevo,jetspec}.


The TCM parameters and their dependences are $\alpha(\sqrt{s})$, $\bar \rho_s(\sqrt{s})$, $[T,n(\sqrt{s})]$, $[\bar y_t,\sigma_{y_t}(n_{ch}',\sqrt{s}),q(n_{ch}',\sqrt{s})]$, where the square brackets enclose soft and hard model-function parameters respectively and $T$ does not vary significantly over a broad range of collision systems. L\'evy exponent $n$ describes the ``power-law'' tail of soft component $\hat S_0(p_t)$ and in the form $1/n(\sqrt{s})$ increases with energy as $\sqrt{\log(s/s_0)}$.  The newly-observed soft-component energy dependence may provide insight on the internal structure of projectile nucleons prior to collision. Two hard-component parameters in the form $1/q(\sqrt{s})$ and $\sigma_{y_t}(\sqrt{s})$ increase linearly with $\log(s/s_0)$ as in Figs.~\ref{enrat1} (left) and \ref{params} (left), just as expected from the QCD $\sqrt{s}$ systematics of an underlying jet energy spectrum and buttressing a jet interpretation for the hard component. The hard-component mode $\bar y_t \approx 2.6$ ($\bar p_t \approx 1$ GeV/c) is approximately independent of collision conditions and may  reflect a universal ``infrared cutoff'' of jet energy spectra near 3 GeV~\cite{jetspec2}. 

Another observation newly derived from high-statistics 200 GeV \pp\ data is multiplicity bias of the hard-component shape. Hard-component parameters $q(n_{ch}')$ and $\sigma_{y_t+}(n_{ch}')$ vary as in Fig.~\ref{check} (left) to determine hard-component evolution with \nch\ {\em above} the hard-component mode. 13 TeV spectrum ratios suggest similar variations although a more limited $n_{ch}'$ interval is spanned. Below the hard component mode $\sigma_{y_t-}(n_{ch}')$ varies as in Fig.~\ref{varyyes} (left) in a manner  (anti)correlated with $\sigma_{y_t+}(n_{ch}')$ above the mode. There is apparently no corresponding mechanism incorporated in current MCs, as in Fig.~\ref{basicrat} (left) for instance. 

The TCM hard component has been extended in  the present study to the largest available collision-energy range, from ISR to LHC energies. The extent of correlation of the inferred spectrum hard component with measured properties of eventwise-reconstructed jets over the same interval can be assessed by comparing Fig.~\ref{enrat3} (right) of this study with Figs.~5 and 9 of Ref.~\cite{jetspec2} which presents a universal curve describing measured jet (scattered parton) energy spectra for \pp\ collision over the same energy interval. Complementary to that quantitative correspondence the TCM soft component, representing the majority of produced hadrons in any system, appears to be essentially independent of the collision system. That property seems to conflict with models based on significant particle rescattering and thermalization that depend on \aa\ centrality and collision energy.

\section{summary} \label{summ}

A two-component (soft + hard) model (TCM) of hadron production near midrapidity from high-energy \pp\ collisions was derived previously from the charge-multiplicity \nch\ dependence of \pt\ spectra from 200 GeV \pp\ collisions at the relativistic heavy ion collider (RHIC). 
Based on comparisons with QCD theory and other forms of data the two components have been interpreted to represent longitudinal projectile-nucleon dissociation and minimum-bias (MB) transverse dijet production.

 The hard  component of the TCM (a peaked distribution on transverse rapidity \yt\ with exponential tail) was previously assumed to be approximately independent of \nch.  Reanalysis of high-statistics 200 GeV spectrum data in the present study reveals that the entire hard-component shape varies significantly and smoothly with \nch. The parameter variations are described by simple functions. The hard-component model with mode near $p_t = 1$ GeV/c is observed to broaden above the mode and narrow below the mode with increasing \nch, suggesting that imposition of a condition on event multiplicity biases the underlying jet spectrum: requiring larger \nch\ biases to a harder jet spectrum with greater fragment yield. 

Spectrum \nch\ dependence at large hadron collider (LHC) energies was presented recently in the form of spectrum {\em ratios} that retain only partial information about spectrum structure. To extrapolate the TCM to LHC energies new analysis techniques are required. The LHC ratio method is applied to high-statistics 200 GeV spectrum data for which spectrum structure is well understood. A method is devised to derive TCM soft and hard components individually from spectrum ratios. The same methods are then applied to LHC spectrum ratios to derive separate soft and hard components for those spectra. The present analysis reveals similar \nch\ evolution of the hard-component shape at 13 GeV. Lower-statistics data at 0.9 TeV are consistent with that trend. 
The \nch\ dependence of popular Monte Carlo models presented in the same ratio format is consistent with a fixed hard component and differs substantially from data (a larger jet-fragment contribution peaking at lower \pt).

Given TCM results at 200 GeV and 13 TeV the general energy dependence of soft and hard TCM spectrum components is determined accurately from 17 GeV to 13 TeV, an energy interval where low-$x$ gluons play a major role in hadron production at midrapidity. The newly-observed energy evolution of the soft component, modeled by a L\'evy distribution on transverse mass \mt, suggests that the growth of soft-component \pt\ fluctuations (measured by L\'evy exponent $n$ in the form $1/n$) may result from Gribov diffusion. The energy dependence of hard-component parameters is controlled by QCD parameter $\log(s/s_0)$ as expected for jet-related structure. Energy trends observed at the RHIC extrapolate to LHC energies in a manner consistent with trends for event-wise-reconstructed jet spectra.

Although TCM components can be recovered from spectrum ratios using the techniques developed in this study, ratio methods are problematic for several reasons: (a) Spectrum ratios alone can only define a hard/soft ratio of TCM components: At least one complete spectrum must be introduced to achieve separation. (b) Without a TCM context spectrum ratios cannot be interpreted physically because they confuse at least two hadron production mechanisms. (c) Spectrum ratios tend to exaggerate structure at higher \pt\ and  suppress important structure at lower \pt\ {\em where most jet fragments appear}. (d) In conventional data/model ratio comparisons significant model discrepancies at lower \pt\ are suppressed. Direct comparison of data-model {\em differences} to statistical errors over the largest possible  \pt\ interval is imperative. 

Determination of an accurate TCM for {isolated} spectra (rather than ratios) over a broad range of event multiplicities and collision energies as in the present study establishes an accurate and efficient representation of a large volume of spectrum data. The TCM provides explanatory power, quantitative links to data manifestations from a variety of collision systems and direct links to QCD theory not provided by any other data model.

\begin{appendix}

\section{200 $\bf GeV$ $\bf p_t$  spectra $\bf vs$ $\bf n_{ch}$} \label{200gevspec}

To establish a context for spectrum-ratio results from Ref.~\cite{alicespec} at LHC energies the same ratio method is applied to 200 GeV \pt\ spectrum data that have a previously-established TCM and physical interpretations~\cite{ppprd,ppquad}. The hard-component model $\hat H_0(p_t)$ is assumed to be fixed for this treatment as in the cited references. In Sec.~\ref{hardev} that constraint is relaxed and a revised $\hat H_0(p_t;n_{ch}')$ varying with uncorrected $n_{ch}'$ to accommodate data is inferred.

\subsection{Spectrum ratios vs $\bf n_{ch}$ and model ratio $\bf T(p_t)$} \label{ppprdspecrat}

The 13 TeV spectrum-ratio data from Ref.~\cite{alicespec} in Fig.~\ref{ratdata} can be better interpreted by applying a similar ratio analysis to the 200 GeV data from Ref.~\cite{ppquad} in Fig.~\ref{ppspec1} where the spectrum structure is well understood. The spectrum-ratio data may be used to estimate TCM model ratio
\bea
T_0(p_t) &\equiv& \frac{\hat H_0(p_t)}{\hat S_0(p_t)}
\eea
assuming {\em fixed} hard-component model $\hat H_0(p_t)$, but the individual model components are not accessible from spectrum ratios alone. The intermediate quantities $X(p_t)$ and $Y(p_t)$ below facilitate inference of TCM {\em data} ratio $T(p_t)$  from measured spectrum ratios. 

Because 200 GeV spectra are normalized by corrected soft-component density $\bar \rho_s$ as in Fig.~\ref{ppspec1} (left) and Refs.~\cite{ppprd,ppquad} the relevant spectrum {\em data} ratio is the first line of
\bea \label{eqnx}
X(p_t;n_{ch1}',n_{ch2}')  &\equiv& \left(\frac{\bar \rho_{s2}}{\bar \rho_{s1} }\right)  \frac{ \bar \rho_0'(y_t;n_{ch1}')}{ \bar \rho_0'(y_t;n_{ch2}')}
\\ \nonumber
X_0(p_t;n_{ch1}',n_{ch2}')&=&  \left(\frac{\bar \rho_{s2}}{\bar \rho_{s1} }\right) 
\frac{\bar \rho_{s1} \hat S_0(p_t) + \bar \rho_{h1} \hat H_0(p_t)}{\bar \rho_{s2} \hat S_0(p_t) + \bar \rho_{h2} \hat H_0(p_t)}
\\ \nonumber
&=& \frac{1 + \alpha \bar \rho_{s1} T_0(p_t)}{1 + \alpha \bar \rho_{s2} T_0(p_t)}
\\ \nonumber
&\rightarrow&  \frac{\bar \rho_{s1} }{\bar \rho_{s2}} ~~~\text{for larger \pt}.
\eea
The corresponding TCM for spectrum ratios based on a fixed hard-component model   is the second line of Eq.~(\ref{eqnx}) assuming inefficiencies cancel in data ratios.

\begin{figure}[h]
  \includegraphics[width=1.65in]{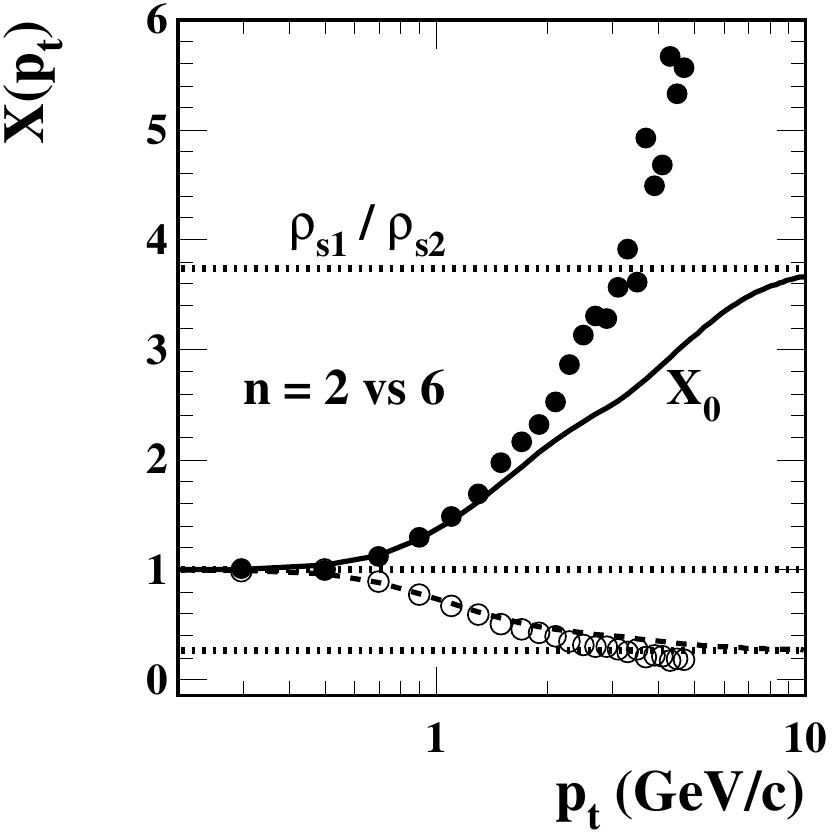}
  \includegraphics[width=1.65in]{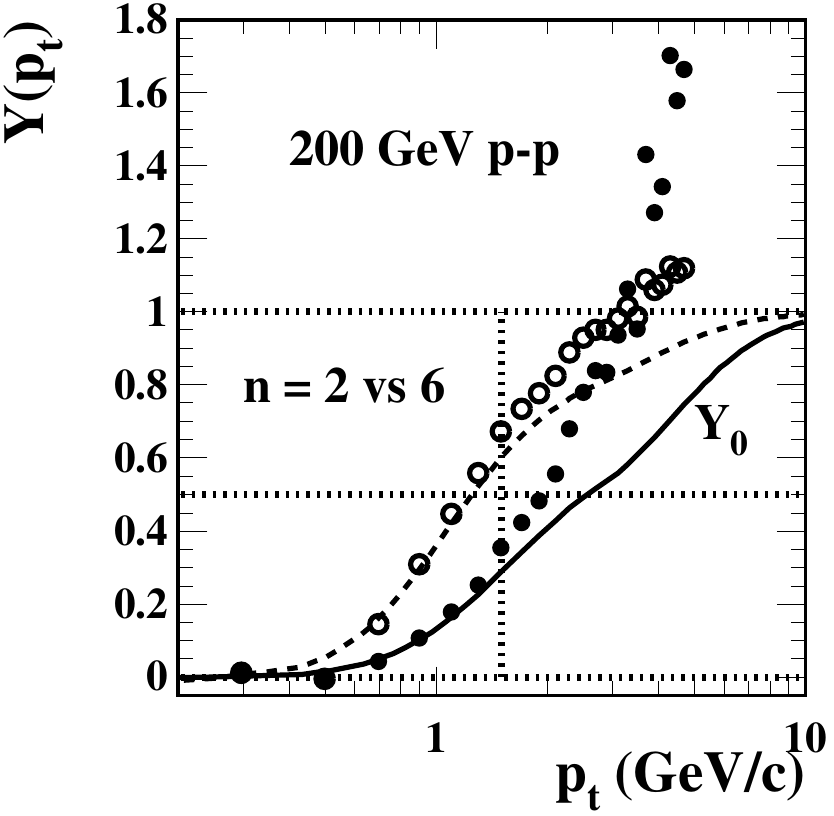}
\caption{\label{specrat1}
Left: Spectrum ratio $X(p_t;n_1,n_2)$ defined in Eq.~(\ref{eqnx}) (first line) for $n_1 = 2$ and $n_2 = 6$ (open points) and the inverse (solid points). The curves represent Eq.~(\ref{eqnx}) (third line) with 200 GeV fixed hard-component parameters from Table~\ref{engparam}.
Right: Quantity $Y(p_t;n_1,n_2)$ defined by Eq.~(\ref{yeq}) (first line) (points).  The curves represent TCM Eq.~(\ref{yeq}) (second line).
 } 
\end{figure}

Figure~\ref{specrat1} (left) shows spectrum ratios for $n_1 = 2$ over $n_2 = 6$ (open circles) and its inverse (solid points). Both cases are included to illustrate that apparently very different trends plotted in this way may be equivalent. Results for other combinations are similar. The soft-component densities are $\bar \rho_{s} = 3.38$ and 12.6 respectively from Table~\ref{multclass} (NSD is $\bar \rho_{s} \approx 2.5$).  The TCM in Eq.~(\ref{eqnx}) (third line) is represented by the solid and dashed curves. The density ratios (fourth line) are represented by upper and lower dotted lines as limiting cases of TCM $X_0(p_t;n_1,n_2)$. The discrepancy between ratio data and the TCM at larger \pt\  is addressed in Sec.~\ref{hcev}.

Figure~\ref{specrat1} (right) shows transformed data (points) as the first line of
\bea \label{yeq}
Y(p_t;n_{ch1}',n_{ch2}') \hspace{-.05in} &\equiv& \hspace{-.05in}  \frac{1}{\bar \rho_{s1} / \bar \rho_{s2} - 1}\left[X(p_t;n_{ch1}',n_{ch2}') \hspace{-.03in}-\hspace{-.03in} 1\right] ~~~~~
\\ \nonumber
&& \hspace{-1.2in} Y_0(p_t;n_{ch1}',n_{ch2}')= \frac{\alpha \bar \rho_{s2} T_0(p_t)}{1 + \alpha \bar \rho_{s2} T_0(p_t)}
= \frac{H_2(p_t)}{S_2(p_t) + H_2(p_t)}.
\eea
The curves $Y_0$ going asymptotically to 1 represent a fixed $\hat H_0(y_t)$ model and intercept 0.5 (dotted line) where $\alpha \bar \rho_{sx}T_0(p_t) = 1$ ($x = 1$ or 2), shifting left or right depending on multiplicity class $n$ in the denominator of $X(p_t)$. Data deviations relative to the TCM at higher \pt\ indicate that a fixed hard-component model does not describe ratio data adequately, but with minor adjustment the TCM becomes an accurate representation as in Sec.~\ref{hardev}.

\begin{figure}[h]
  \includegraphics[width=1.65in]{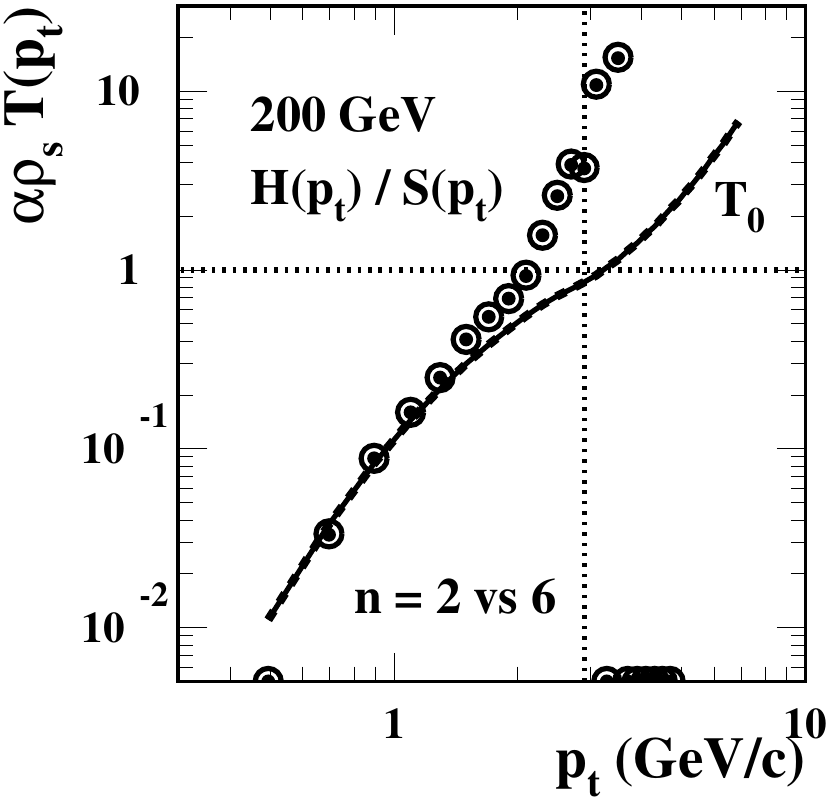}
  \includegraphics[width=1.65in]{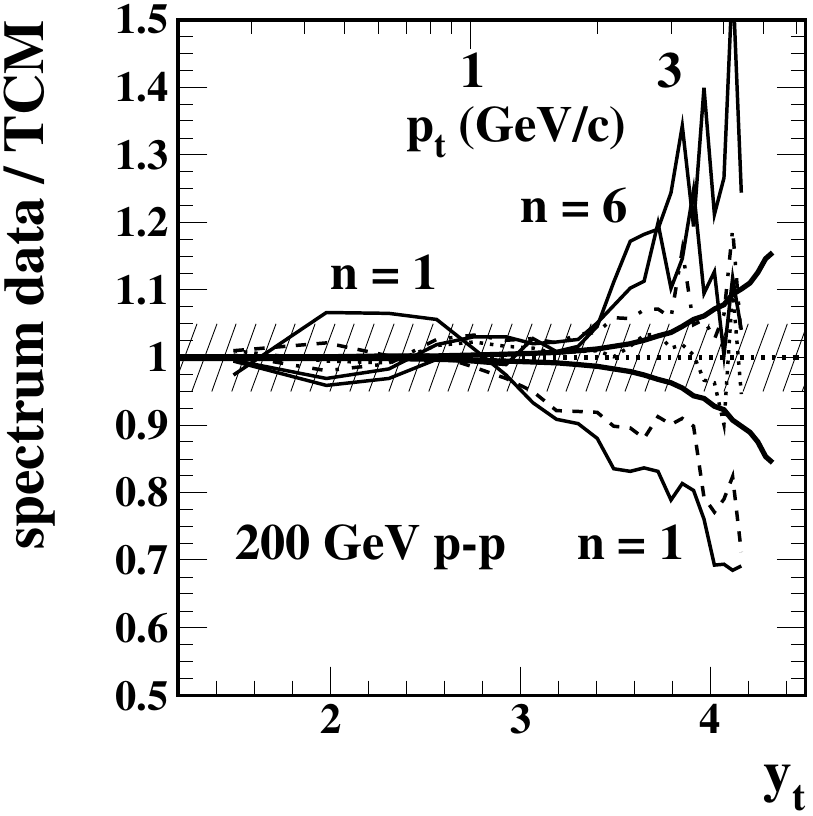}
\caption{\label{specrat2}
Left: Hard-soft ratio $T(p_t)$ (points) for two spectrum ratios. Since the ratios are inverses the data systems coincide here. The 200 GeV TCM model ratio $T_0(p_t) \equiv \hat H_0(p_y) / \hat S_0(p_t)$ (curve) is defined by parameters in Table~\ref{engparam} except hard-component $\sigma_{y_t}= 0.465$ and $q = 5.0$.
Right: Ratios of data spectra to TCM equivalents with {\em fixed} hard component for six multiplicity classes (curves with several line styles). The hatched band indicates $\pm5$\% fractional deviations. Deviations from unity at larger \yt\ suggest systematic bias of jet production with increasing $n_{ch}'$. The two smooth solid curves symmetric about unity represent one-sigma statistical errors and thus define an uncertainty band for the data.
 } 
\end{figure}

Figure~\ref{specrat2} (left) shows data (points) in the form
\bea \label{wrongy}
\alpha \bar \rho_s T(p_t) &\equiv&  \frac{Y(p_t)}{1-Y(p_t)} 
\eea
scaled to 200 GeV NSD \pp\ collisions ($\bar \rho_{s} \approx 2.5$).  Model $\alpha \bar \rho_s T_0(p_t)$ (curves) intercepts unity near $p_t = 3$ GeV/c ($y_t \approx 3.75$) corresponding to the hard-soft crossover in Fig.~\ref{ppspec1} (left). Data with $Y(p_t) > 1$ are undefined. The open and solid points coincide in this format. The superposed solid and dashed curves represent a common TCM $T_0(p_t)$ with fixed hard component corresponding to the bold dashed curves in Fig.~\ref{ppspec1}, to the dash-dotted and dotted lines in Fig.~\ref{check} (left) and approximately to convolution of a measured 200 GeV MB jet spectrum with measured \pp\ fragmentation functions~\cite{fragevo}. As noted, the strong discrepancies between data and TCM ratios indicate that the assumption of a fixed hard-component model should be revisited. However, ratio data alone do not provide sufficient information to reformulate the TCM.
 
Figure~\ref{specrat2} (right) shows ratios of spectrum data from Fig.~\ref{ppspec1} (left) to the corresponding TCM expression in Eq.~(\ref{ppspec}) with fixed hard component. The hatched band indicates $\pm 5$\% deviations. Above $y_t = 3$ significant systematic variation (10\% increase per multiplicity class at 4 GeV/c) suggests the requirement for a decreasing power-law exponent $q$ with increasing \nch, as might be expected if demand for larger event multiplicities biases the jet spectrum. Although the ratio deviations from unity at lower \yt\ are smaller in absolute magnitude they are {\em statistically more significant} as discussed in Sec.~\ref{significance}. Hard-component \nch\ dependence (model parameter variations) for 200 GeV \pp\ collisions  is established in Sec.~\ref{hardev} via direct comparisons between {isolated spectra} and the TCM.

\section{0.9 TeV spectrum TCM} \label{900gev}

The TCM energy parametrization summarized in Table~\ref{engparam} and multiplicity dependences determined in Sec.~\ref{hardev} can be tested by comparison with additional data from an intermediate energy. Reference~\cite{alice9} presents a \pt\ spectrum from 0.3M 0.9 TeV  INEL \pp\ collisions (Fig.~3) and corresponding multiplicity-dependent ratios (Fig.~6, lower), the ratios as in Fig.~\ref{basicrat} (left) of the present study.

\subsection{Predicting a 0.9 TeV spectrum TCM}

\begin{figure}[h] 
  \includegraphics[width=1.67in]{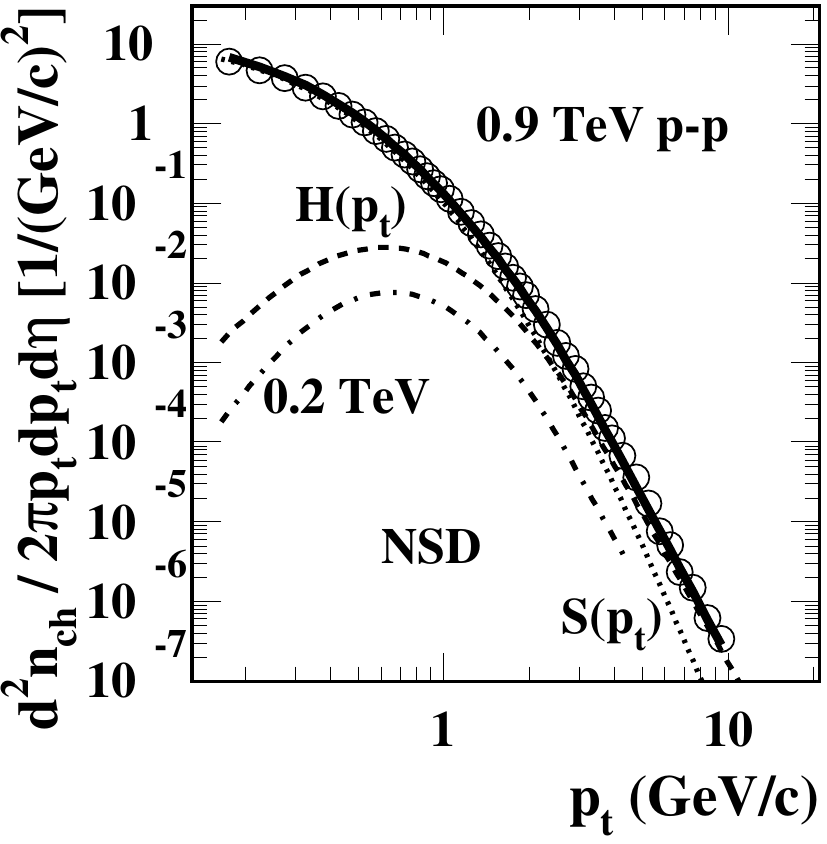}
  \includegraphics[width=1.63in]{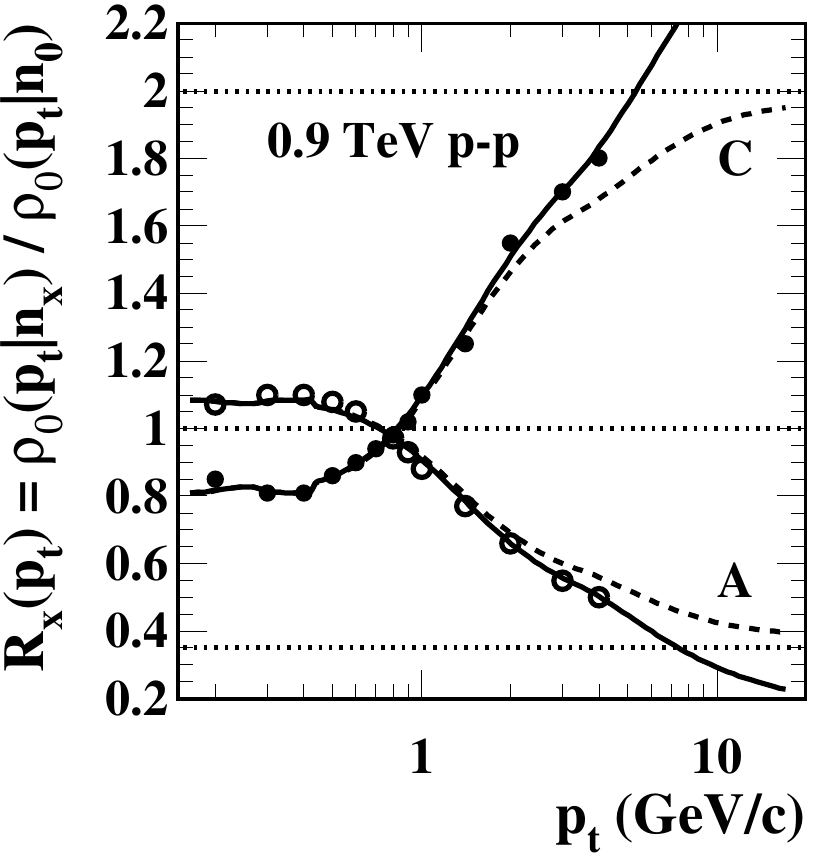}
\caption{\label{spec9}
Left: Spectrum data for a 0.9 TeV NSD \pt\ spectrum from Fig.~3 of Ref.~\cite{alice9} (points). The 0.9 TeV curves are predictions based on interpolated TCM parameters from Table~\ref{engparam}. The dash-dotted curve is the corresponding 200 GeV hard component for comparison.
Right: Spectrum ratio data derived from  Fig.~6 (lower panel) of Ref.~\cite{alice9} (points) for two multiplicity bins relative to a reference. Dashed curves are fixed-hard-component TCM predictions determined by parameters in Table~\ref{engparam}. Solid curves are obtained by small adjustments to hard-component parameters $\sigma_{y_t}$ and $q$ as for 13 TeV data. This panel can be compared with  Fig.~\ref{basicrat} (left).
 } 
\end{figure}

Figure~\ref{spec9} (left) shows spectrum data (points) from Fig.~3 of Ref.~\cite{alice9}. The solid curve is the corresponding TCM defined by  the parameters in Table~\ref{engparam} with no adjustments.  Also shown are predicted soft $S(p_t)$ (dotted) and hard $H(p_t)$ (dashed) TCM spectrum components for 0.9 TeV NSD collisions and the hard component for 200 GeV NSD collisions (dash-dotted) for comparison. The plot format is the same as for Fig.~\ref{specfit} (left).

Figure~\ref{spec9} (right) shows spectrum ratio $R(p_t;n_{ch}')$ data (points) from Fig.~6 (lower panel) of Ref.~\cite{alice9}. The reported bin-mean \nch\ ($n_{acc}$) values for three multiplicity bins are 3, 7 and 17. The data for central bin B convey no significant information (the spectra in ratio are too similar) and are omitted. The common reference for three ratios is the INEL spectrum. Based on the bin-B ratio data in Ref.~\cite{alice9} $\bar n_{acc} \approx 6$ is the effective INEL value.
Given the asymptotic ratio limits at larger \pt\ (dotted lines) the {\em effective} bin means should then be $\bar n_{acc} \approx 0.35 \times 6 = 2.1$ (A), $\approx 1.15 \times 6 \approx 7$ (B) and  $\approx 2 \times 6 = 12$ (C).

The dashed curves (approaching the dotted lines) represent Eq.~(\ref{eqnr}) with TCM parameters for 0.9 TeV and fixed $\hat H_0(p_t)$ from Table~\ref{engparam} unaltered. The data description is generally good with deviations from the TCM at larger \pt\ similar to those noted in Figs.~\ref{specrat1} (left) and \ref{basicrat} (left). The limited event number precludes detailed analysis of hard-component variation with \nch\ via spectrum ratios. The solid curves result from $q = 4.7$ and 4.3 and $\sigma_{y_t} = 0.525$ and 0.535 for bins A and C respectively. The variations are consistent with interpolation of the parameter  trends in Fig.~\ref{check} (left) between 200 GeV and 13 TeV

This comparison confirms that the overall spectrum TCM described in the present study serves as an accurate representation of \pt\ spectrum data over large \pt, energy and multiplicity ranges, with isolation of two components representing distinct hadron production mechanisms that may  be compared {\em independently} with relevant theory.

\subsection{Model fits with a power-law spectrum model}

Table~2 of Ref.~\cite{alice9} does provide significant information on spectrum multiplicity dependence in the form of model fits to spectra for various multiplicity classes. The model function [Eq.~(1) of Ref.~\cite{alice9}] is described as a ``modified Hagedorn function'' (i.e.\ power-law model)
\bea \label{s0other}
\bar \rho(p_t;n_{ch}') &=& \frac{p_t}{m_t} \frac{A(p_{t0},b)}{(1 + {p_t}/{p_{t0}})^b}.
\eea
Factor $p_t / m_t$ makes no difference to the fits described below and is omitted for simplicity. Parameter $p_{t0}$ can be expressed as $p_{t0} = b\, T$ with $b$ ($\leftrightarrow n$) and $T$ obtained independently for comparison with Eq.~(\ref{s0}). An indirect comparison can be made between 0.9 TeV data and TCM predictions via those fit results: The same fit model is applied to the 0.9 TeV TCM defined by Table~\ref{engparam} and the parameter trends are compared with those from  Ref.~\cite{alice9}. 

\begin{figure}[h]
  \includegraphics[width=3.3in]{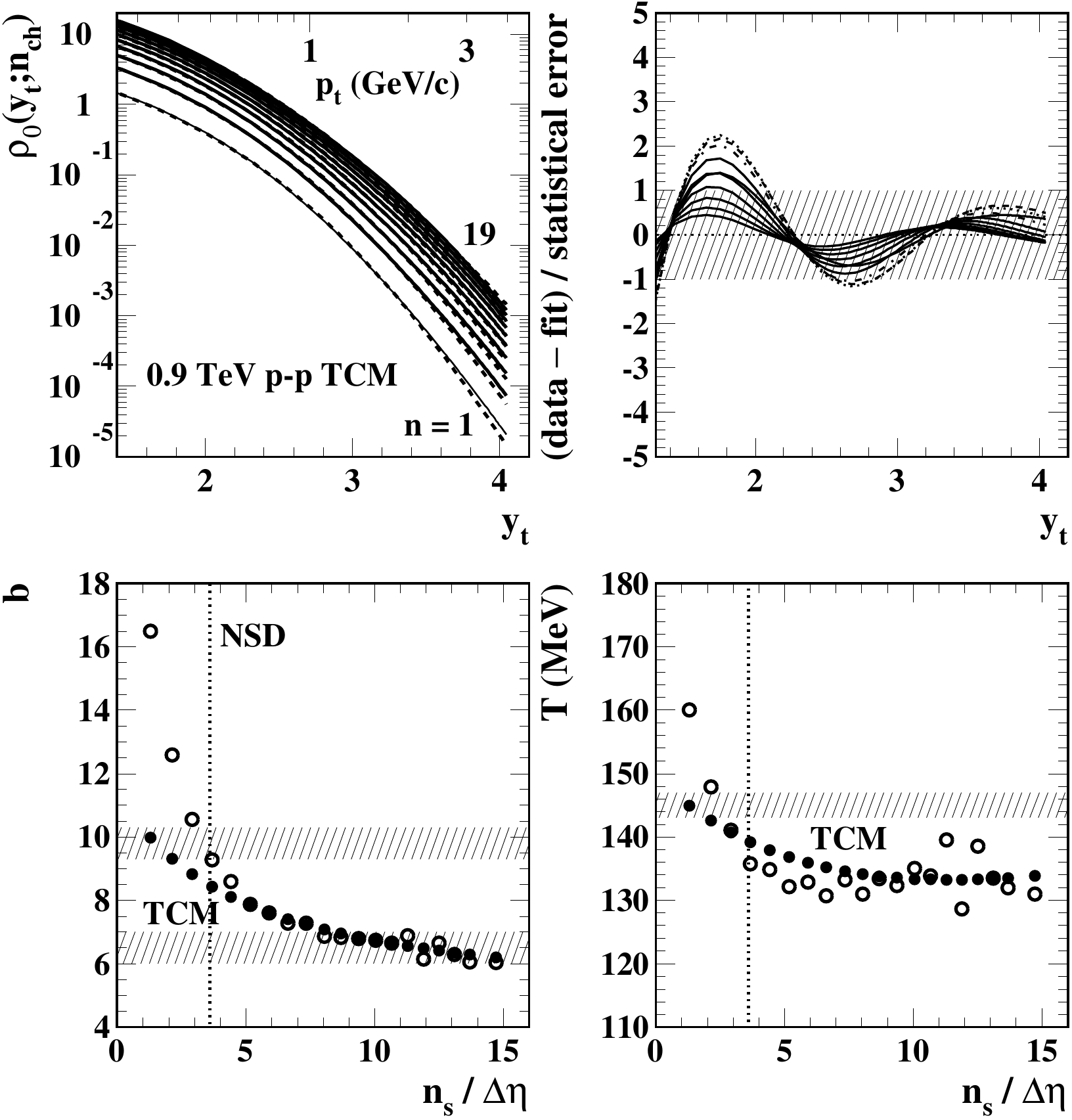}
\put(-142,215) {\bf (a)}
\put(-23,215) {\bf (b)}
\put(-142,85) {\bf (c)}
\put(-23,85) {\bf (d)}
\caption{\label{tcm9}
(a) TCM spectra for 0.9 TeV \pp\ collisions (solid) and fits to those spectra with Eq.~(\ref{s0other}) (dashed) for twenty multiplicity classes (odd-number classes are plotted) as defined in Ref.~\cite{alice9}.
(b) Fit residuals relative to statistical errors for spectrum fits in (a). The hatched band represents a large r.m.s.\ statistical error for the small event number.
(c) Fitted exponent $b$ from fits of Eq.~(\ref{s0other}) to TCM spectra (solid) vs those from fits to 0.9 TeV data (open) from Ref.~\cite{alice9}.
(d) Fitted slope parameter $T = p_{t0} / b$  from  fits to the TCM (solid) vs those from fits to data (open).
}
\end{figure}
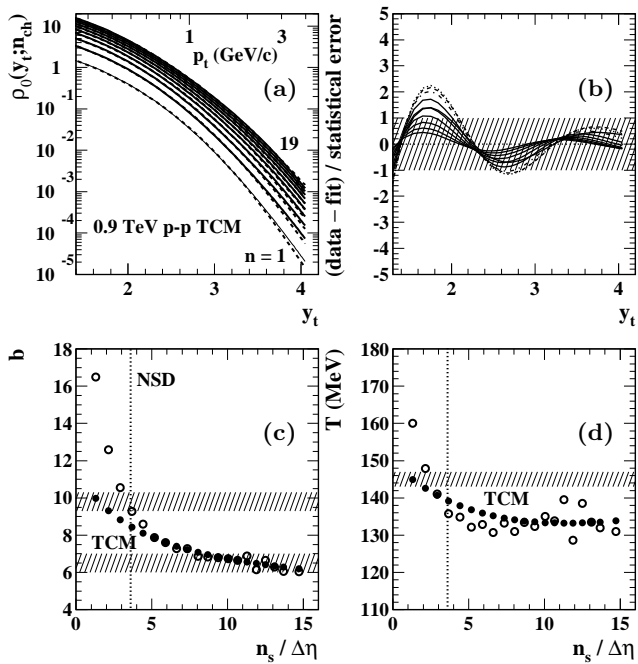

Note that Eq.~(\ref{s0other}) defined on \pt\ is significantly different at lower \pt\ from L\'evy distribution Eq.~(\ref{s0}) defined on $m_t - m_h$, and neither function can {\em in isolation} meaningfully represent the \nch\ dependence of an intact \pt\ spectrum because they cannot describe the spectrum hard component as predicted by measured jet properties~\cite{hardspec,fragevo,jetspec2}. Also note that Fig.~\ref{tcm9} compares Eq.~(\ref{s0other}) to the full spectrum TCM, not to Eq.~(\ref{s0}) alone.

Figure~\ref{tcm9} (a) shows TCM spectra with fixed hard component based on 0.9 TeV NSD parameters from Table~\ref{engparam} (solid) for 20 multiplicity bins as defined in  Ref.~\cite{alice9}. For each corrected $n_{ch}$ value the corresponding $\bar \rho_s$ was obtained using $\alpha = 0.01$ from Fig.~\ref{params} (left).  The power-law model Eq.~(\ref{s0other}) with parameters $(A,T,b)$ (dashed) was fitted to the TCM spectra.

Figure~\ref{tcm9} (b) shows fit residuals as spectrum differences {\em relative to \pt-bin statistical errors}~\cite{ppprd} in the form 
\bea \label{info}
\frac{\Delta \rho_0}{\sqrt{\rho_{0,ref}}}  &\rightarrow & \frac{\Delta n_{ch}'(p_t)}{\sqrt{n_{ch}'(p_t)}}
\\ \nonumber
&=&  \sqrt{N_{evt} p_t dp_t 2\pi \Delta \eta} \frac{\rho_{0,dat}(p_t) - \rho_{0,ref}(p_t)}{\sqrt{\rho_{0,ref}(p_t)}} 
\eea
 assuming Poisson bin errors, and $n_{ch}'(p_t)$ is the {\em accepted} charge multiplicity in a \pt\ bin summed over the event ensemble. Equation~(\ref{info}) measures  bin-wise information conveyed by a data spectrum relative to a reference (fit model, Monte Carlo).  Panel (b) shows fit residuals in the form relevant to $\chi^2$ minimization. The power-law model seems to describe the 0.9 TEV TCM adequately  (no excess beyond  statistics shown by the hatched band) but the event number is only 0.3M. For the spectrum data in Ref.~\cite{ppquad} the hatched band  would be reduced 7-fold and the same residuals then become very significant (see Secs.~\ref{significance} and \ref{tcmenergy}). The apparently-oscillating residuals from fits with the power-law model function have been described as ``log-periodic oscillations'' in Ref.~\cite{wilkosc}.

Figure~\ref{tcm9} (c) and (d) show fit results from Table 2 of Ref.~\cite{alice9} for two model parameters (open circles) representing $b$ and $T \equiv p_{t0} / b$. The solid points represent fits of Eq.~(\ref{s0other}) applied to the TCM as described above.  
In panel (c) the upper hatched band represents fixed $\hat S_0(m_t)$ L\'evy exponent $n \approx 9.8$ and the lower hatched band represents fixed $\hat H_0(y_t)$ parameter $q \approx 4.5$ (with $q + 2 = 6.5$ as the relevant exponent for \pt). As event multiplicity increases from left to right Eq.~(\ref{s0other}) attempts to accommodate in turn first the soft component alone then  the hard component. In panel (d) the hatched band represents fixed $\hat S_0(m_t)$ parameter $T = 145$ MeV. For larger multiplicities the fitted parameter drops away from the TCM value as the power-law model attempts to accommodate the increasing hard-component amplitude. The good agreement between most open and solid points indicates that the TCM is a satisfactory representation of 0.9 TeV spectrum data for all but the smallest multiplicities.

The data from Ref.~\cite{alice9} (open circles) rise substantially above fits to the TCM with fixed hard component (solid points) for the lowest multiplicity classes. Such differences should arise if the TCM does not model low-\pt\ excursions of the hard component for smaller event multiplicities, as in Fig.~\ref{ppspec1} (right). That similar excursions may appear over a range of collision energies is indicated by Fig.~4 (left) of Ref.~\cite{tomalicempt} where ensemble-mean-\pt\ hard-component trends  for all collision energies are reduced for the lowest multiplicity classes.

\section{$\bf p$-$\bf p$ Charge Multiplicities} \label{lhcmult}

In this appendix negative-binomial-distribution (NBD) models of multiplicity distributions (MDs) for several energies are summarized and energy dependence of charge yields relevant to the TCM is inferred. This material is presented to provide context for the analysis of LHC spectrum ratios in Secs.~\ref{13tevspecc} and \ref{edepp}.

The LHC spectrum study in Ref.~\cite{alicespec} refers to imposed charge-multiplicity conditions not fully specified. To better provide a comparison with RHIC data information on LHC charge multiplicities from Ref.~\cite{alicemult} can be used. In particular there are issues of consistency between direct $dn_{ch}/d\eta$ density measurements and multiplicity-distribution (MD) mean values $\bar n_{ch}$ that should coincide.  Charge densities on $\eta$ are reported in Table 7 of Ref.~\cite{alicemult} for three event classes and several energies. In this study results for NSD events and $|\eta| < 0.5$ are emphasized as a common reference. Results for other conditions are scaled accordingly. Table 12 of Ref.~\cite{alicemult} includes mean values $\bar n_{ch}$ from parametrizations of data MDs that may be contrasted with the direct density measurements.


\subsection{Multiplicity distributions and binnings} \label{multclass1}

Probability distributions on \pp\ event multiplicity can be described by a single negative binomial distribution (NBD) at lower collision energies, but at higher energies a double NBD is required by data~\cite{nbds}.  The relevant definitions are provided by Eqs.~(14) and (16) from Ref.~\cite{alicemult} 
\bea
P(n_{ch};\bar n,k) \hspace{-.05in} &=& \hspace{-.05in} \frac{\Gamma(n_{ch} + k)}{\Gamma(k)\Gamma(n_{ch} + 1)} 
\left( \frac{\bar n}{\bar n + k} \right)^{n_{ch}} \hspace{-.05in} \left( \frac{k}{\bar n + k} \right)^k~~~
\eea
\vspace{-.2in}
\bea \nonumber
P(n_{ch}) \hspace{-.05in} &=&\hspace{-.05in}  \lambda \left[ \alpha\, P(n_{ch};\bar n_1,k_1) + (1-\alpha)\, P(n_{ch};\bar n_2,k_2) \right].~~~
\eea
The NBD parameters for four energies are taken from Table 10 of Ref.~\cite{alicemult} for $|\eta| < 0.5$ ($\Delta \eta = 1$) and reproduced in Table~\ref{nbdtab} below. 
The parameter values for 13 TeV are extrapolated from the lower energies.

\begin{table}[h]
  \caption{Double-NBD parameters for NSD \pp\ collisions at several energies and $|\eta| < 0.5$, from Table 9 of Ref.~\cite{alicemult}. The 13 TeV (starred) entries are extrapolated from lower-energy data. The $\bar \rho_{00}$ entries are the actual means of the NBD models. The $\bar \rho_{s0}$ entries are taken from Fig.~\ref{edep} (dotted).
}
  \label{nbdtab}
\begin{center}
\begin{tabular}{|c|c|c|c|c|c|c|c|c|} \hline
 Energy (TeV) & $\lambda$ & $\alpha$ & $\bar n_1$ & $k_1$ & $\bar n_2$ & $k_2$  & $\bar \rho_{00}$  & $\bar \rho_{s0}$ \\ \hline
 0.9  & 0.94  & 0.55 & 2.4 &  2.6 & 6.0 & 3.3 & 3.8 & 3.61 \\ \hline
 2.76 & 0.93 & 0.51  & 2.5 & 2.6  & 8.0 &  3.1 & 4.8 & 4.55 \\ \hline
 7  & 0.94 & 0.70 & 3.6 & 1.8 & 12 &  4.1 & 5.6 & 5.35 \\ \hline
8  & 0.93 & 0.57 & 3.1 & 2.0 & 11 & 3.2 & 5.8 & 5.46 \\ \hline
 13$^*$ & 0.935 & 0.58 & 3.3 & 2.0 & 12 & 3.5 & 6.2 & 5.87 \\ \hline
\end{tabular}
\end{center}
\end{table}

\begin{figure}[h]
  \includegraphics[width=3.3in,height=1.6in]{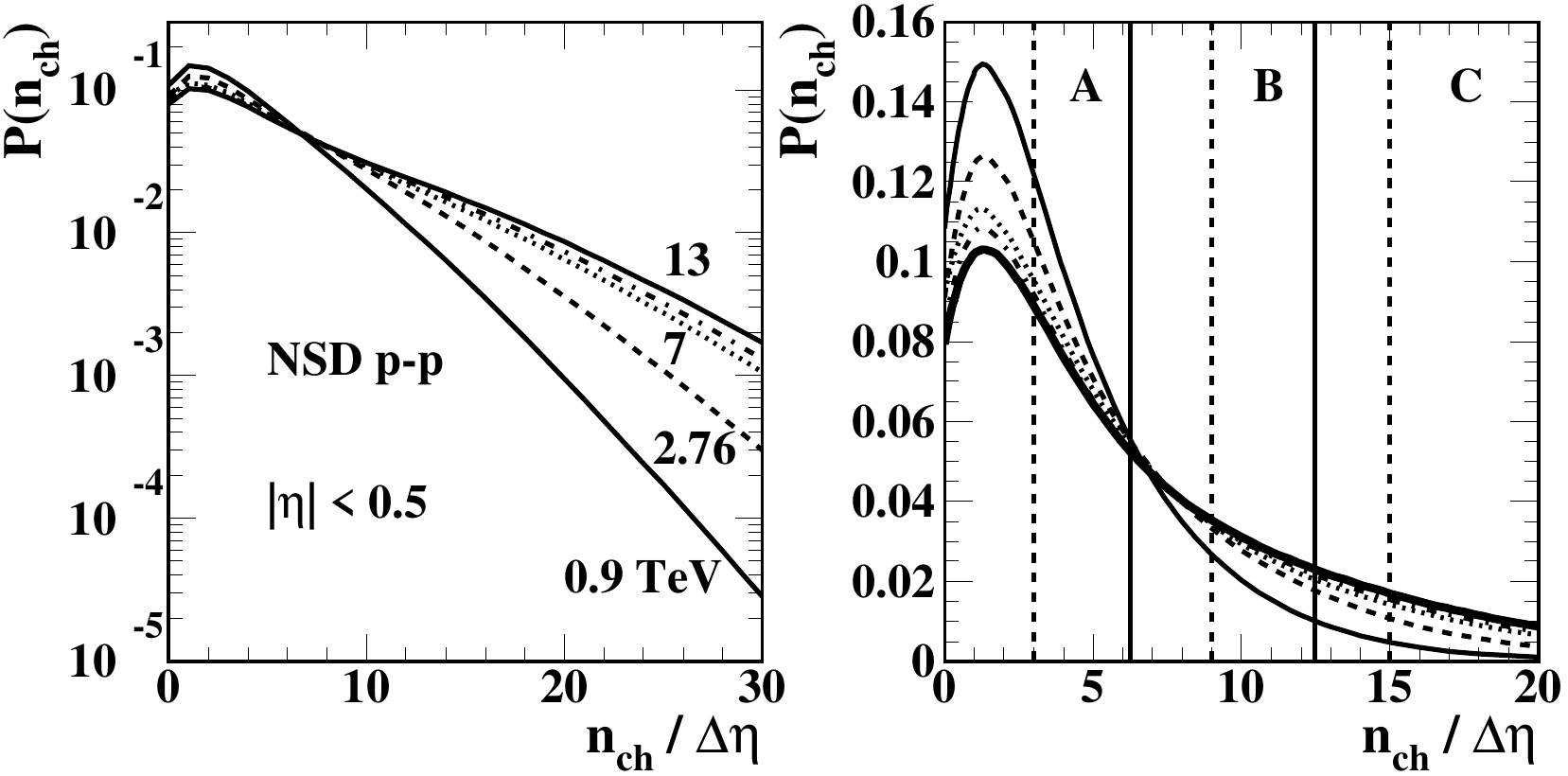}
\caption{\label{nbd}
Left: Fitted parametrization of measured probability distributions on \pp\ event multiplicity (MDs) for five collision energies~\cite{alicemult} based on a double negative binomial distribution (NBD) with parameters from Table~\ref{nbdtab}.
Right: Distributions from the left panel plotted on a linear scale. The vertical solid lines indicate multiplicity bin boundaries defined in Ref.~\cite{alicespec}. The vertical dashed lines are bin means determined in the present study from the fitted 13 TeV NBD distribution (bold).
 } 
\end{figure}

Figure~\ref{nbd} (left) shows  double NBDs from Ref.~\cite{alicemult} fitted to NSD data MDs on \nch. The fit residuals are at the percent level. Up to $n_{ch} / \Delta \eta$ there is no significant difference between single- and double-NBD fits. The double-NBD mean values are denoted by $\bar \rho_{00}$ in Table~\ref{nbdtab}.  Entries for $\bar \rho_{s0}$ are inferred from the dotted curve in Fig.~\ref{edep}.

Figure~\ref{nbd} (right) shows the same distributions on a linear scale with the multiplicity bin system defined in Ref.~\cite{alicespec}. Referring to the 13 TeV distribution (bold solid curve) the left solid vertical line marks the 13 TeV distribution mean $\bar \rho_{00} = 6.2$. The right solid line is at twice that value, together defining  three multiplicity bins A, B and C in Ref.~\cite{alicespec}. Those definitions are based on the accepted multiplicity $N^\text{acc}_{ch}$ within $\Delta \eta = 1.6$, but a single common efficiency factor should apply to all values.
The vertical dashed lines indicate three bin means located approximately at 3, 9 and 15 compared to ensemble mean $\bar \rho_{00} = 6.2$. The ratios to $\bar \rho_{00}$ are then approximately 3/6.2 = 0.48 for A,  9/6.2 = 1.45 for B and 15/6.2 = 2.4 for C.

\subsection{Event-multiplicity energy dependence} \label{nsedep}

 The energy dependence of \pp\ event multiplicities at collision energies up to 13 TeV can be inferred from several sources, including INEL $>0$ data from Ref.~\cite{alicespec} and data for several trigger conditions  from Ref.~\cite{alicemult}.

\begin{figure}[h]
  \includegraphics[width=3.3in]{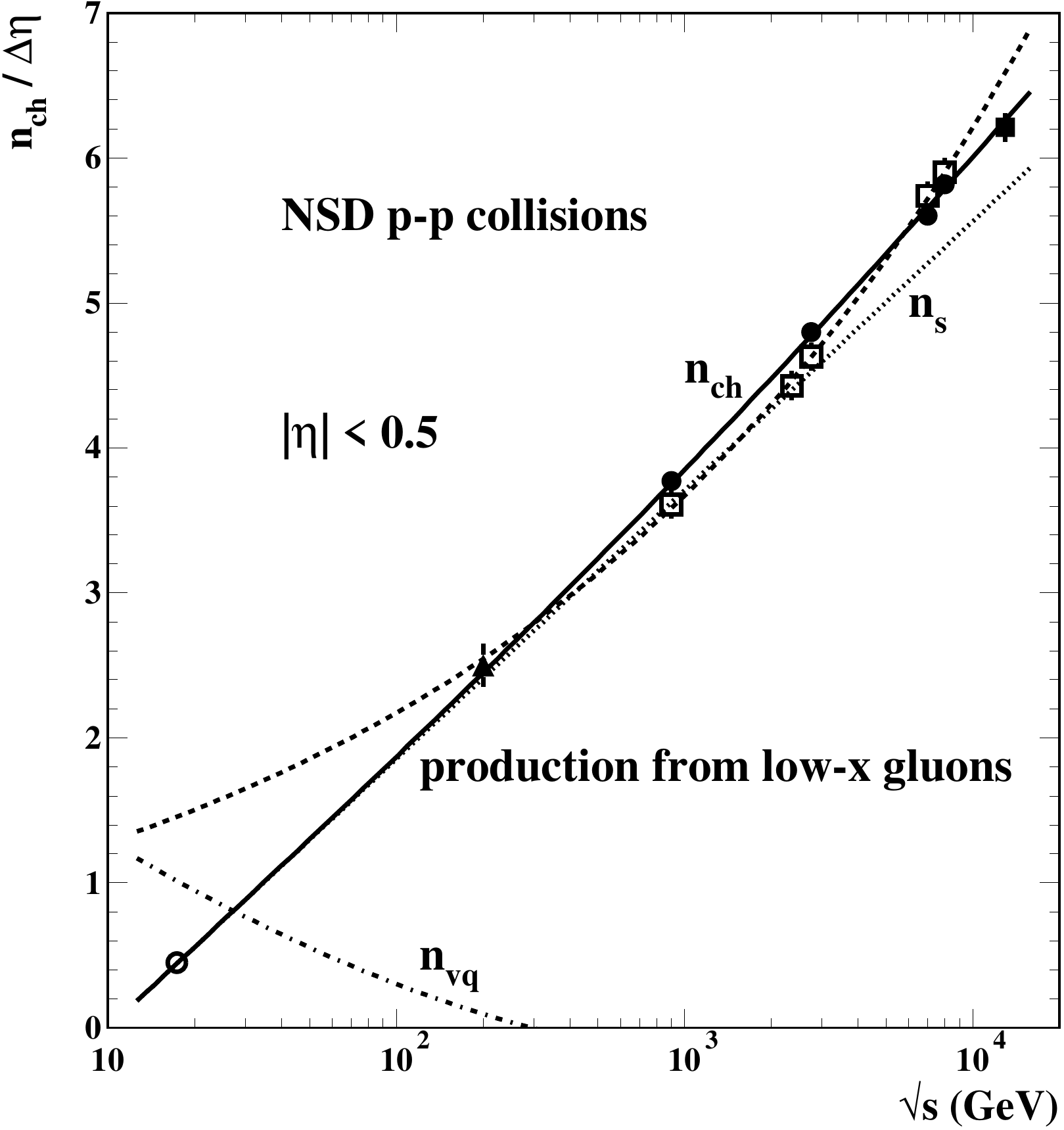}
\caption{\label{edep}
{\it p-p} collision-energy dependence of the charged-hadron angular density near $\eta = 0$. The open squares are from Table 7 of Ref.~\cite{alicemult}.  The solid square is 13 TeV INEL $>0$ datum 6.46 from Ref.~\cite{alicespec} scaled down by 1.04 to NSD value 6.21.  The solid dots are weighted means of NBD distributions in Fig.~\ref{nbd} from fits to measured MDs in Ref.~\cite{alicemult} reported as $\bar \rho_{00}$ in Table~\ref{nbdtab}. The solid triangle is an estimate of the NSD value for 200 GeV, and the open circle is an extrapolation to 17.2 GeV. Curves are described in the text.
 }  
\end{figure}

Figure~\ref{edep} shows data (open squares) for five energies from Table 7 of Ref.~\cite{alicemult} scaled to NSD (yields in Table 7 scale as  INEL:NSD:INEL $>0$ = 0.81:1.00:1.04).  The solid square is 13 TeV INEL $>0$ (inelastic events with at least one charged particle accepted) datum 6.46 from Ref.~\cite{alicespec} scaled down by 1.04 to NSD value 6.21. The solid dots are NBD means $\bar \rho_{00}$ from Table~\ref{nbdtab} (except 13 TeV).
The dotted curve is the soft-component estimate $\bar \rho_s \approx 0.81 \ln(\sqrt{s} / \text{10 GeV})$ interpreted to represent participant low-$x$ gluons from projectile dissociation. That trend determines the $\bar \rho_s$ values in Table~\ref{engparam}. The intercept 10 GeV is inferred from jet-related  energy trends at RHIC energies~\cite{anomalous}, and coefficient 0.81 was adjusted so that $\bar \rho_0 = n_{ch} / \Delta \eta$ best accommodates the open squares. The TCM trend (solid curve) is $\bar \rho_0 = \bar \rho_s + \bar \rho_h$ with $\bar \rho_h = \alpha \bar \rho_s^2$ and $\alpha(\sqrt{s})$ defined in Sec.~\ref{parsum}. The difference between solid and dotted curves is the jet-related contribution $\bar \rho_h$.

 The ``power law'' trend $0.76 s^{0.114}$ (dashed) reported in Ref.~\cite{alicemult} provides an approximate empirical description of the total charge density near $\eta = 0$ down to lower energies (especially data below 100 GeV), but that result may be misleading. One may question whether the power-law trend represents a single production mechanism when two or three mechanisms could contribute with very different energy dependences. The dotted and solid TCM curves are interpreted to represent hadron production from low-$x$ gluons that must fall to zero at lower collision energies, whereas hadron production relating to valence quarks {\em near  $\eta = 0$} (possibly the difference $n_{vq}$ between dashed and solid curves) should fall to zero at higher energies but may dominate at lower energies. 

Higher-energy data appear to support that picture. The solid dots are weighted means of NBD distributions in Fig.~\ref{nbd} that describe accurately the NSD MDs in Ref.~\cite{alicemult}. Those NBD-based estimates, systematically displaced from (but consistent with) ALICE NSD data from Table 7 of Ref.~\cite{alicemult} (open squares), are better described by the TCM trend than by the power-law trend.  

\subsection{Accepted vs corrected multiplicities and ratios}

In App.~\ref{ppprdspecrat} accurate knowledge of the $\bar \rho_s$ values for different event classes was essential to process spectrum ratios and interpret the results. The corresponding information in Ref.~\cite{alicespec} seems incomplete.
The relation to accepted multiplicities $N^{acc}_{ch}$ should be as follows. The acceptance factor corresponding to a \pt\ acceptance cutoff near 0.15 GeV/c is $0.80\pm 0.02$ according to Fig.~2 (right) of Ref.~\cite{tomalicempt}. That factor should change little with collision energy since it is determined mainly by fixed slope parameter $T \approx 145$ MeV. The factor corresponding to mean tracking efficiency is  $0.70\pm0.03$ averaged over the accepted \pt\ interval, based on Sec.~3 of Ref.~\cite{alicespec}. Collision-energy dependence should again be small. The overall acceptance factor should then be $0.56\pm0.04 \approx 0.6$. 

In Ref.~\cite{alicespec} the reported fully-corrected charge density for the INEL $>0$ event class is  $dn_{ch} / d\eta$ = 6.46 within $|\eta| < 0.5$ and 6.61 within $|\eta| < 1.0$. For the spectrum study the {\em accepted} charge density within $|\eta| < 0.8$ is $n_{ch}' / \Delta \eta = 6.73 / 1.6 = 4.2$ while the density (possibly) corrected for tracking efficiency but within $p_t > 0.15$ GeV/c is  $n_{ch}'' / \Delta \eta = 9.41 / 1.6 = 5.88$ (``from the spectrum in Fig.~3''). Interpolated to $|\eta| < 0.8$  the fully-corrected charge density should be $dn_{ch} / d\eta \approx 6.55$, and the implied overall acceptance factor is then $4.2 / 6.55 = 0.64$, somewhat higher than the expected 0.57. The ratio 5.88 / 6.55 = 0.90 is also substantially higher than the expected \pt-acceptance factor 0.80. However, 4.2/5.88 = 0.71 is consistent with the expected tracking efficiency 0.70. Although not self-consistent the acceptance factors should cancel in the multiplicity {\em ratios} 0.48 (A), 1.45 (B) and 2.4 (C) calculated above. However, there seem to be substantial differences between those values and what can be inferred directly from spectrum ratios in Sec.~\ref{13tevspecc}.

\end{appendix}


\end{document}